\documentclass[useAMS,usenatbib]{mnras}
\usepackage{graphicx}
\usepackage{amssymb}
\usepackage{amsmath}
\usepackage[authoryear]{natbib}
\usepackage{color}
\usepackage{ctable}
\usepackage{multirow}
\usepackage {threeparttable}
\usepackage{caption}
\usepackage{comment}

\newcommand{\mc}[3]{\multicolumn{#1}{#2}{#3}}

\include{jdf}
\bibpunct{(}{)}{;}{a}{}{,}
\title[LARGESS data catalogue and optical spectroscopy]{The Large Area Radio Galaxy Evolution Spectroscopic
  Survey (LARGESS): Survey design, data catalogue and GAMA/WiggleZ spectroscopy} 
\author[John H. Y.  Ching et al.]{
\parbox[t]{\textwidth}{ John H. Y. Ching$^1$, Elaine M. Sadler$^1$, Scott M. Croom$^1$, Helen M. Johnston$^1$, Michael B. Pracy$^1$, Warrick J. Couch$^2$, A. M. Hopkins$^2$, Russell J. Jurek$^3$, K. A. Pimbblet$^{4,5}$}
\vspace*{6pt}\\
$^1$Sydney Institute for Astronomy, School of Physics, University of Sydney, NSW 2006, Australia \\
$^2$Australian Astronomical Observatory, PO Box 915, North Ryde, NSW 1670, Australia \\
$^3$Australia Telescope National Facility, CSIRO, Epping, NSW 1710, Australia \\
$^4$Milne Centre for Astrophysics, University of Hull, Cottingham Road, Kingston-upon-Hull, HU6 7RX \\
$^5$School of Physics and Astronomy, Monash University, Clayton, VIC 3800, Australia \\
}

\begin{document}

\date{Received 0000; Accepted 0000}

\pagerange{\pageref{firstpage}--\pageref{lastpage}} \pubyear{2016}

\maketitle

\label{firstpage}

\begin{abstract}
We present the Large Area Radio Galaxy Evolution Spectroscopic Survey (LARGESS), a spectroscopic catalogue of radio sources designed to include the full range of radio AGN populations out to redshift $z\sim0.8$. 
The catalogue covers $\sim800$\,deg$^2$ of sky, and provides optical identifications for 19,179 radio sources from the 1.4\,GHz Faint Images of the Radio Sky at Twenty-cm (FIRST) survey down to an optical magnitude limit of $i_{\rm mod} < 20.5$ in Sloan Digital Sky Survey (SDSS) images. Both galaxies and point-like objects are included, and no colour cuts are applied. 
In collaboration with the WiggleZ and Galaxy And Mass Assembly (GAMA) spectroscopic survey teams, we have obtained new spectra for over 5,000 objects in the LARGESS sample. 
Combining these new spectra with data from earlier surveys provides spectroscopic data for 12,329 radio sources in the survey area, of which 10,856 have reliable redshifts. 
85\% of the LARGESS spectroscopic sample are radio AGN (median redshift $z=0.44$), and 15\% are nearby star-forming galaxies (median $z=0.08$).  Low-excitation radio galaxies (LERGs) comprise the majority (83\%) of LARGESS radio AGN at $z<0.8$, with 12\% being high-excitation radio galaxies (HERGs) and 5\% radio-loud  QSOs. Unlike the more homogeneous LERG and QSO sub-populations, HERGs are a heterogeneous class of objects with relatively blue optical colours and a wide dispersion in mid-infrared colours.  This is consistent with a picture in which most HERGs are hosted by galaxies with recent or ongoing star formation as well as a classical accretion disk. 
\end{abstract}


\section{Introduction}
Over the past fifteen years, large surveys at optical, infrared and radio wavelengths have allowed us to make significant progress in understanding the typical radio properties of galaxies in the local and distant Universe. Two large-area radio surveys carried out by the  Very Large Array (VLA) operated by the National Radio Astronomy Observatory (NRAO),  the Faint Images of the Radio Sky at Twenty-cm \citep[FIRST;][]{becker95} and the NRAO VLA Sky Survey \citep[NVSS;][]{condon98} have been particularly influential. Both are 1.4 GHz continuum radio surveys  covering a large fraction of the sky down to milli-Jansky flux densities. The high resolution and positional accuracy of the FIRST survey is complemented by the lower resolution of NVSS, which has better surface brightness sensitivity. Several studies \citep[e.g.][]{sadler02,hopkins03,best05,mauch07,best12} have matched NVSS and FIRST radio sources to counterparts in the optical or infrared. These optical/infrared identifications, combined with spectroscopic information such as redshifts, emission line and absorption line measurements, have advanced our understanding of the physical processes responsible for radio emission from nearby galaxies.

For extragalactic radio sources, the radio continuum emission may arise from either an active galactic nucleus (AGN) or processes related to star formation. 
In star-forming galaxies, the observed radio emission is usually dominated by synchrotron emission from relativistic electrons accelerated by supernova remnants in \textsc{Hii} regions, with a smaller contribution from thermal free-free emission \citep{condon92}. The short-lived massive stars in the \textsc{Hii} regions of star-forming galaxies photoionize the surrounding gas and produce a characteristic pattern of emission lines in the observed spectrum. 

Spectroscopic studies of radio AGN reveal two main populations: those with prominent optical emission lines, and those with weak or no emission lines
\citep{longair79,laing94}. We follow current practice and refer to the first (strong emission-line) population as high-excitation radio galaxies (HERGs) and the second as low-excitation radio galaxies (LERGs). The difference between these two populations is thought to reflect differences in the accretion efficiency of gas onto the central black hole \citep{hardcastle07}. A comprehensive review of the properties of the two classes of radio AGN is given by \cite{heckman14}. 
 
In the current paradigm, the HERGs undergo {\em cold-mode} (also known as {\it radiative mode}) accretion, characterised by a high accretion efficiency such that gas is accreted rapidly onto the galaxy's central black hole. This allows the formation of a radiatively-efficient accretion disk that photoionizes the surrounding gas to produce the observed high-excitation emission lines. The term cold-mode refers to the past temperature of the gas, which in this case has never reached the virial temperature of the halo \citep{keres09}. The LERGs on the other hand undergo {\em hot-mode} (also known as {\it jet-mode}) accretion, where the gas has at least reached the virial temperature in the past and is generally cooling from a surrounding hot X-ray corona. This is an inefficient accretion process without a radiatively efficient accretion disk, so the optical spectra of LERGs show weak or no emission lines.

Hot-mode accretion is expected to occur in high halo-mass systems \citep[$>2-3 \times 10^{11} M_{\sun}$;][]{keres09}, particularly at low redshift, and cold-mode accretion in lower-mass systems over a wider range in redshift \citep{hardcastle07,van-de-voort11}. Recent observational studies of the properties \citep{best05a,smolcic09a,janssen12}, environments \citep{best04,bardelli10,gendre13,sabater13} and evolution \citep{smolcic09,best12} of HERGs and LERGs appear to confirm this picture, showing that HERGs are typically found in lower-mass galaxies with younger stellar populations, and in poorer environments than the LERGs, which are typically in the most massive galaxies, with an old stellar population, and found in rich environments. 

%Table 1
% Table of survey regions
\ctable[
  notespar,
  star,
  cap = {Spectroscopic survey regions},
  caption={Regions covered by optical spectroscopic surveys. The coverage and overlap was calculated using the Virtual Observatory footprint service \citep[][http://www.voservices.net/footprint]{budavari07}.},
  label={tab:regions},
  ]{l r rr rr r rc cc}%
{ \tnote[a]{The official WiggleZ limit for this region has a maximum $\delta=8.0$; the additional 0.1 was mistakenly put into the original search. However, since the WiggleZ pointings included  this extra small area, we include these objects as well.}\\
\tnote[b]{The actual 2SLAQ regions are several small strips along the equatorial region, but for simplicity we adopt the two large pseudo-2SLAQ strips shown here.} }
{     \FL
	& \multicolumn{1}{c}{Field} & \multicolumn{2}{c}{R.A. (deg)} & \multicolumn{2}{c}{$\delta$ (deg)} & Total Area & \multicolumn{2}{c}{FIRST-SDSS overlap} &\multicolumn{2}{c}{Spectral completeness to $i=20.5$}\\
	\cline{8-9} \cline{10-11} 
    	Survey & \multicolumn{1}{c}{ID} & min & max & min & max & (deg$^{2}$) & (deg$^{2}$) & Fraction & Spectrum  & Redshift \\
       &&&&&&&&& observed & success rate \\
	\hline\hline
     	WiggleZ & & \\
    	& 0h & 350.1 &  359.1  & $-$13.4 & $+$1.8 & 136.0 & 44.7 & 0.33 & 55\%  & 89\% \\
    	& 1h & 7.5 &  20.6       & $-$3.7 &  $+$5.3 & 118.3 & 32.7 & 0.28 & 64\% & 95\%  \\
    	& 3h & 43.0 &  52.2     & $-$18.6 & $-$5.7 & 116.0 & 7.2 & 0.06 & 55\% & 86\%  \\
    	& 9h & 133.7 & 148.8 & $-$1.0 & $+$8.1\rlap{$^{a}$} & 137.8 & 136.7 & 0.99 & 71\% & 84\% \\
    	& 11h & 153.0 &  172.0  & $-$1.0 &  $+$8.0 & 172.1 & 172.1 & 1.00 & 66\% & 84\% \\
        & 15h & 210.0 & 230.0 & $-$3.0 & $+$7.0 & 201.7 & 200.0 & 0.99  & 63\% & 88\% \\
    	& 22h & 320.4 & 330.2  & $-$5.0 &  $+$4.8 & 96.2 & 24.5 & 0.25  & 73\% & 88\% \\
    	GAMA &  &  \\
    	& 9h & 129.0 & 141.0 & $-$1.0 &  $+$3.0 & 48.2 & 48.2 & 1.00  & 91\% & 94\% \\
    	& 12h &  174.0 & 186.0 & $-$2.0 &  $+$2.0 & 48.2 & 48.2 & 1.00  & 86\% & 88\% \\
    	& 15h &  211.5 & 223.5 & $-$2.0 &  $+$2.0 & 48.2 & 48.2 & 1.00 & 86\% &  89\% \\
    	2SLAQ\tmark[b] &  &  \\
    	& - & 123.0 & 230.0 & $-$1.259 & $+$0.840 & 325.0 & 301.6 & 0.93 & 65\% & 91\% \\
    	& - &  309.0 & 59.70 & $-$1.259 & $+$0.840 & 347.9 & 224.3 & 0.64 & 57\% & 93\%
\LL
}

The most powerful radio sources are known to undergo strong cosmic evolution, with their volume density at redshift $z\sim2$\ being up to a thousand times higher than it is today  \citep[e.g.][]{longair66,dunlop90}.  The cosmic evolution of  lower-power radio AGN appears to be much less rapid \citep{sadler07,donoso09,simpson12}, but is only just starting to be mapped out separately for the HERG and LERG sub-populations beyond the local Universe \citep{best14}. 

Our aim in undertaking the work described in this paper was to produce a new, large and complete spectroscopic radio-source catalogue that would allow us to track the HERG and LERG populations in detail over a wide range in radio luminosity back to at least redshift $z\sim0.8$ (i.e. a lookback time almost half the age of the Universe) as well as studying the radio galaxy and radio-loud QSO populations across a common range in redshift. There is growing evidence that the redshift evolution of the HERG and LERG populations is very different \citep[e.g.][]{best12,simpson12,best14}, and that observed luminosity-dependent cosmic evolution of the radio luminosity function is driven mainly by the different cosmic evolution of these two populations \citep{heckman14}. 

One key motivation for this new study arose from earlier work on the evolving radio AGN luminosity function carried out by \cite{sadler07} and \cite{donoso09}. These authors used relatively large samples of radio-detected AGN (391 objects in the \cite{sadler07} spectroscopic sample; 14,453 objects with photometric redshifts in the larger-area \cite{donoso09} sample) to measure radio luminosity functions in the redshift range $0.4<z<0.7$ with unprecedented accuracy.  Both samples were photometrically selected to target luminous red galaxies \citep[LRGs; ][]{eisenstein01} but exclude blue galaxies with ongoing star-formation. \cite{sadler07} explicitly noted that the rate of cosmic evolution measured for low-power radio galaxies in their study was only a lower limit, since the LRG sample they used had a strict colour cut-off, whereas no such colour restriction was applied to the $z\sim0$\ radio galaxy sample used as the local benchmark. 
By compiling a new sample of distant radio AGN without any pre-selection on colour, we wanted to find out whether there is indeed a significant population of `blue' radio galaxies in the distant Universe and (if so) how their properties compare with the better-studied population of `red' radio galaxies. 

The data catalogue presented in this paper includes over 10,000 spectroscopically-observed radio sources, with a median redshift of $z\sim0.44$ for the radio AGN which make up $\sim85$\%\ of the sample. Our sample of 2281 radio-source spectra at $0.5<z<1$ represents an order-of-magnitude increase over previous spectroscopic samples in this redshift range. For example, the recent  \cite{best14}\ measurement of the radio luminosity function out to $z=1$ used a catalogue of 211 radio-loud AGN at $0.5<z<1.0$, while the \cite{simpson12} measurement used $\sim$100 spectroscopically-observed objects in the same redshift range (supplemented by a similar number of photometric redshift estimates). A companion paper by \cite{pracy06}  uses the dataset presented here to make new measurements of the evolving radio luminosity functions of HERGs and LERGs out to redshift $z\sim0.8$. 

We describe the optical and radio catalogues used to compile our sample in \S\ref{sec:catalogues} and the radio-optical matching process in \S\ref{sec:sdss_first_matching} and \S\ref{sec:nvss_matching}.  The spectroscopic follow-up program is discussed in \S\ref{sec:spec_data}, and the related completeness analysis presented in \S\ref{sec:spec_completeness}. \S\ref{sec:spec_class} describes the identification of star-forming galaxies and the classification of high- and low- excitation radio galaxies, and \S8 presents the full data catalogue. The full sample and some sub-samples are characterised in \S\ref{sec:sample_character},  
while \S10 compares the properties of a matched sample of HERG and LERG host galaxies. 
Finally, we present a summary of the LARGESS sample properties in \S\ref{sec:summary1}. 

Throughout this paper we adopt the cosmological parameters $H_{0}=70$ km s$^{-1}$ Mpc$^{-1}$, $\Omega_{\Lambda}=0.7$ and $\Omega_{m}=0.3$. All optical magnitudes are corrected for Galactic dust extinction and {\em k}-corrected using the \textsc{kcorrect} code \citep{blanton07}. An analysis of a subset of the LARGESS-GAMA (Galaxy And Mass Assembly) sample \citep{hardcastle13} found a mean radio spectral index (between 325 MHz and 1.4 GHz) of $\alpha = -0.7$ (where $S_{\nu} \propto \nu^{\alpha}$), and we adopt this value to calculate a radio {\em k}-correction. 

%Figure 1
\begin{figure*}
\centering
\includegraphics[width=0.85\textwidth]{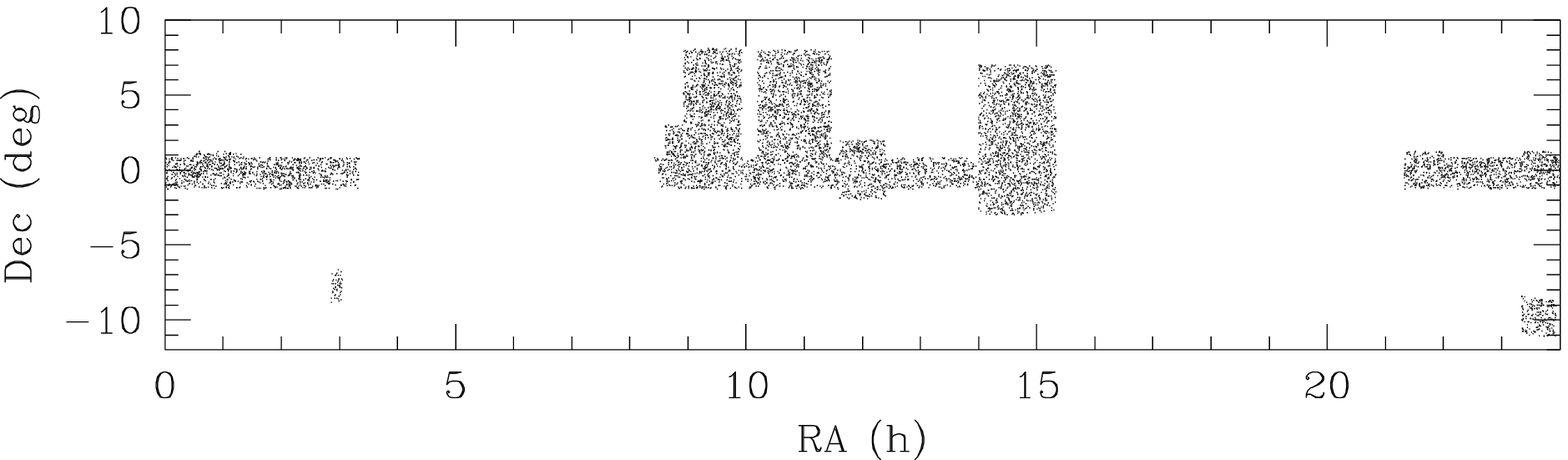}
\includegraphics[width=0.85\textwidth]{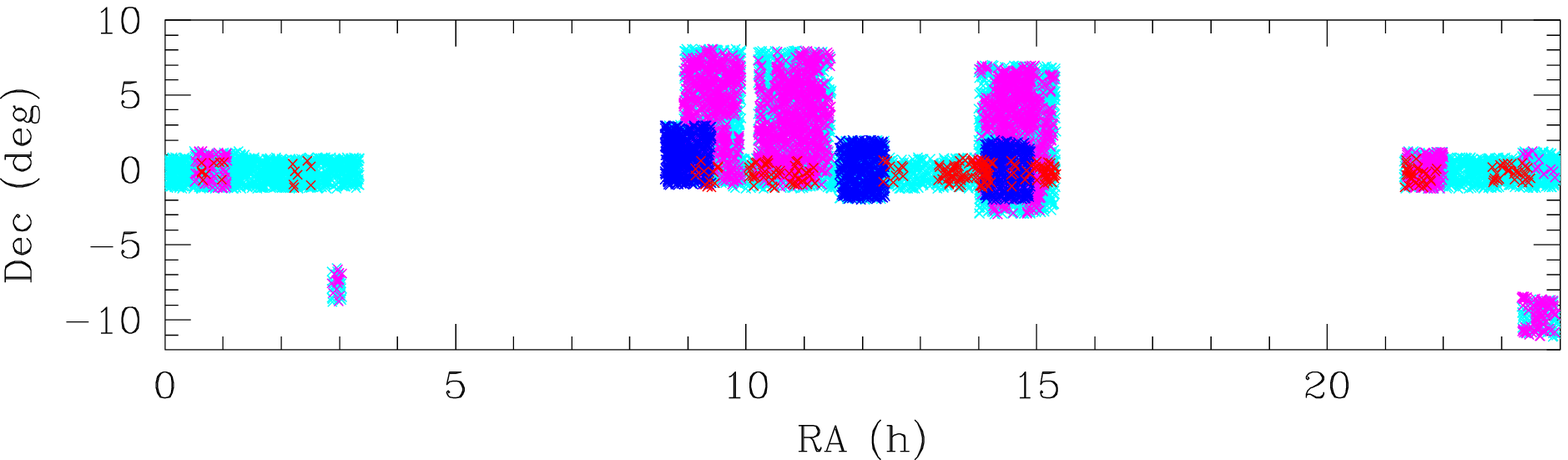}
\caption[
The LARGESS sky coverage]{Sky area covered by the LARGESS radio sample: (top) sky distribution of the full photometric catalogue of 19,179 radio-sources matched to optical objects with $i<20.5$\,mag, (bottom) distribution of the 10,764 objects in the final spectroscopic catalogue that currently have good-quality optical spectra and redshifts. Points are colour-coded according to the source of the redshift measurement: GAMA (dark blue), WiggleZ (magenta), SDSS (cyan), 2SLAQ (red).}
\label{fig:coverage:specradio}
\end{figure*}

\section{Optical and Radio Catalogues}\label{sec:catalogues}

\subsection{The SDSS photometric sample}\label{subsec:sdss_sample}
The Sloan Digital Sky Survey \citep[SDSS;][]{york00} is a large imaging and spectroscopic survey, covering five optical bands: $ugriz$ \citep{fukugita96}. We used the Sixth Data Release of the SDSS \citep[SDSS DR6,][]{adelman-mccarthy08}, which contains images and parameters for about 287 million objects over an area of 9,583 deg$^2$.

The SDSS 95\% detection repeatability for stars in the $i$-band is at 21.3 mag \citep{stoughton02}. We adopted a more conservative limiting $i$-band extinction-corrected magnitude limit of $i_{\rm mod}\leq 20.5$ for our optical sample, since observational constraints made it difficult for us to obtain reliable redshifts for objects fainter than this. 

Our optical catalogue covers the sky area defined in Table \ref{tab:regions} and shown in Figure \ref{fig:coverage:specradio}. 
This region contains over 8 million SDSS DR6 sources, along with SDSS optical spectra for objects brighter than the SDSS spectroscopic survey limit of $r_{\rm pet}<17.7$\,mag. It also overlaps with several other large spectroscopic surveys that probe to fainter magnitude limits (and higher redshifts) than SDSS: 2SLAQ \citep{cannon06}, GAMA \citep{driver11} and WiggleZ \citep{drinkwater10}.  

In collaboration with the GAMA and WiggleZ teams, we were able to make additional spectroscopic observations (beyond the planned public surveys) for radio-selected objects in the GAMA and WiggleZ survey regions, as discussed in more detail in \S5. 

\subsection{The FIRST and NVSS radio surveys}

FIRST \citep{becker95} and NVSS \citep{condon98} are 1.4 GHz continuum surveys carried out on the Very Large Array (VLA). The FIRST survey covered over 9,000 deg$^{2}$ of the Northern (8,444 deg$^{2}$ ) and Southern (611 deg$^{2}$ ) Galactic caps, mainly overlapping with the SDSS coverage. 

The FIRST survey used the VLA B-configuration, which provides a resolution of 5 arcsec full width at half maximum (FWHM), with a typical root-mean-square noise ($\sigma_{\rm rms}$) of $\sim$0.15 mJy. The positional accuracy of FIRST sources is $<$1 arcsec at the survey threshold.
The typical detection threshold of the FIRST survey is $\sim1$ mJy, though co-added observations at two epochs along the equatorial region (R.A. = 21\fh3 to 3\fh3, $\delta$ = $-1$\degr~to $+1$\degr) enabled the detection threshold to drop to $\sim0.75$ mJy. We use the July 2008 release of the FIRST catalogue, which only contains sources with peak flux density (after correcting for CLEAN bias) greater than five times the local $\sigma_{\rm rms}$ at that point (i.e. $S^{\rm FIRST}_{\rm peak}-0.25 > 5 \sigma_{\rm rms}$) and peak flux density $S^{\rm FIRST}_{\rm peak}\geq$ 0.75 mJy.

The NVSS covers the sky north of $\delta=-40$\degr. The NVSS observations were carried out in the D and DnC configurations to provide a resolution of 45 arcsec FWHM. The lower resolution provides better surface-brightness sensitivity than the FIRST survey, but with poorer positional accuracy. The typical rms noise in the NVSS images is $\sim$0.45 mJy beam$^{-1}$ with a catalogue completeness limit of $\sim$2.5 mJy. 

\section{FIRST-SDSS matching}\label{sec:sdss_first_matching}

The techniques for matching FIRST and SDSS sources are now well-established  at low redshift \citep[e.g.][]{best05,sadler07,best12}.  Our approach is similar, except that 
we are matching to a fainter optical limit than earlier studies. For example, the surface density of galaxies in our $i_{\rm mod}\leq20.5$ optical sample is $\sim$9,300 deg$^{-2}$, i.e. over 50 times higher than the $\sim$170 deg$^{-2}$ surface density of the \cite{best12} sample. 

%Figure 2
\begin{figure*}
\centering
\begin{minipage}\textwidth
\includegraphics[height=0.41\textwidth,angle=-90]{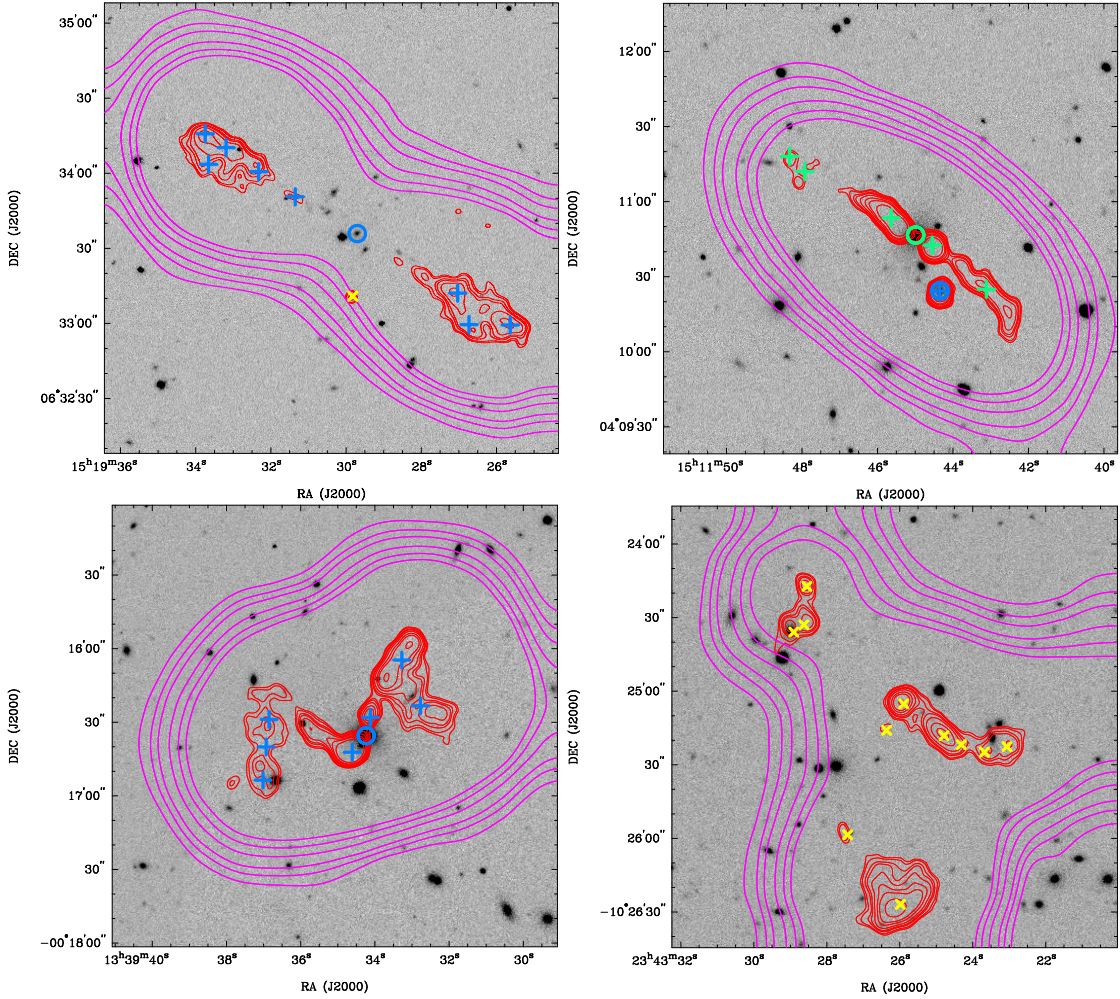}
\includegraphics[height=0.41\textwidth,angle=-90]{./figures/9267.eps}
\end{minipage}
\begin{minipage}\textwidth
\includegraphics[height=0.41\textwidth,angle=-90]{./figures/3715.eps}
\includegraphics[height=0.41\textwidth,angle=-90]{./figures/67.eps}
\end{minipage}

\caption[Examples of complex FIRST groups that warranted visual inspection]{Four examples of complex FIRST source groups that were matched visually. SDSS $i$-band greyscale images are overlaid with FIRST (red) and NVSS (magenta) contours. Open circles show the positions of reliable optical matches to the FIRST components (plus symbols), and correspond in colour if they belong to the same source group. FIRST components without an optical counterpart are shown by yellow crosses. {\it Top left:}\ A reliable optical match ($\emph{P}=3$, as defined in \S\ref{subsubsec:vis_match}) to a group of eight FIRST components, with one additional unmatched FIRST source (yellow cross). {\it Top right:}\ Two reliable optical matches ($\emph{P}=4$ for both), one to a group of five FIRST components and the other to a single FIRST source. {\it Bottom left:}\  An optical match ($\emph{P}=3$) to a complex group of seven FIRST sources.  
{\it Bottom right:}\  A complex group of FIRST sources where there is no optical counterpart with $i_{\rm mod}\leq20.5$.}
\label{fig:radio_complex}
\end{figure*}

\subsection{Identifying multi-component FIRST sources} 
Around 10\% of FIRST radio sources have complex, extended radio morphology resolved into several components in the FIRST catalogue \citep[e.g.][]{ivezic02}. 
To identify the optical counterparts of these extended sources, we combine the collapsing technique introduced by \cite{cress96} and refined by \cite{magliocchetti98} with the tiered algorithm used by several authors \citep{best05,sadler07,donoso09}. We start by identifying the most complex multi-component sources, and then work down to simpler systems with fewer radio components, where an optical identification is more straightforward. 

\cite{cress96} showed that about 30\% of all FIRST sources lie within 72\,arcsec of another FIRST source, and considered these to be mainly genuine associations. 
\cite{magliocchetti98} later showed that some of the \cite{cress96} groups were actually unrelated sources that happen to lie close in projection on the sky. To reduce the number of spurious matches, \cite{magliocchetti98} applied additional constraints to decide whether or not a group of FIRST sources was part of a single system. Their constraints were motivated by known properties of radio sources, such as the ratio of the integrated flux density between the lobes and the flux-separation relation \citep{Oort87} for extended sources.

Following \cite{cress96}, we identified and grouped all FIRST sources with a separation of $\leq 72$ arcsec on the sky (groups can span $>72$ arcsec in total). We then applied a range of further tests to groups of two or more sources to determine whether they were likely to be associated with a common optical counterpart. We used Monte Carlo techniques both to set appropriate selection parameters and to estimate the reliability of our final set of matches, as described in \S\ref{subsubsec:monte_carlo}.
 
\subsection{Visual matching of complex sources}\label{subsubsec:vis_match}
We visually inspected all groups of {\it four or more FIRST sources}, since these are too complex for reliable automated matching. Figure \ref{fig:radio_complex} shows some examples of these complex source groups. 
As noted below, visual matching was also used for some groups with two or three FIRST components.

The information used for visual matching included the SDSS $i$-band image, FIRST contours and/or greyscale image, NVSS contours and/or greyscale image, positions of FIRST sources in the field and positions of SDSS sources with $i_{\rm mod}\leq20.5$. The user selected the most appropriate optical counterpart (which may or may not be in our $i_{\rm mod}\leq20.5$  optical catalogue) for each FIRST source, and assigned a quality code ({\em P}-value), ranging from 1 to 4, to quantify the confidence of each match. Optical identifications with $\text{\em P}\geq3$ are considered reliable enough to use in later analysis. 

We used a blind test to estimate the confidence of visual matches with $\text{\em P}\geq3$. To do this we took a sample of 135 FIRST radio groups and conducted a visual analysis where a random half of the sources was matched with the real sky at the position of the source and the other half matched with a random sky image at a different position. In all, we identified 25 optical counterparts with $\text{\em P}=3$, and 35 optical with $\text{\em P}=4$. For the visual matches with $\text{\em P}=3$, 5/25 identifications came from the random sky image rather than the real one. For visual matches with $\text{\em P}=4$, only 2/35 identifications were from the random image. From this, we estimate rough confidence levels of about 80\% (for $\text{\em P}=3$) and 94\% (for $\text{\em P}=4$) for our visual identifications of the most complex FIRST sources. 

%Figure 3
\begin{figure*}
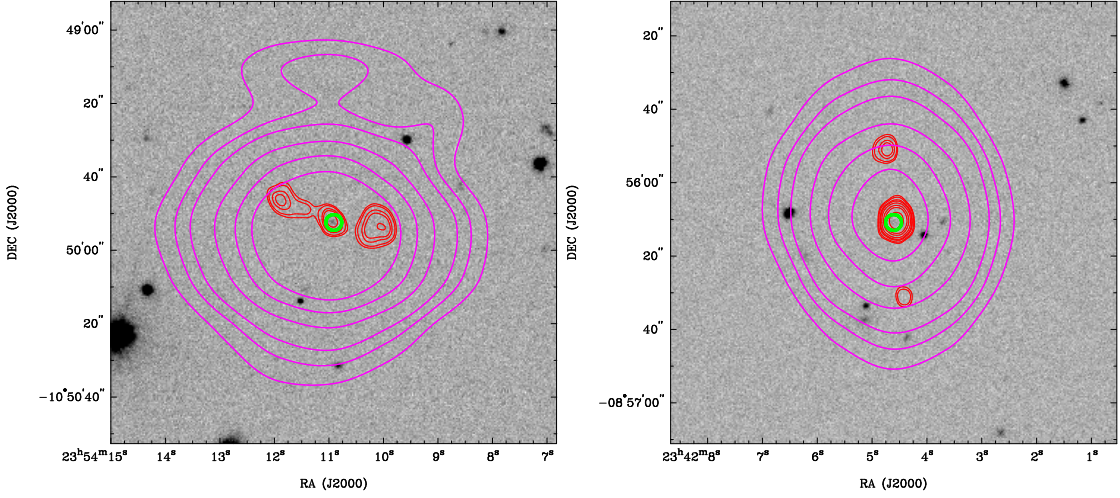

\includegraphics[height=0.41\textwidth,angle=-90]{./figures/34-FJ235410.911-104952.43.ps}
\includegraphics[height=0.41\textwidth,angle=-90]{./figures/202-FJ234204.609-085610.86.ps}
\caption[Two examples of FIRST groups with three members that appear to be genuinely associated, but the optical counterpart is fainter than the LARGESS optical limit]{Two examples of FIRST groups (red) with three radio components where the optical counterpart is identified (green) but is fainter than our optical limit of $i_{\rm mod} \leq 20.5$.  Objects like this are not included in the final LARGESS catalogue. }\label{fig:ftrip_noopt}
\end{figure*}

\subsection{Automated matching of groups of FIRST sources}\label{subsubsec:vis_auto23}
\subsubsection{Groups of three FIRST sources} 
Groups of three FIRST components associated with a single host galaxy are likely to contain a core and two lobe components. We chose to define the core (middle) component as the one with the smallest angular distance from the other two FIRST members.  The remaining two members were assumed to be lobe components (F1 and F2). 

To accept the group as a single source, we required the total integrated FIRST flux densities ($S_{int}$) of the lobe components F1 and F2 to be within a factor of three of each other, i.e. 
\begin{equation}\label{eq:flux_ratio}
1/3\leq {S_{\rm int}}({\rm {F1})/{S_{\rm int}}(F2)} \leq 3
\end{equation}
This is tighter than the factor of four limit used by \cite{magliocchetti98}, and increases the reliability of the group as a genuine double-lobe plus core radio galaxy. For groups that satisfied this test, we assigned the position of the group as the position of the core component and matched this position to the SDSS catalogue. Groups that did not satisfy the flux-ratio test were reclassified as candidate double sources after removing the lobe component with the largest difference in flux density from the middle component. 
Matches to an SDSS optical object were automatically accepted at this stage if: 

\vspace*{-0.2 cm}
\begin{enumerate}
  \item 
  $\theta_{\rm match} < 3$\,arcsec (where $\theta_{\rm match}$ is the offset between the radio centroid and the closest SDSS object), 
  \item 
  neither lobe component has an SDSS object within 2.5\, arcsec, and 
  \item 
  the shortest component separation ($\theta_{c}$) is at least one-third of the longest value ($\theta_{a}$), i.e. $\theta_{c} \geq 0.33\times\theta_{a}$, to ensure that the core component is reasonably close to the radio centroid.
\end{enumerate}

\vspace*{-0.1cm}
We also visually inspected all triple sources that satisfied the following slightly looser criteria: 
\begin{enumerate}
  \item $3 < \theta_{\rm match} < 5$\,arcsec and neither lobe component has an SDSS object within 2.5\, arcsec; or
  \item $\theta_{\rm match} < 3$\,arcsec and neither lobe component has an SDSS object within 2.5\, arcsec, but $\theta_{c} < 0.3\times\theta_{a}$; or
  \item  $\theta_{\rm match} < 2.5$\,arcsec and  $\theta_{c} \geq 0.3\times\theta_{a}$, but one lobe component has an SDSS source within 2.5\,arcsec. 
\end{enumerate}
\vspace*{-0.1cm}
This visual matching added 113 triple-source matches to the 266 found by automated matching. In addition, we identified some FIRST triple groups that were genuine double lobe-core systems with an optical counterpart fainter than our survey limit of $i_{\rm mod} = 20.5$. 
Figure \ref{fig:ftrip_noopt} shows two examples.

\subsubsection{Groups of two FIRST sources}
Two FIRST sources associated with a single host galaxy are likely to be either a pair of lobes or a core and hotspot. We accepted pairs of FIRST sources (F1 and F2) as a genuine association if the integrated flux density ratio of the two components was within a factor of three (i.e. satisfied equation\,\ref{eq:flux_ratio}\ above) 
and the pair also satisfied an additional test set out by \citet{magliocchetti98}, i.e. 
\begin{equation}\label{eq:mag1} 
\theta_{\rm pair} \leq 100\times \sqrt{S_{\rm tot}/100}, 
\end{equation}
where $\theta_{\rm pair}$ is the separation between the two FIRST sources in arcsec, and S$_{\rm tot}$ is the sum of the integrated flux densities of the two components S$_{\rm int}$(F1) and S$_{\rm int}$(F2) in mJy.  Adding this constraint allows us to combine bright subcomponents even at relatively large separation, while keeping faint sources as single objects. 

Matches to an SDSS optical object were automatically accepted at this stage if either:
\vspace*{-0.2 cm}
\begin{enumerate}
  \item $\theta_{\rm match} < 3$\,arcsec, neither FIRST component has an SDSS object within 2.5\, arcsec and $\theta_{\rm match}\leq \theta_{\rm pair}/2$ (where $\theta_{\rm match}$ is the angular separation between the radio centroid and the closest optical object and $\theta_{\rm pair}$ is the separation of the two radio components), or 
 \item the matched SDSS object is within 2.5 arcsec of one FIRST component, the other FIRST component has no optical counterpart within 2.5 arcsec, and $\theta_{\rm match} \leq \theta_{\rm pair}/2$; 
\end{enumerate}
The first of these criteria picks out double-lobe radio galaxies, while the second identifies core-lobe systems. 

We visually inspected pairs of sources where: 
\vspace{-0.2cm}
\begin{enumerate}
  \item 3 arcsec $< \theta_{\rm match}\leq$ 5 arcsec, neither neither FIRST component has an SDSS object within 2.5\, arcsec and $\theta_{\rm match} \leq \theta_{\rm pair}/2$, or
  \item the matched optical source is within 2.5 arcsec of one FIRST components, the other FIRST component has no optical counterpart within 2.5 arcsec, and $\theta_{\rm match} > \theta_{\rm pair}/2.$
\end{enumerate}
This visual matching added 224 double-source matches to the 981 found by automated matching. 

\subsection{Automated matching of single FIRST sources} 
Finally, we carried out automated matching of the large number of FIRST sources not already identified as part of a multi-component system. 
For these sources, we accepted the closest SDSS optical match within 2.5\,arcsec.  This 2.5\,arcsec cutoff is more restrictive than the 3.0\,arcsec value adopted by 
\cite{sadler07} and \cite{helfand15}, but was chosen on the basis of Monte Carlo tests to optimize the completeness and reliability of the final catalogue, taking into account the high surface density of optical objects down to our magnitude limit of $i_{\rm mod} \leq 20.5$. 

\subsection{Summary of the cross-matching process}\label{subsec:fs_matchresults}

Table \ref{tab:n_first} summarizes the results of the cross-matching process for the full survey area (as well as the sub-area covered by the three GAMA fields listed in Table\ref{tab:regions}).  There are 19,179 optical identifications in the final catalogue of FIRST-SDSS matches across the full survey area, with a total of 22,438 FIRST components. 
These 19,179 radio-source IDs all have optical photometry and morphological parameters from the SDSS in five ($ugriz$) photometric bands, and comprise the main LARGESS sample with $i_{\rm mod} \leq 20.5$\,mag. 
The great majority of LARGESS objects (89.5\%) are single-component FIRST sources, with multi-component sources making up 10.5\% of the sample. This is similar to the fraction of FIRST-SDSS matches with complex morphology found by \cite{ivezic02}. 

For the three GAMA fields, which have complete overlap between the SDSS and FIRST surveys (see Table \ref{tab:regions}), we find SDSS matches with $i_{\rm mod}\leq20.5$\,mag for 3168 radio sources made up of 3727 FIRST components. 

Overall, 28.6\% of FIRST sources were matched with an SDSS object brighter than $i_{\rm mod}=20.5$, and this appears consistent with the matching rate of 32.9\% quoted by \cite{helfand15}\  for the full SDSS photometric catalogue (which has a slightly fainter optical limit). 

%Table 2
\ctable[
notespar,
cap = {Number of matched optical counterparts with different number of FIRST components},
caption = {Number of matched optical counterparts in the final catalogue, and in the GAMA sub-region used for completeness and reliability estimates.},
label = {tab:n_first}
]{l rr}%
{ %\tnote[a]{footnote}
 }
{ \FL 
\multicolumn{1}{c}{\multirow{2}{*}{Number of FIRST components}} & \multicolumn{2}{c}{Optical counterparts}\\
\cline{2-3}
 & \multicolumn{1}{c}{All} & \multicolumn{1}{c}{GAMA fields }\\
\hline\hline
One             & 17,163 & 2,803\\
Two             & 1,294  &   237\\
Three          &   454  &     88\\
Four or more &  269 &      40 
\LL
}

\subsection{Completeness and reliability of the matched catalogue}\label{subsubsec:monte_carlo}

We used Monte Carlo tests in the three GAMA fields (which are fully covered by all three imaging surveys: SDSS, FIRST and NVSS) 
to estimate the completeness and reliability of our matching technique. 

We generated five pseudo-random optical catalogues by offsetting the GAMA catalogue in declination using shifts of $\Delta\delta$=[0.3, 0.5, 1.0, 1.5, 2.5] deg. Objects shifted outside the GAMA regions in this process were wrapped around to the other side of each region, so that the total number and coverage of the random catalogues is the same as the test sample and the random catalogue retains most of the projected clustering of the original catalogue. The random catalogues were matched against the FIRST radio data in the same way as the real optical catalogue in the GAMA fields, and the matching results were scaled up by the ratio of GAMA to total sky areas (see Table \ref{tab:regions}) for comparison with the full sample. 

\subsubsection{Reliability} 
The reliability R of the final catalogue, i.e. the probability that a matched radio source is genuinely associated with an SDSS object rather than being a random projection on the sky, is calculated as: 
\begin{equation}\label{eq:reliab} 
\text{R}= ( 1 - \langle N^{\rm match}_{\rm rand}\rangle/N^{\rm match}_{\rm true}),
\end{equation}
where $N^{\rm match}_{\rm true}$ is the number of matches from the true catalogue, and $\langle N^{\rm match}_{\rm rand}\rangle$ is the average number of matches using the random catalogues. 

Comparing the final number of radio sources in our catalogue (19,179) with the average number of accepted matches ($\sim$1,247) from the random sample 
gives us an overall reliability of 93.5\% for the LARGESS sample. 

This is lower than the  value of $\sim$98\% for lower-redshift samples \citep[e.g.][]{best05,sadler07} because we are matching to fainter optical objects than previous studies (and also matching with both galaxies and stellar objects) and the higher surface density of optical objects means that the probability of a chance association is increased. At these faint magnitudes, a higher level of reliability could only be achieved by sacrificing completeness, and our final matching strategy was chosen to give a reasonable compromise between completeness and reliability. 

%Figure 4
\begin{figure}
\centering
\includegraphics[width=0.5\textwidth]{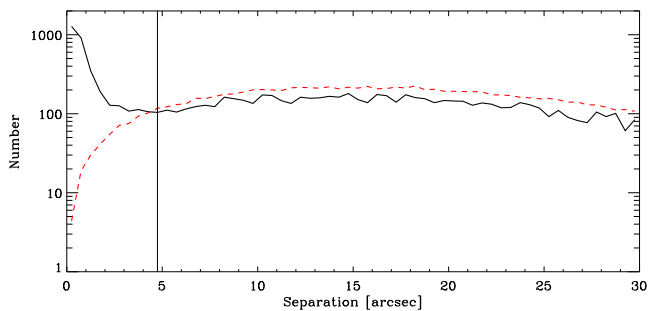}
\caption[Single FIRST-SDSS matches compared to FIRST-random positions as a function of separation]{FIRST-SDSS matches (black; solid line) for single-component sources, compared to the average number of random matches (red; dashed line) in the GAMA fields as a function of separation out to 30\,arcsec. 
The vertical line shows the separation at which the average number of FIRST-SDSS matches approaches the number of matches expected by chance.}
\label{fig:single_match}
\end{figure}

\subsubsection{Completeness} 
We also used the GAMA fields to estimate the completeness of the full LARGESS sample, i.e. the fraction of all genuine associations that are identified by our matching process. 

Figure \ref{fig:single_match} shows the number of FIRST-SDSS matches for single-component sources out to 30 arcsec separation, compared to the normalised random rate from Monte-Carlo catalogues. 
The ratio of the areas under the two curves gives a 5\% chance that a match with separation $<2.5$\,arcsec is coincidental, and there is also a small excess of genuine matches out to separations of 4.8 arcsec. 

The main source of incompleteness in our sample is the loss of genuine matches with radio-optical separations larger than our 2.5\,arcsec matching radius. Our Monte Carlo tests show that $\sim140$ genuine GAMA-FIRST associations will be missed by our 2.5\,arcsec cutoff, implying a completeness of $\sim$95\% for LARGESS sources with a single FIRST component.

For objects with multiple FIRST components, Monte Carlo tests give a slightly higher completeness level (97\%) for the optical matching due to the larger cutoff radius used for matches (3 \,arcsec, compared to 2.5\,arcsec for the single sources).  Set against this, the linking process used to associate multiple FIRST components may miss some genuine associations - though we expect this incompleteness to be small because of the high level of visual inspection used in checking the results. We therefore estimate that the completeness of the final sample ($\sim95$\%) is similar for single and multi-component FIRST sources, which is comparable to the completeness of previous radio samples \citep[e.g.][]{best05}.

\section{NVSS matching}\label{sec:nvss_matching}
We now have our final photometric catalogue of 19,179 FIRST radio sources identified with SDSS optical objects brighter than $i_{\rm mod} = 20.5$\,mag.  
Since the FIRST measurements may underestimate the total flux density of extended radio sources (see \S2.2), our next step was to cross-match with the NVSS catalogue to get a more reliable measurement of the integrated radio flux density for each object. 

\subsection{NVSS counterparts to FIRST sources}
 We used the FIRST components associated to each source in the LARGESS catalogue to cross-match between the FIRST and the NVSS catalogues, using a similar methodology to earlier studies \citep[e.g.][]{best05,sadler07,kimball08}. The cross-matching process is generally straightforward for objects associated with a single FIRST source. For more complex sources, the matching was done as follows: 

For NVSS sources with two or more FIRST matches within 45\,arcsec, we need to ensure that the NVSS flux density assigned to a FIRST-optical match is not artificially boosted by contributions from unrelated sources within the larger NVSS beam. To do this, we summed the integrated flux density from all FIRST components within 45 arcsec of an NVSS source. For each FIRST component, we then used its fractional contribution to the total FIRST flux density to assign a scaled proportion of the NVSS flux density to that component. 
For NVSS-FIRST matches where the FIRST source is associated with an optical counterpart, the NVSS flux density assigned to that FIRST source is now considered to be associated with the corresponding optical counterpart. 

\subsection{NVSS sources without a FIRST match} 
Around 8,000 NVSS sources in our survey regions did not have a FIRST source within 45 arcsec. We visually inspected the 1,299 NVSS components that lay within 3\,arcmin of one of our FIRST sources. As noted by \cite{best05}, this 3 arcmin radius is large enough to pick up any extended NVSS components, but smaller than the typical separation of unrelated NVSS sources (8--10\,arcmin). We found a further 159 NVSS components associated with 121 LARGESS objects (30 with two NVSS components and 4 with three NVSS components). In addition, 259 NVSS components were found to be associated with 252 (7 with two NVSS components) SDSS ($i_{\rm mod} \leq20.5$) objects that were not previously in the LARGESS catalogue. 

From this, we estimate that $\sim 20\%$ (252/1,299) of the $\sim8,000$ NVSS sources without a FIRST detection in our survey area will be associated with a SDSS ($i_{\rm mod} \leq20.5$) counterpart that is not already part of our final LARGESS sample. In other words, our survey area includes $\sim1,600$ faint radio sources that are too diffuse to be detected by the FIRST survey. 
A comparison with the \cite{best12} sample showed that excluding objects with a NVSS detection but no FIRST detection mainly excludes star-forming galaxies at low redshift, so does not significantly affect the completeness of our catalogue for radio AGN (HERGs and LERGs).

\subsection{Comparison of FIRST and NVSS flux densities} 
\begin{figure}
\centering
\includegraphics[width=0.5\textwidth]{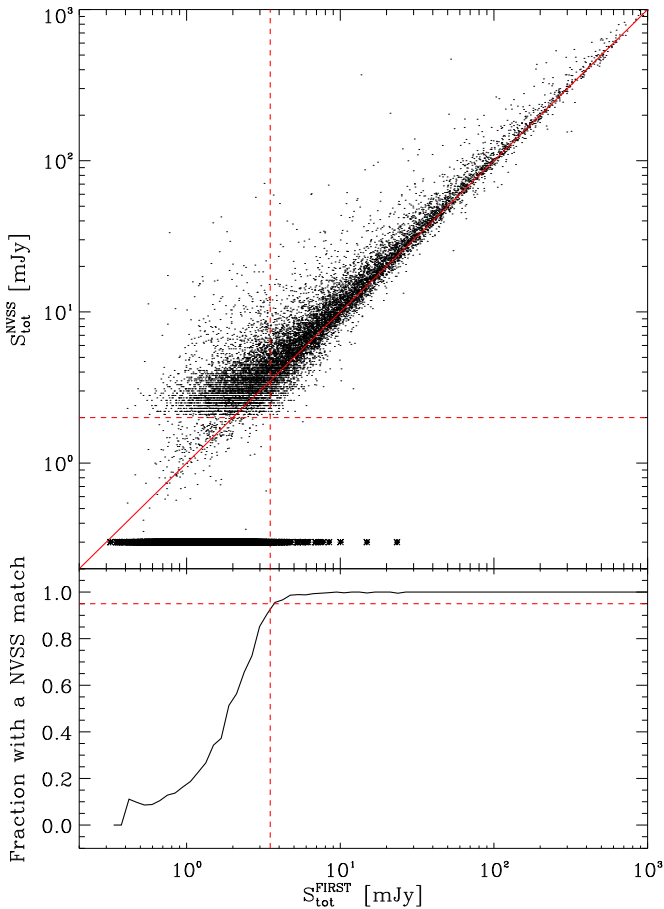}
\caption[Comparison of the total FIRST and NVSS flux densities, and completeness of FIRST-NVSS matches as a function of the total FIRST flux density]{ {\it Top panel:} Comparison of FIRST ($S^{\rm FIRST}_{\rm tot}$) and NVSS ($S^{\rm NVSS}_{\rm tot}$) total flux densities for LARGESS objects. The horizontal dashed line is at $S^{\rm NVSS}_{\rm tot}=2$ mJy --- some points lie below this line because their total NVSS flux density was adjusted to take into account multiple FIRST matches as discussed in the text. 
FIRST sources without a NVSS match are plotted at $S^{\rm NVSS}_{\rm tot}=0.3$ mJy, and the solid line shows a one-to-one relation in flux density. 
{\it Bottom panel:}  The fraction of objects with an NVSS match as a function of $S^{\rm FIRST}_{\rm tot}$. The horizontal dashed line is at 0.95, and a vertical dashed line marks the value of $S^{\rm FIRST}_{\rm tot}$ at this 95\% matching level. }
\label{fig:first_nvss}
\end{figure}

\begin{figure*}
\includegraphics[width=0.49\textwidth]{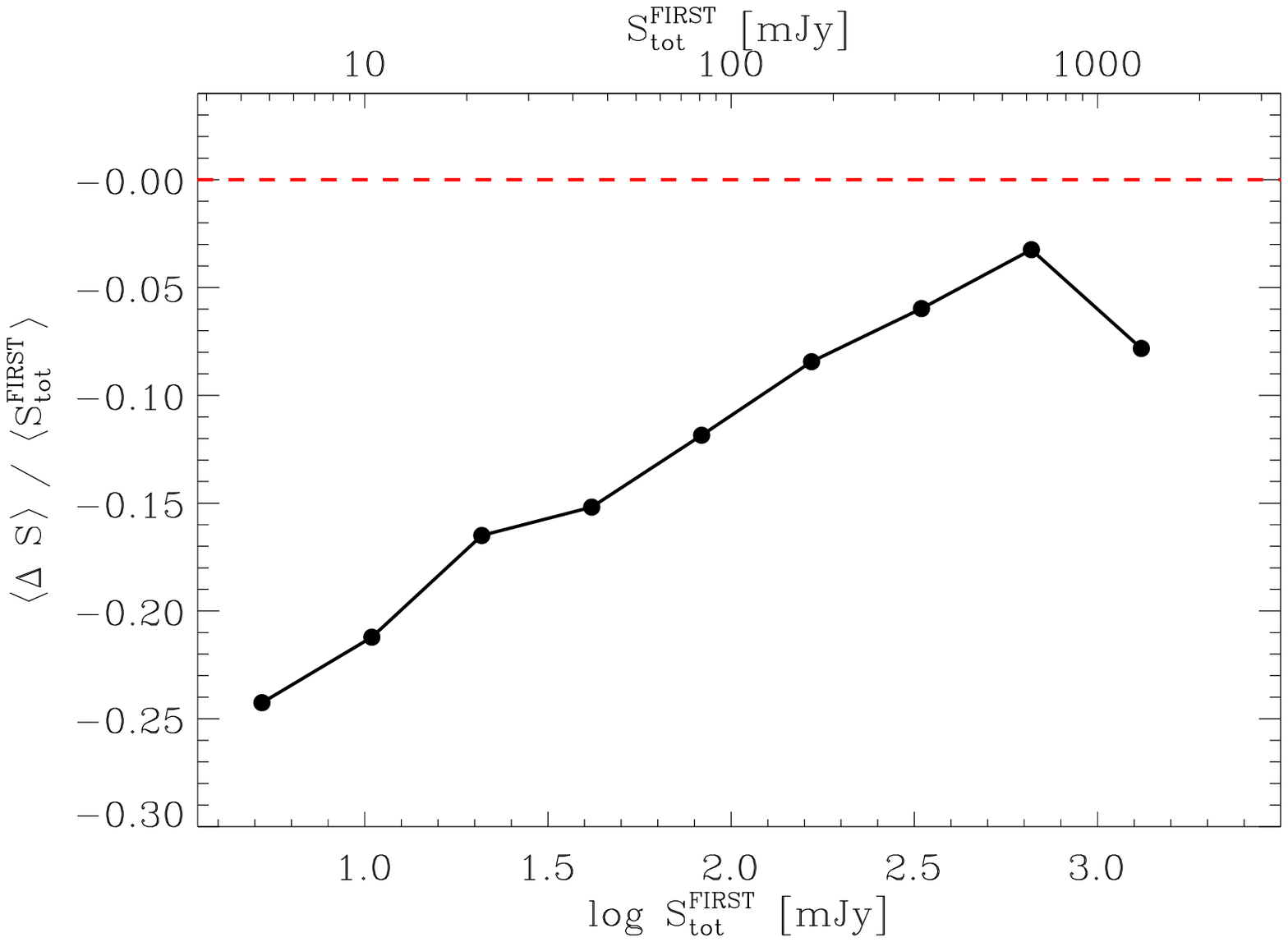}
\includegraphics[width=0.49\textwidth]{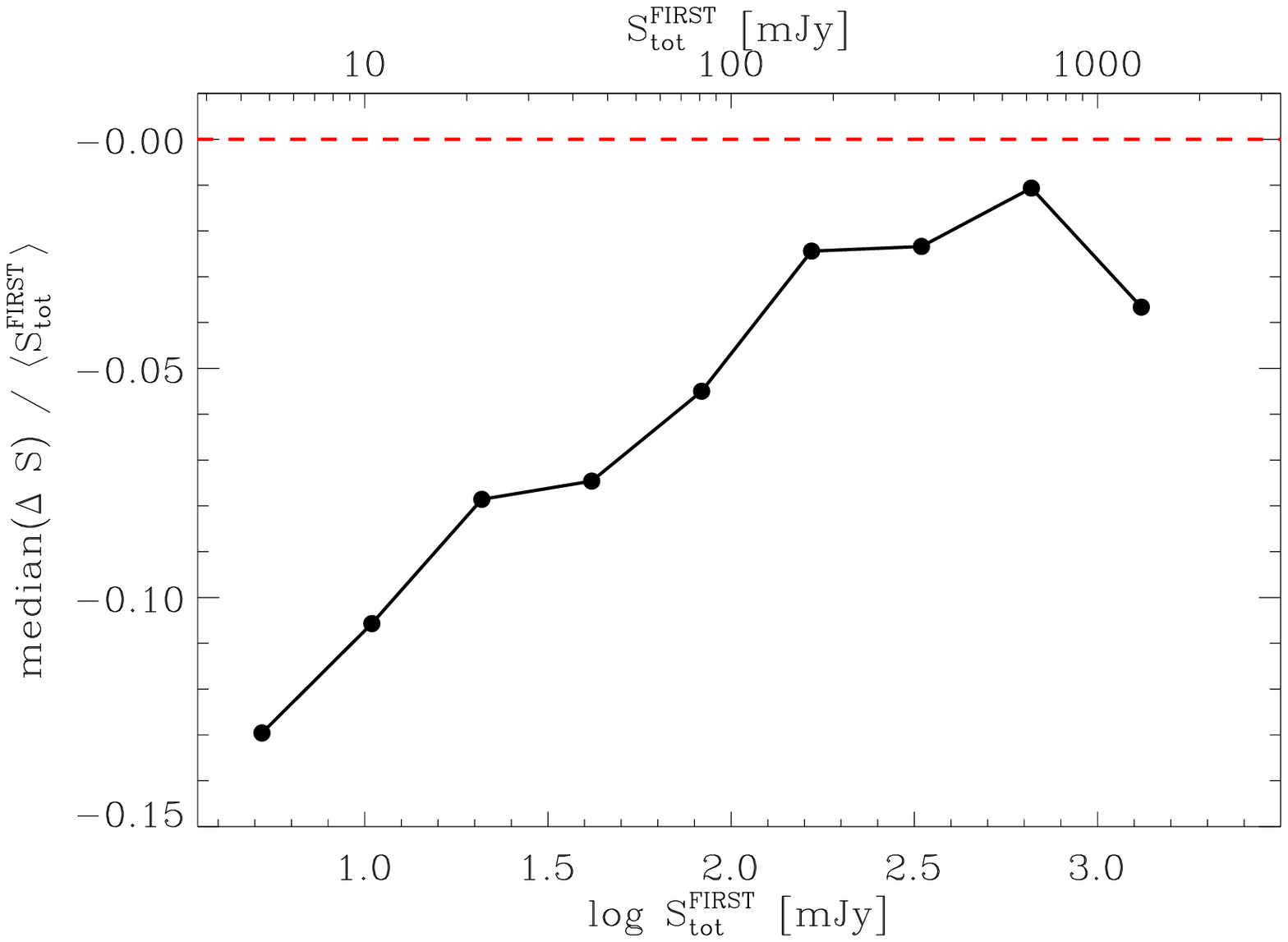}
\caption[Fractional uncertainty between the total FIRST and NVSS flux densities]{The fractional difference between the total flux densities of individual sources measured by NVSS and FIRST, in bins of FIRST flux density. Both plots are for sources that have $S^{\rm FIRST}_{\rm tot}>3.5$mJy and a NVSS match. The left panel shows the mean difference between the FIRST and NVSS flux densities ($\langle\Delta S\rangle=\langle S^{\rm FIRST}_{tot}-S^{\rm NVSS}_{\rm tot} \rangle$) divided by the mean FIRST flux density ($\langle S^{\rm FIRST}_{\rm tot} \rangle$).  The right panel uses the median difference rather than the mean. }
\label{fig:radio_pairwise_rms}
\end{figure*}

Figures \ref{fig:first_nvss} and \ref{fig:radio_pairwise_rms} compare the FIRST and NVSS flux densities for matched objects in the LARGESS sample. 
As can be seen from the bottom panel of Figure \ref{fig:first_nvss}, 95\% of the LARGESS sample with $S^{\rm FIRST}_{\rm tot} \geq 3.5$\,mJy have an NVSS match. 
Any analysis requiring accurate flux densities for extended radio sources should therefore impose a $S^{\rm FIRST}_{\rm tot} \geq 3.5$ mJy limit for the LARGESS sample. 

%Figure 5

The left panel of Figure \ref{fig:radio_pairwise_rms} shows the mean difference between the FIRST and NVSS flux densities ($\langle\Delta S\rangle=\langle S^{\rm FIRST}_{\rm tot}-S^{\rm NVSS}_{\rm tot} \rangle$) divided by the mean FIRST flux density in logarithmically spaced bins of FIRST flux density. Since the distribution of $\Delta S$ is slightly asymmetric (see Figure \ref{fig:first_nvss}), we also apply the same analysis using the median difference instead of the mean difference (right panel of Figure \ref{fig:radio_pairwise_rms}). Large discrepancies are only seen at the lowest flux densities, where the average difference is $\sim25\%$ (or $\sim 13\%$ for the median difference) of the average FIRST flux density. 

From these results, we estimate that between 5\% and 25\% of the 1.4\,GHz flux density of a typical LARGESS source is in a diffuse component detected by NVSS but missed by the FIRST survey. As can be seen from Figure \ref{fig:radio_pairwise_rms}, the discrepancy between the NVSS and FIRST flux density measurement increases at lower flux density levels.

%Figure 6

\section{Spectroscopic Data }\label{sec:spec_data}
In compiling the LARGESS catalogue, we aimed to achieve as high a level of spectroscopic completeness as possible across a large area of sky. To do this, we combined  existing data from earlier spectroscopic surveys with new spectra obtained in collaboration with the WiggleZ and GAMA teams (mainly using `spare' fibres not assigned to the main WiggleZ/GAMA survey targets). The final spectroscopic completeness for the catalogue as a whole is 64\% (i.e. 12,329 of the 19,179 objects in the LARGESS catalogue have at least one spectroscopic observation) and the redshift completeness is currently 57\% (10,856 objects have a reliable optical redshift).  As can be seen from Table \ref{tab:regions}, the completeness varies with sub-region and is highest ($>80$\% spectroscopic completeness) in the three GAMA regions. 

\subsection{Spectra and redshifts from earlier surveys} 
We incorporated spectra and redshifts from earlier surveys into our catalogue by cross-matching our sample with the 2dF and 6dF QSO Redshift Surveys \citep[ 2QZ and 6QZ;][]{croom04}, the 2dF-SDSS LRG And QSO survey \citep[2SLAQ;][]{cannon06,croom09} and the SDSS DR6 spectroscopic catalogue \citep{adelman-mccarthy08}. 

We set up a uniform quality classification system for redshifts from these earlier surveys. The interactive redshift code \textsc{runz} \citep{saunders04} was designed for use in the 2dF Galaxy Redshift Survey \citep[2dFGRS;][]{colless01} and also used by the 2SLAQ-LRG, GAMA and (in a modified version) WiggleZ spectroscopic surveys \citep{drinkwater10}. The user estimates a redshift by cross-correlating an observed spectrum with a set of template spectra and separate emission line fits, then inspects the estimated redshift and has the option to adjust the measurement. They then assign a quality flag {\em Q} to indicate the reliability of the final redshift. These surveys all used the same criteria for {\em Q} = 1 to 4, where higher values of {\em Q} indicate a higher confidence level for the redshift measurement. Redshifts with $\emph{Q}\geq 3$ are considered to be reliable. We adopted a range of 0 to 6 for {\em Q} (GAMA only uses {\em Q} = 1 to 4), where the additional {\em Q} = 0 identifies a poor-quality (or missing) spectrum; {\em Q} = 5 indicates an extremely reliable redshift from a good-quality  spectrum and {\em Q} = 6 is reserved for spectra classified as Galactic stars. 

We converted the redshift quality codes from the 2SLAQ-QSO, 2QZ, 6QZ and SDSS surveys to new values ({\em Q}$_{\rm initial}$) as outlined in Table \ref{tab:zflag2q}.  As a check, we also re-redshifted a subset of spectra from these surveys  as described below. For most objects, {\em Q}$_{\rm initial}$ was unchanged after re-redshifting. Of the 6,325 LARGESS sources with spectroscopic observations from one or more of these earlier surveys, 5,798 were initially classified as having reliable redshifts (i.e. $\emph{Q}_{\rm initial} \geq 3$).

%Table 3
\ctable[
star,
notespar,
cap = {Initial redshift quality assigned to redshift obtained from earlier surveys},
caption = {Conversion between redshift quality codes {\tt zconf} for SDSS and {\tt zflag} for 2QZ/2SLAQ-QSO) and the initial quality code ({\em Q}$_{\rm initial}$) used in the LARGESS catalogue.  The last three columns show the results from our re-redshifting of SDSS spectra for objects in the GAMA fields. $N(\text{re-redshift})$ is the number of SDSS spectra re-redshifted, and  $N(\text{agree})$ is the number of spectra where the re-redshift and SDSS redshifts agree within 0.01 (i.e. $|\Delta z| <0.01$) and the quality is considered reliable ($\emph{Q}>2$). The final column is the percentage that agree ($N(\text{agree})/N(\text{re-redshift})$) for each {\em Q}$_{\rm initial}$ bin.},
label = {tab:zflag2q}
]{l rr rrr}%
{ %\tnote[a]{footnote}
}{
\FL
Survey & SDSS & 2SLAQ-QSO and 2QZ  & \multicolumn{3}{c}{SDSS re-redshift}\\
\cline{4-6}
{\em Q}$_{\rm initial}$ &  {\tt zconf}          & {\tt zflag} & $N(\text{re-redshift})$ & $N(\text{agree})$ & \% agree\\
\hline\hline
6 & --                                    & --               &    -- &    -- &     --\\
5 & $>0.99$                          & --               & 576 & 575 &  99.8\\
4 & $>0.95$ and $\leq 0.99$ & --                & 172 & 169 &   98.2\\
3 & $>0.80$ and $\leq 0.95$ & 11               & 143 & 132 &  92.3\\
2 & $>0.50$ and $\leq 0.80$ & 12, 21 (1 case) or 22   &  66 & 35 &   53\\
1 & $\leq 0.50$                      & $>22$        &  37 &    8 &    22\\
0 & --                                    & --                &    -- &  -- &     --
\LL
}

\subsection{New spectroscopic observations}\label{subsec:spec_obs}

Our goal was to obtain new optical spectra for  all LARGESS objects with $i_{\rm mod}<20.5$ which did not already have a reliable redshift from the spectroscopic catalogues mentioned above. To do this, we carried out piggyback observations in conjunction with two large spectroscopic observing programs, the GAMA \citep{driver11,hopkins13} and WiggleZ \citep{drinkwater10} surveys, both of which used the 3.9 m Anglo-Australian Telescope (AAT) at the Siding Spring Observatory (SSO). All  our new spectra were taken using the fibre-fed AAOmega spectrograph with the two-degree field fibre positioner (2dF). Objects that were not already part of the main GAMA or WiggleZ target sample were assigned as lower-priority filler targets in the survey fields (see \cite{driver11} and \cite{drinkwater10} for priority listing).

Piggybacking on these large spectroscopic surveys 
allows us to obtain spectra efficiently for a large sample of radio galaxies whose surface density is too low to make effective use of the 400 fibres available in a  2dF field. We can also use the parent samples of GAMA and WiggleZ galaxies to measure environments and to build well-defined non-radio control samples to compare with. 

The spectra taken by the GAMA survey team covered the wavelength range 3720--8850$\AA$, with a typical integration time of 3000--5000 seconds.  
There were two phases to the GAMA survey, both of which are now complete. The first phase (GAMA-I) formed the basis for our spectroscopic target selection. 
A second phase (GAMA-II) extended the first by adding two southern fields, expanding the equatorial regions and also including objects with $r_{\rm pet}<19.8$\,mag\  in all regions. The GAMA-II spectroscopic targets come from the SDSS seventh data release (DR7) photometric catalogue. We did not add any additional objects to the LARGESS sample to reflect the boundary changes in GAMA-II, but our original GAMA targets remained in the GAMA-II spectroscopic target list and we have used both GAMA-I and -II spectra in our final data catalogue.

The spectra taken by the WiggleZ survey team covered the wavelength range 4700--9500$\AA$, with an integration time of 3600 seconds \citep{drinkwater10}. The WiggleZ main survey targets were restricted to objects with $r$-band magnitudes in the range $20<r<22.5$ (the bright cutoff was applied to avoid observing low-redshift galaxies, since the WiggleZ target redshift range was $0.4<z<1.0$). The WiggleZ survey team observed 3,674 radio targets, of which only 203 ($\sim 6\%$) were main WiggleZ targets. 

\subsection{Re-redshifting}\label{subsec:rez}

The GAMA and WiggleZ survey teams both measured redshifts on-the-fly at the telescope after each observation. This first-pass redshift was used to select targets for observing on following nights, and in most cases (especially for WiggleZ) became the final redshift of the target. Although the first-pass redshifts and quality codes were usually reliable, they had some inhomogeneities caused by different observers (with various expertise/experience) assigning redshifts, and in the case of WiggleZ, \textsc{runz} was optimized to measure redshifts from emission lines. To control and homogenise the data quality, we re-redshifted five sets of spectra: 

\noindent
(i) those observed by WiggleZ; \\
(ii) those observed by GAMA; \\
(iii) SDSS spectra in the GAMA regions; \\
(iv) all SDSS spectra with $\emph{Q}_{\rm initial}\leq2$; \\
(v) all other spectra with $\emph{Q}_{\rm initial}\leq2$ and ${\rm SNR}>8.5$. 

We re-redshifted sets (i) and (ii) to homogenise the redshifts and qualities between the GAMA and WiggleZ survey.  This is a different and independent re-redshifting from the one carried out by the GAMA survey team \citep{driver11,liske15}. 
Set (iii) was re-redshifted to compare the automatically assigned redshifts and quality codes  from SDSS to those assigned by manual inspection of the spectra. Re-redshifting of the last two sets was done to identify and correct redshifts of objects which had low redshift confidence in the GAMA or WiggleZ survey, often because their optical spectra showed unusual  features. 

For WiggleZ observations prior to 2010, four of the authors (JHYC, SMC, EMS and HMJ) re-redshifted all the radio target spectra by eye. JHYC re-redshifted all the 2011 WiggleZ observations, as well as sets (ii) to (v) above. 

\subsection{Final Redshifts}\label{subsec:final_z}
12,329 objects in the LARGESS sample have at least one spectroscopic observation.  
For each of these objects, we defined a single {\em best} redshift and quality ({\em QOP}) by comparing all available redshifts for that source as described below. 
Table \ref{tab:nQOP} gives a breakdown of the final number of redshift measurements in each {\em QOP} quality bin.

The LARGESS sample is based on the SDSS DR6, which was the most recent SDSS release at the start of the project. Since then, the SDSS-II survey has been completed with the release of SDSS DR7. For objects without an existing redshift measurement, we adopted the SDSS DR7 redshift where available.  
The quality codes for the SDSS DR7 spectra were converted as shown in Table \ref{tab:zflag2q}. This added an additional 1,130 spectra to the final catalogue. 

%Table 4
\ctable[
notespar,
cap = {Number of source for each {\em QOP} value.},
caption = {Number of sources in each redshift reliability bin {\em QOP}, shown separately for the GAMA/WiggleZ areas and the remaining lower-completeness regions (see Table \ref{tab:regions}). Higher {\em QOP} values indicates a more reliable redshift, and $\emph{QOP}\geq3$ is taken to be reliable enough for scientific analysis. {\em QOP} = 6 is reserved for Galactic stars.},
label = {tab:nQOP}
]{l rrr}%
{ %\tnote[a]{footnote}
 }
{ \FL 
{\em QOP} &            GAMA/WiggleZ    &           Other & Total\\
                  &             regions  &         regions & \\ 
\hline\hline	
0 & 11 & 0 & 11\\
1 & 631 & 78 & 709\\
2 & 703 & 50 & 753\\
3 & 1669 & 428 & 2097\\
4 & 4422 & 537 & 4959\\
5 & 2679 & 1030 & 3709\\
6 & 88 & 3 & 91\\
Total & 10203 &  2126 & 12329\\
Not observed & 4447  & 2403 & 6850 \\
\hline
Total including \\
unobserved 
objects & 14650 & 4529 & 19179 %\\
\LL
}

Almost half the sample (7,595 objects) have a single spectroscopic observation and only a single redshift measurement.  
Most of these are located in parts of the 2SLAQ strips which do not overlap the WiggleZ or GAMA area, or are objects with a $\emph{Q}\geq3$ redshift from SDSS or an earlier survey. 
We automatically accepted this redshift and quality code and included it in the final spectroscopic catalogue. 

1,442 objects have a single spectrum from the WiggleZ survey, but multiple redshifts for that spectrum from the re-redshifting process. In this instance, we took the redshift value with the most agreements. If there were no agreements, one of the authors (JHYC) selected a final redshift and quality. 

For the 997 sources with two or more spectroscopic observations and a single redshift measurement for each observation, we started by identifying reliable redshifts that agreed with each other ($|\Delta z| < 0.01$ and $\emph{Q}\geq3$) and accepted the value with the most agreements. For redshifts with the same number of agreements, we took the set of agreements with the highest {\em Q} assigned, and within this set we selected the redshift and quality assigned to the spectrum with the highest signal-to-noise ratio (SNR) as the best redshift. If there were no agreements, we accepted the redshift with the highest {\em Q}.

There are 1,165 sources with both multiple spectra and multiple redshift measurements from re-redshifting. 
For objects in this category, we again identified reliable redshifts that were in agreement ($|\Delta z| < 0.01$ and both have $\emph{Q}\geq3$) and defined the {\em best} redshift as the one with the most agreements. If there were multiple sets of redshifts with the same number of agreements, then we accepted the one with the highest {\em Q}, and if {\em Q} was the same, we adopted the redshift assigned using the highest-SNR spectrum.

\subsection{Redshift reliability}
To assess the reliability of our redshift measurements, we used a similar technique to previous studies \cite[e.g.][]{colless01,croom04,drinkwater10} and compared our final {\em QOP} = 3--5 redshifts with a repeat observation of the same object with $\emph{Q}_{\rm rep}\geq3$ (where {\em QOP} is the quality associated with our final redshift and $\emph{Q}_{\rm rep}$ is the redshift quality associated with the repeated observation).  

We found that redshift measurements with $Q\geq3$ are generally highly reliable, with implied single-measurement blunder (i.e. $|\Delta z| \geq 0.01$) rates of 8.5\% and 1\% for {\em QOP} = 3 and 4 respectively. For pairs where the two redshifts agree, the pairwise rms dispersion of redshift differences is also small (with a typical value of  $|\Delta z| \leq 0.0025$). The {\em QOP} = 5 pairs have a high fraction of broad emission-line QSOs, where the broad peaks sometimes make it difficult to determine a consistent redshift.  As a  result, the {\em QOP} = 5 redshift measurements have a slightly higher dispersion on average than the {\em QOP} = 4 measurements.

%Table 5
\ctable[
notespar,
cap = {Emission line wavelength definitions from MPA-JHU.},
caption = {Emission line wavelength definitions from MPA-JHU for the [OIII]$\lambda 5007$, [OIII]$\lambda 4959$ and H$\beta$ emission lines.},
label = {tab:emline_MPAJHU}
]{l rrr}%
{ %\tnote[a]{footnote}
}{
\FL
Line                        & \multicolumn{1}{c}{Line}                & \multicolumn{1}{c}{Lower}           & \multicolumn{1}{c}{Upper}\\
name                      & \multicolumn{1}{c}{centre (\AA)} & \multicolumn{1}{c}{bound (\AA)} & \multicolumn{1}{c}{bound (\AA)}\\
\hline\hline
H$\beta$                      & 4861.325 & 4851.0 & 4871.0\\
{[OIII]$\lambda 4959$} & 4958.911 & 4949.0 & 4969.0\\
{[OIII]$\lambda 5007$} & 5006.843 & 4997.0 & 5017.0
\LL
}

\subsection{Emission-line measurements}

Our catalogue includes emission-line flux measurements for galaxies with good-quality spectra observed by the WiggleZ, GAMA and/or SDSS teams. Spectra from earlier surveys  like 2SLAQ and 2dFGRS generally lack the accurate flux calibration, spectral resolution or wavelength coverage needed for reliable emission-line measurements.

\subsubsection{GAMA and SDSS emission-line measurements}

The GAMA survey provides emission-line measurements (internal data: GandalfSpecAnalysis v08.3; Steele et al. in preparation) using the \textsc{gandalf} \cite[Gas AND Absorption Line Fitting;][]{sarzi06} code. For galaxies with SDSS spectra, we used the Max-Planck-Institut f\"ur Astrophysik -- Johns Hopkins University (MPA-JHU) emission-line measurements\footnote{http://www.mpa-garching.mpg.de/SDSS}.  
In both cases we ensured that the redshifts used for emission-line measurements agreed with those in our final catalogue. The \textsc{gandalf} and the MPA-JHU emission line measurements both apply a template fit to account for stellar absorption before measuring emission line fluxes. Thus we expect both measurements to be comparable and robust \cite[for a more detailed comparison, see][]{hopkins13}.  

\subsubsection{WiggleZ emission lines}

The WiggleZ spectra have poorer spectrophotometry than the GAMA/SDSS spectra, since the method adopted for spectroscopic curvature correction in WiggleZ spectra 
makes it difficult to subtract the stellar continuum accurately (the WiggleZ survey was designed to measure faint objects with strong optical emission lines and little or no visible continuum). 
We therefore chose to make only a single measurement of the [OIII]$\lambda5007$ emission-line flux and equivalent width for the WiggleZ spectra. The  [OIII]$\lambda5007$ line is not significantly affected by stellar absorption features, and measuring this line allows us to distinguish between low-and high-excitation radio galaxies. 

For these measurements, we used the same wavelength definitions as MPA-JHU (see Table \ref{tab:emline_MPAJHU}) and estimated the continuum at the position of [OIII]$\lambda5007$ by fitting a second-order polynomial to the local continuum.
The emission-line flux was then measured by integrating over the continuum-subtracted [OIII]$\lambda5007$ region. We used the covariance matrix provided by the \textsc{svdfit} routine together with the individual pixel variances to calculate the [OIII]$\lambda5007$ flux error and the continuum error.

The [OIII]$\lambda5007$ equivalent width (EW([OIII]$\lambda5007$)) is measured by dividing the integrated line flux by the continuum flux at the line centre. The estimated continuum flux at the line centre can sometimes equal to or fall below zero due to systematic sky subtraction errors for the faintest objects, but there may still be a prominent [OIII]$\lambda5007$ emission line. To avoid a non-physical value for the EW([OIII]$\lambda5007$), we instead derived a minimum equivalent width by dividing the minimum value of the [OIII]$\lambda5007$ flux (i.e. [OIII]$\lambda5007$ flux minus the error) by the error in the continuum at the line centre.

%Figure 7
\begin{figure*}
\centering
\includegraphics[height=1.04\textwidth,angle=90]{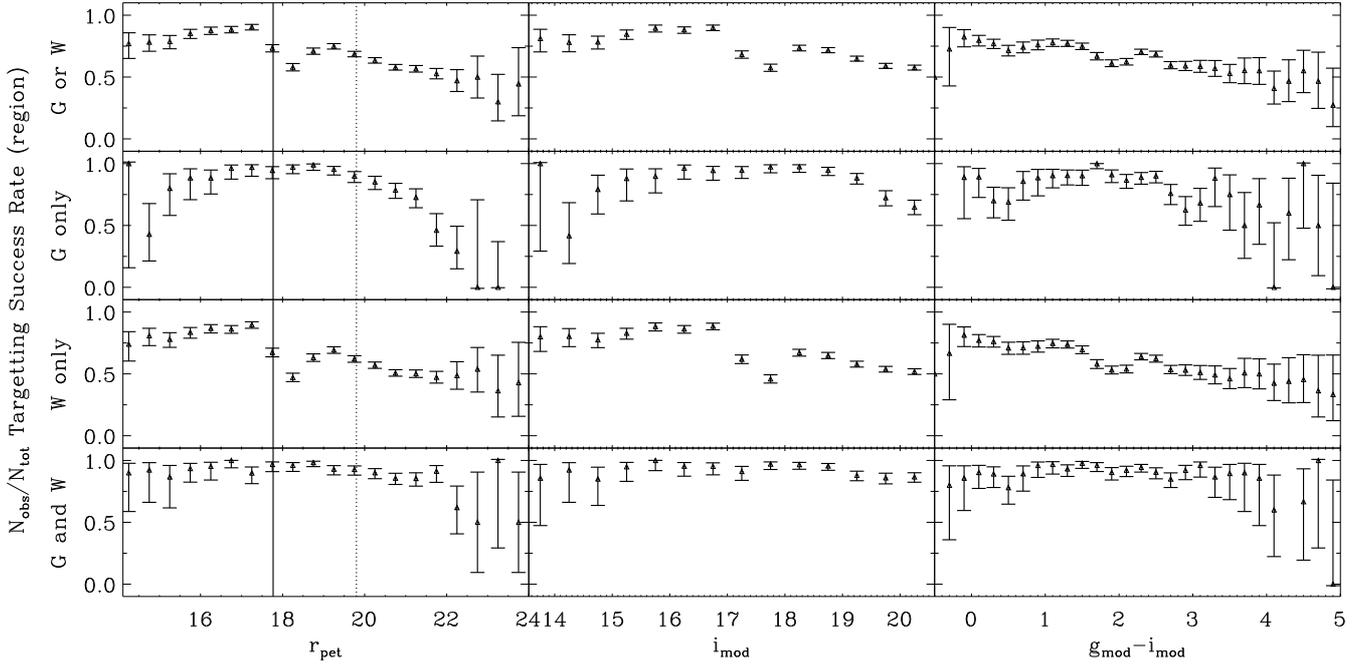}
\caption[Spectroscopic targeting completeness as a function of $r_{\rm pet}$, $i_{\rm mod}$ and $(g_{\rm mod}-i_{\rm mod})$, separated by survey region]{The spectroscopic targeting success rate ($N_{\rm obs}/N_{\rm tot}$) for LARGESS objects in the WiggleZ and GAMA regions (see Table 1) as a function of $r_{\rm pet}$, $i_{\rm mod}$ and $(g_{\rm mod}-i_{\rm mod})$. $N_{\rm tot}$ is the number of sources within a specific region, while $N_{\rm obs}$ is the number with a spectroscopic observation. We only count an object as observed if the best redshift comes from the SDSS, WiggleZ or GAMA survey. The four sets of plots are for different sky regions as described in \S6.1 of the text. 
The vertical solid and dashed lines in the left-hand panel show the  $r_{\rm pet}$ magnitude limits for the SDSS and GAMA-II surveys respectively.  The WiggleZ survey observed objects in the magnitude range $20<r<22.5$.  
}
\label{fig:targeting}
\end{figure*}

\section{Survey Completeness}\label{sec:spec_completeness}

We now quantify the spectroscopic completeness of the catalogue as a function of apparent magnitude and colour. Here, the {\it targeting completeness}\ refers to the fraction of LARGESS objects that have been spectroscopically observed, and the {\it redshift completeness}\ refers to the fraction of objects with a reliable redshift. In this section we focus on the GAMA and WiggleZ regions listed in Table \ref{tab:regions}, which have the highest completeness and so are more likely to be used for follow-up studies. These regions (which are also covered by the SDSS survey) contain a total of 14,650 catalogue sources, of which 10,203 have spectroscopic observations (see Table \ref{tab:nQOP}). 

\subsection{Targeting completeness}

Figure \ref{fig:targeting} shows the fraction of LARGESS sources for which a spectroscopic observation has been made, as a function of $r_{\rm pet}$, $i_{\rm mod}$, and $(g_{\rm mod}-i_{\rm mod})$ colour.  These are separated by survey region as follows: 
\begin{enumerate}
\item
The top row in Figure \ref{fig:targeting} shows the full set of 14,650 objects included in either the GAMA or the WiggleZ fields (Region {\tt G or W}).  
The overall targeting completeness of 70\% for this region closely resembles the targeting completeness for objects in the WiggleZ fields, since the WiggleZ survey area is roughly four times larger than the GAMA area and so contains many more targets. The dip in completeness seen in the left-hand panels at $18<r_{\rm pet}<20$ arises from the bright limit of the WiggleZ sample, as discussed in point (iii) below. 
\item
The second row shows objects in the 67\,deg$^2$\ of sky covered by the GAMA survey area but not the WiggleZ area (region {\tt G only}), for which the overall targeting completeness is 88\%. These spectra come mainly from SDSS at the bright end, with GAMA observations becoming increasingly important at the faint end (the GAMA survey observed secondary targets out to $r_{\rm pet} = 22.5$, with uniform sampling for objects with $18<r_{\rm pet}<21.5$\,mag).  
The targeting rate here is fairly uniform in both $i_{\rm mod}$ magnitude and colour, but shows a gradual fall-off in $r$-band completeness beyond the GAMA-II main sample limit of $r_{\rm pet}<19.8$. 
\item
The next row {\tt W only}\ is for objects in the $\sim540$\, deg$^2$ of sky covered by the WiggleZ survey, but falling outside the GAMA survey regions. Here, the spectroscopic observations are dominated by SDSS at the bright end ($r_{\rm pet} < 17.77$) and WiggleZ observations at the faint end. 
The WiggleZ survey team applied an additional bright limit of $i_{\rm mod} > 18.0$ for our piggyback targets, to 
reduce light contamination from brighter sources due to cross-talk between fibres on the spectrograph CCD (since the WiggleZ main survey targets were faint galaxies with $20<r<22.5$\,mag). The combination of this bright limit and the SDSS faint limit causes a dip in the targeting completeness which can be seen in both the $i_{\rm mod}$ and $r_{\rm pet}$\ plots. The dip does not go to zero, because the SDSS also observed some secondary targets beyond the main survey limit (though the targeting completeness for the SDSS secondary targets drops off quickly for $r_{\rm pet} > 17.77$). 
We also note that bluer objects appear to be preferentially targeted in the {\tt W only} plot of Figure \ref{fig:targeting}. This is mainly because the targeting completeness is higher for SDSS than WiggleZ, and the SDSS observations target brighter objects that are at lower redshifts ($z<0.3$) and so have bluer colours than the fainter objects targeted by the WiggleZ spare-fibre program. 
This effect can be corrected in any follow-up analysis by taking into account the apparent magnitude completeness and limits.
\item
The final row is for objects in the 77\,deg$^2$\ region of sky covered by both GAMA-I and WiggleZ surveys (region {\tt G and W:} in Figure \ref{fig:targeting}; see also Table 1 and Figure 1). Region {\tt G and W} has the most uniform targeting completeness in all three optical parameters ($r_{\rm pet}$, $i_{\rm mod}$ and $g_{\rm mod} - i_{\rm mod}$). The combination of observations from the SDSS, GAMA and WiggleZ surveys ensures that many objects are observed, and washes out the individual magnitude and colour limits from these surveys. 
\end{enumerate} 

\subsection{Redshift completeness}
Not all spectroscopic observations result in a reliable redshift measurement. The overall redshift success rate for spectroscopic observations of the LARGESS sample is 88\, per cent, but this varies with target properties such as brightness, colour and the presence or absence of emission lines. 

\begin{figure*}
\centering
\includegraphics[height=1.04\textwidth,angle=90]{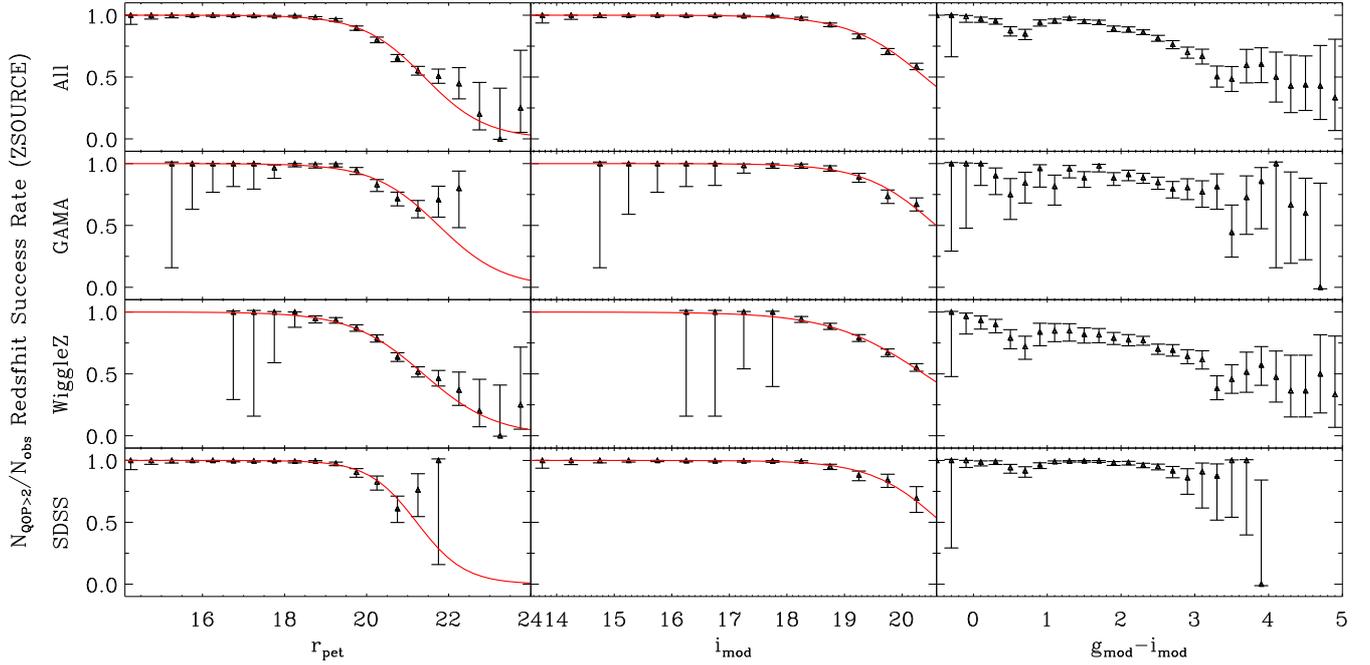}
\caption[Redshift success rate as a function of $r_{\rm pet}$, $i_{\rm mod}$ and $g_{\rm mod}-i_{\rm mod}$]{Redshift success rate ($N_{\emph{QOP}>2}/N_{\rm obs}$, where $N_{\emph{QOP}>2}$ is the number with a reliable redshift and $N_{\rm obs}$ is the number observed) as a function of $r_{\rm pet}$, $i_{\rm mod}$ and $g_{\rm mod}-i_{\rm mod}$. This is only shown for objects with {\tt ZSOURCE} = GAMA, WiggleZ or SDSS. The red line is the best fit to a sigmoid function using maximum likelihood estimation, which is only applied to the apparent magnitude comparisons. Errors are estimated using the method described by \citet{cameron11} for a 95\% confidence interval.}
\label{fig:zsuccess}
\end{figure*}

Figure \ref{fig:zsuccess} shows the redshift success rate as a function of $r_{\rm pet}$, $i_{\rm mod}$, and $g_{\rm mod} - i_{\rm mod}$ colour. These plots show a drop in the redshift success rate towards fainter magnitudes, as well as a decreasing success rate for redder objects with $g_{\rm mod} - i_{\rm mod} \gtrsim 2$. 

\begin{figure}
\centering
\includegraphics[width=0.52\textwidth]{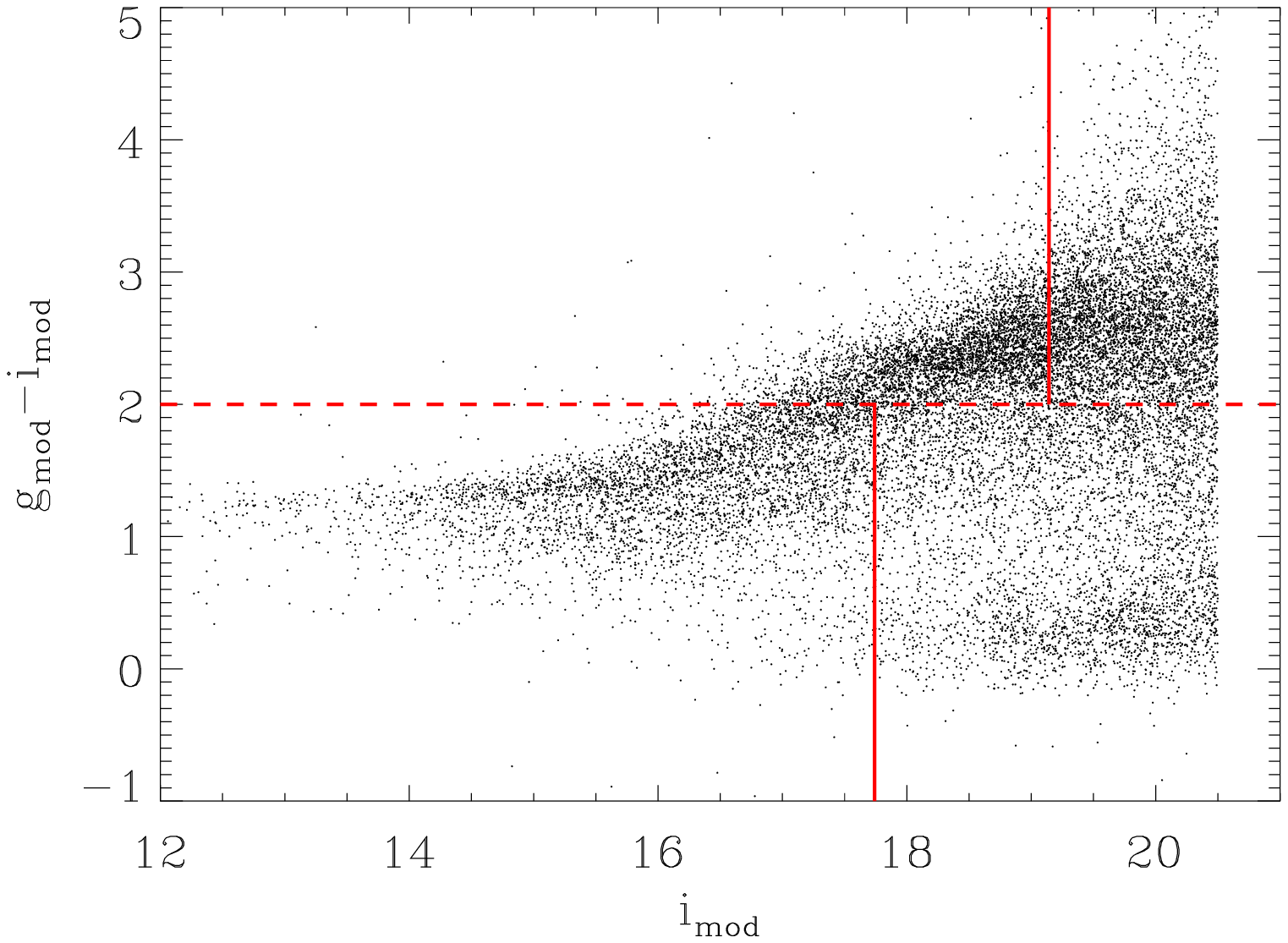}
\caption[$(g_{\rm mod}-i_{\rm mod})$ vs. $i_{\rm mod}$ for the LARGESS sample]{$(g_{\rm mod}-i_{\rm mod})$ colour versus $i_{\rm mod}$ magnitude for the full sample. The horizontal dashed line marks $(g_{\rm mod}-i_{\rm mod})=2$, above which the redshift success rate drops for objects with redder colour. Vertical lines show the mean $i_{\rm mod}$ magnitude for objects redder and bluer than $(g_{\rm mod}-i_{\rm mod})=2.0$.}
\label{fig:gi_i}
\end{figure}

Figure \ref{fig:gi_i} shows the $g_{\rm mod} - i_{\rm mod}$ colour as a function of $i_{\rm mod}$ apparent magnitude for the full LARGESS sample. Objects with $g_{\rm mod} - i_{\rm mod} > 2$ are on average fainter (mean $i_{\rm mod}=19.1$ mag) than objects with $g_{\rm mod} - i_{\rm mod} < 2$ (mean $i_{\rm mod}=17.7$ mag). Therefore the lower redshift success rate for objects with $g_{\rm mod} - i_{\rm mod} > 2$ may be explained, in part, by the fact they are fainter objects. In general bluer objects are also more likely to have emission lines, which increase the chance of measuring a reliable redshift. 

We can model the overall redshift completeness as a function of magnitude using the sigmoid function \citep[e.g.][]{loveday12,ellis07}:
\begin{equation}
 y(x)=1/[1+e^{a(x-b)}],
\end{equation}
 where $x$ is the photometric magnitude, $a$ is the stiffness of the function and $b$ is the magnitude at a redshift success rate of 50\% (i.e. $y(b) = 0.5$).  The best fits are shown as red lines in Figure \ref{fig:zsuccess}, and the parameters are listed in Table \ref{tab:sigmoid_fit}. 

%Figure 8 

%Table 6
\ctable[
%star,
notespar,
cap = {Sigmoid parameters for redshift completeness},
caption = {Sigmoid parameters for redshift completeness from maximum likelihood estimation, as plotted in Figure \ref{fig:zsuccess} and discussed in \S6.2 of the text. } ,
label = {tab:sigmoid_fit}
]{l rr rr }%
{ %\tnote[a]{footnote}
}{
\FL
 & \multicolumn{2}{c}{$r_{\rm pet}$} & \multicolumn{2}{c}{$i_{\rm mod}$}\\
 \cline{2-3}\cline{4-5}
{\tt ZSOURCE} & $a$ & $b$ & $a$ & $b$\\
\hline\hline
All & 1.33$^{+0.04}_{-0.04}$ & 21.41$^{+0.56}_{-0.55}$ & 1.54$^{+0.05}_{-0.05}$ & 20.39$^{+0.45}_{-0.44}$ \\
GAMA & 1.27$^{+0.08}_{-0.07}$ & 21.75$^{+1.43}_{-1.34}$ & 1.62$^{+0.13}_{-0.12}$ & 20.59$^{+1.06}_{-1.01}$ \\
WiggleZ & 1.14$^{+0.04}_{-0.04}$ & 21.38$^{+0.80}_{-0.78}$ & 1.23$^{+0.05}_{-0.05}$ & 20.38$^{+0.69}_{-0.67}$ \\
SDSS & 1.71$^{+0.12}_{-0.11}$ & 21.22$^{+1.16}_{-1.10}$ & 1.64$^{+0.10}_{-0.10}$ & 20.69$^{+1.17}_{-1.11}$ 
\LL
}

%Figure 9

\section{Spectral Classification}\label{sec:spec_class}

In this section, our goal is to use the optical spectra of LARGESS sources to determine the dominant physical process (either an active galactic nucleus (AGN) or star formation) responsible for the radio emission in each individual object within our sample. In other words, we are classifying the {\it radio source}\ rather than the optical spectrum itself.
 
In most cases, the classification of the radio source can be deduced directly from the optical spectrum.  However, as discussed by \citet{best12}, it is important to be able to identify objects where the optical spectrum is dominated by strong emission lines from a radio-quiet AGN\footnote{In this paper, we use the terms 'radio-quiet' and 'radio-loud' to refer to objects in which the radio continuum emission is predominantly powered by star-formation and AGN processes respectively.}   
but the radio emission is powered mainly by star-formation processes. Here, the optical star-formation signature may be obscured by the AGN lines in the single-fibre spectra we are using. As discussed in S7.2, we make a quantitative comparison between the observed radio luminosity and the star-formation rate estimated from the H$\alpha$ emission line to identify such objects. 

We classified the optical spectra of LARGESS sources in two ways. A first-pass {\it visual classification} (described in \S\ref{subsec:VISCLASS}) allows us to identify  BL Lac candidate and emission-line objects where the lines have broad wings (class AeB below), as well as providing a useful comparison with earlier work and a series of checks on the automated classification process. We also make an automated {\it quantitative classification}\ (see \S\ref{subsec:AUTOCLASS}) for objects where the  [OIII]$\lambda5007$ emission line falls within the GAMA/WiggleZ/SDSS spectral range.

\subsection{Qualitative (visual) spectral classification}\label{subsec:VISCLASS}

\begin{figure*}
\centering
\begin{minipage}\textwidth
\includegraphics[width=0.49\textwidth]{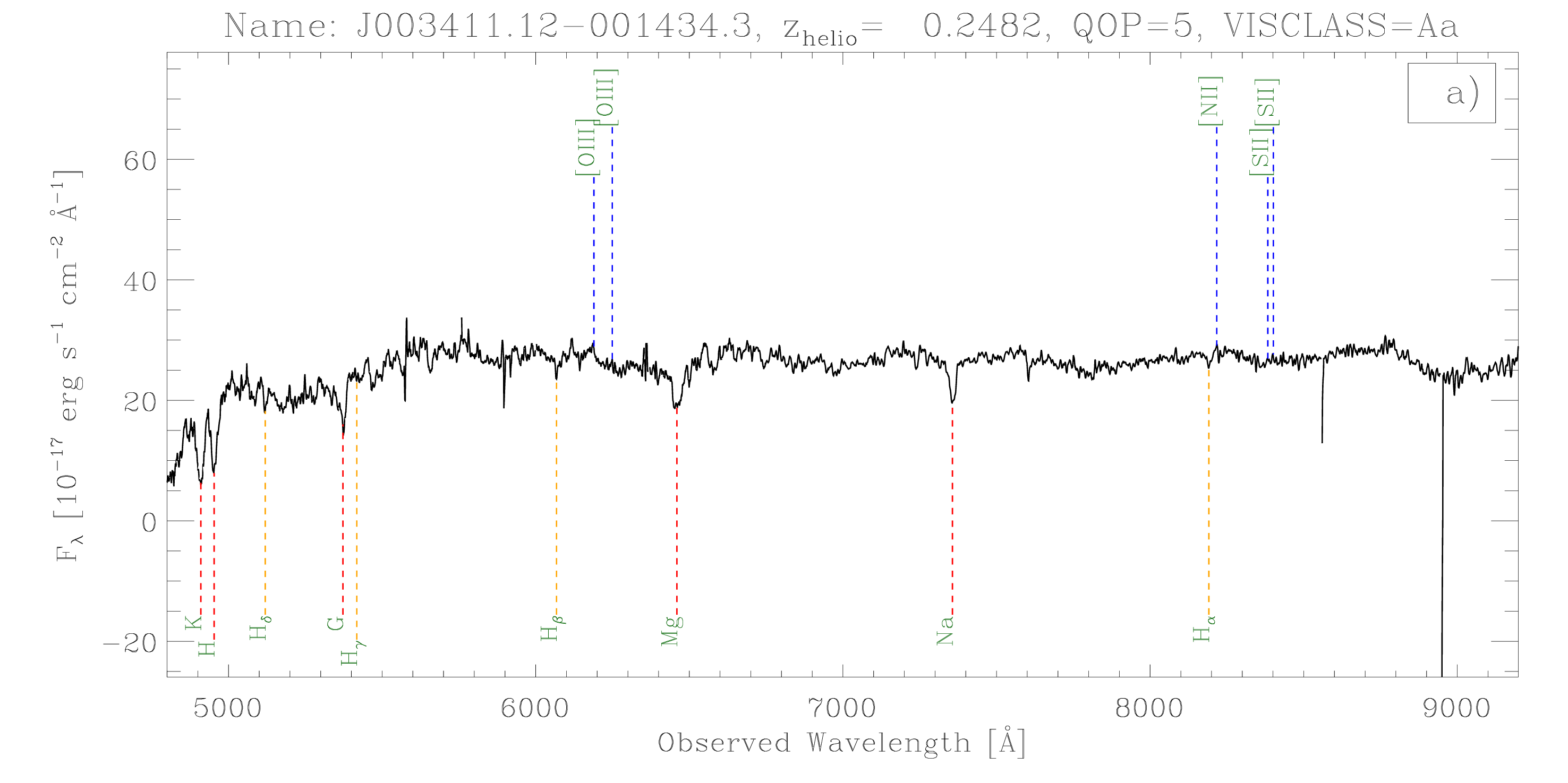}% Aa
\includegraphics[width=0.49\textwidth]{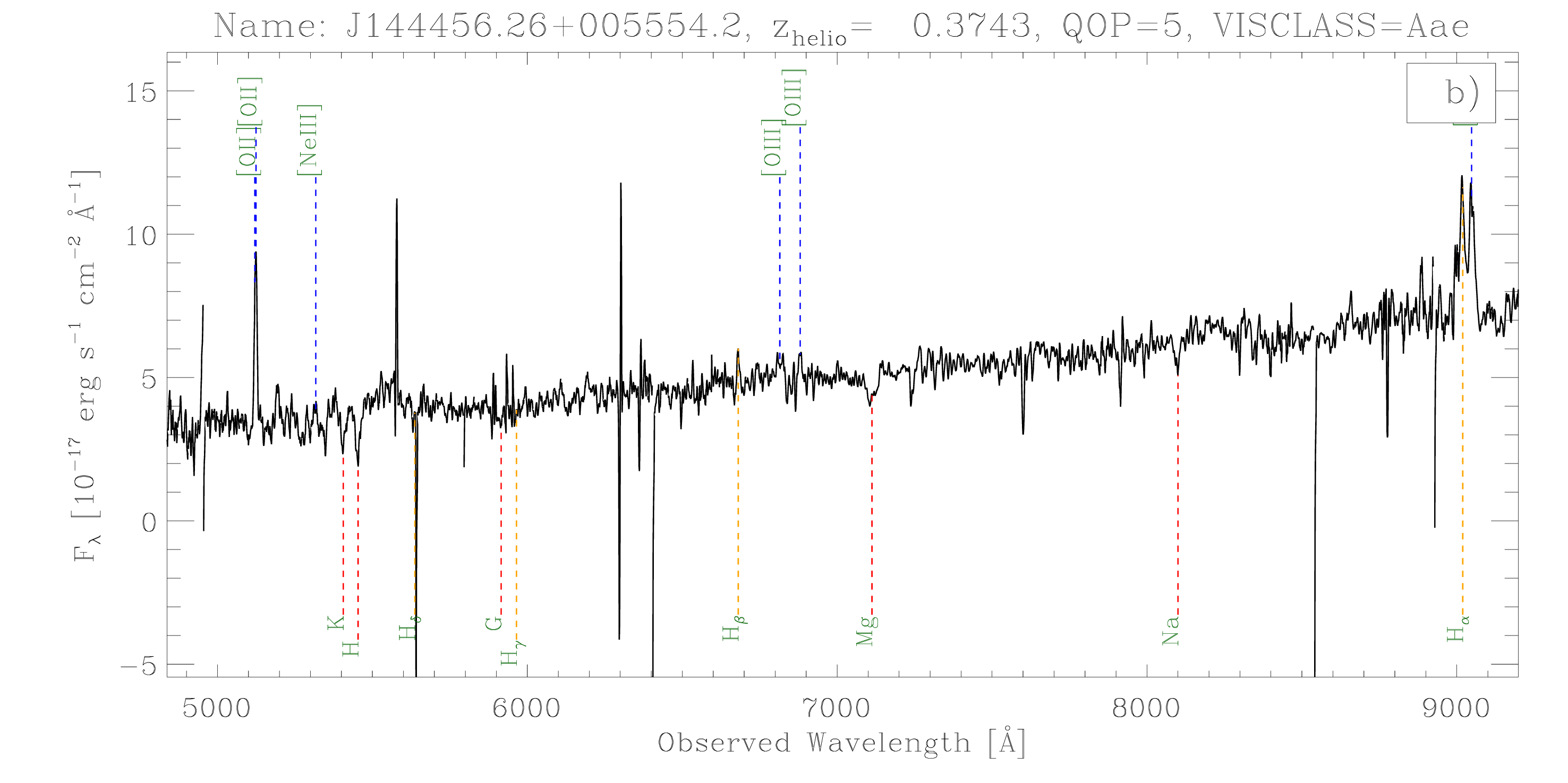}% Aae
\end{minipage}
\begin{minipage}\textwidth
\includegraphics[width=0.49\textwidth]{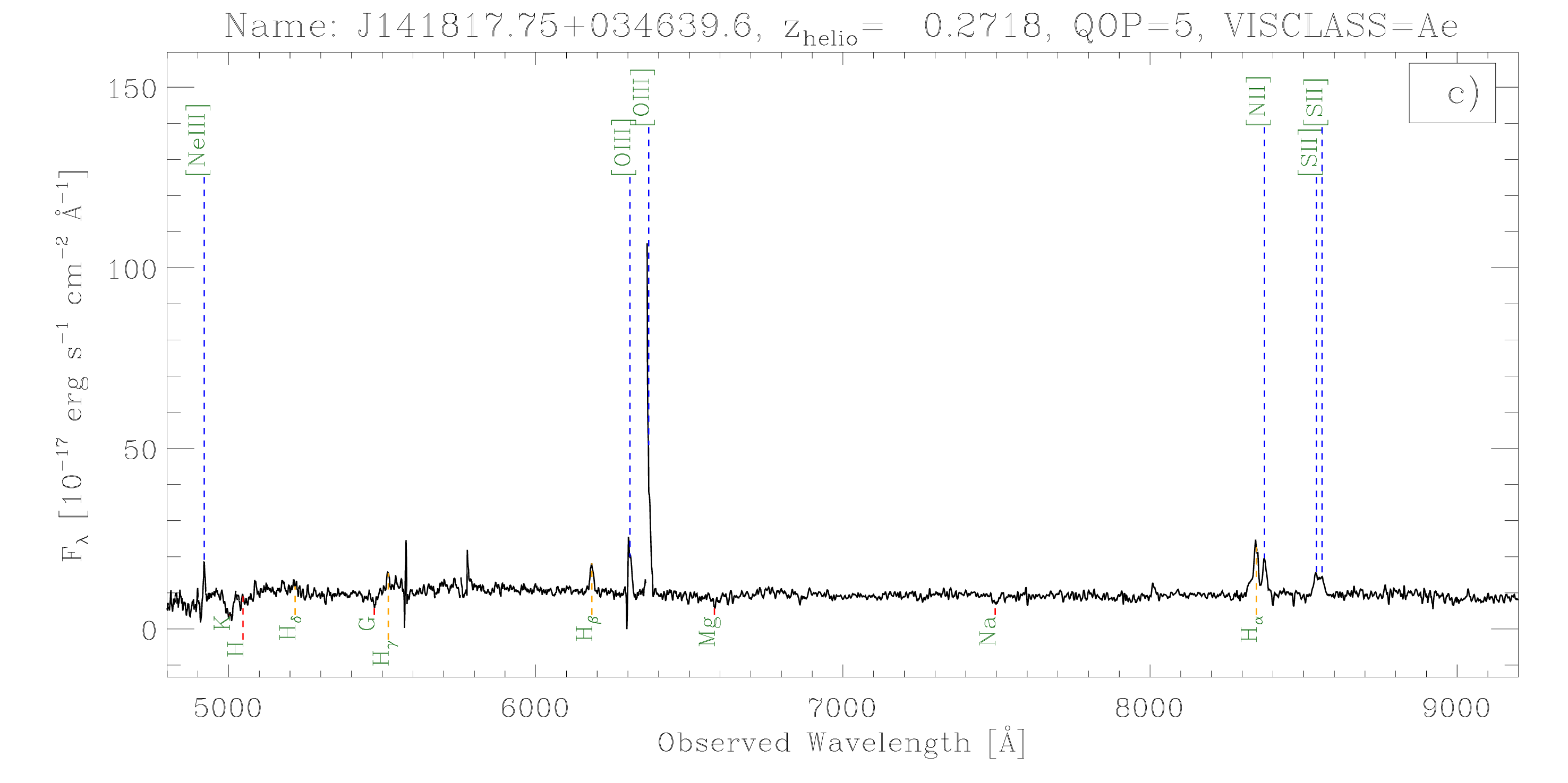}% Ae
\includegraphics[width=0.49\textwidth]{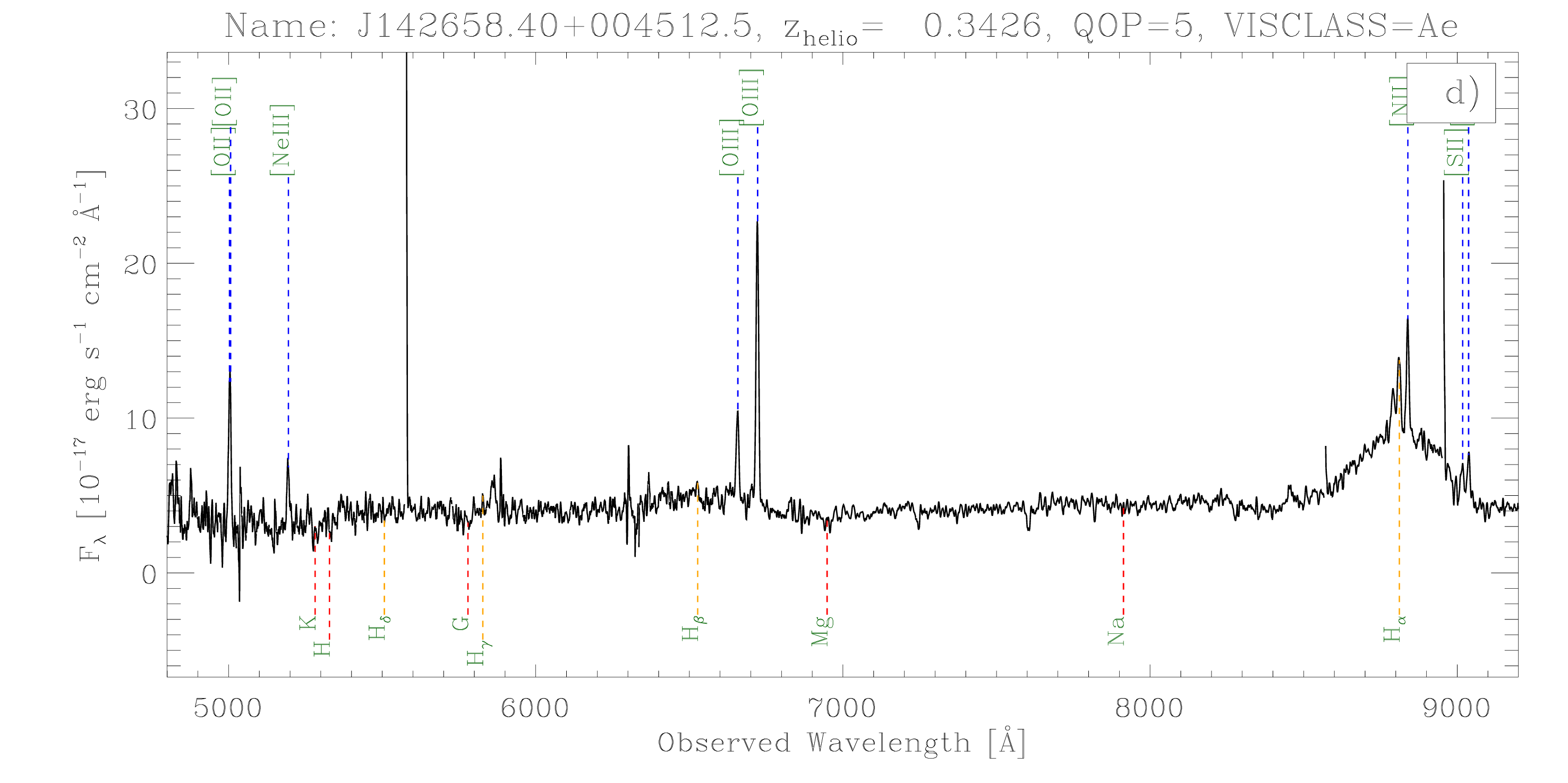}%AeB
\end{minipage}
\begin{minipage}\textwidth
\includegraphics[width=0.49\textwidth]{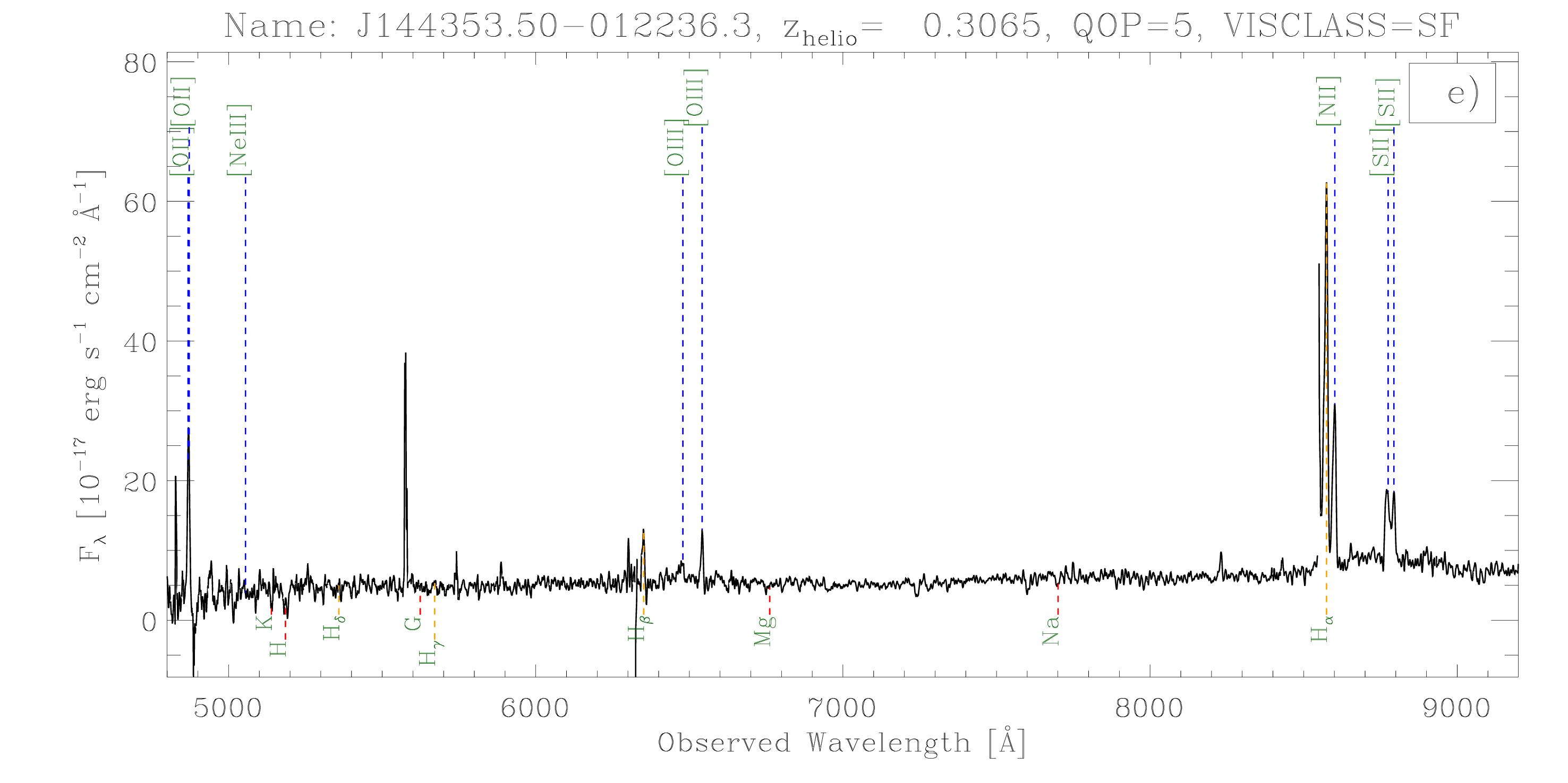}%SF
\includegraphics[width=0.49\textwidth]{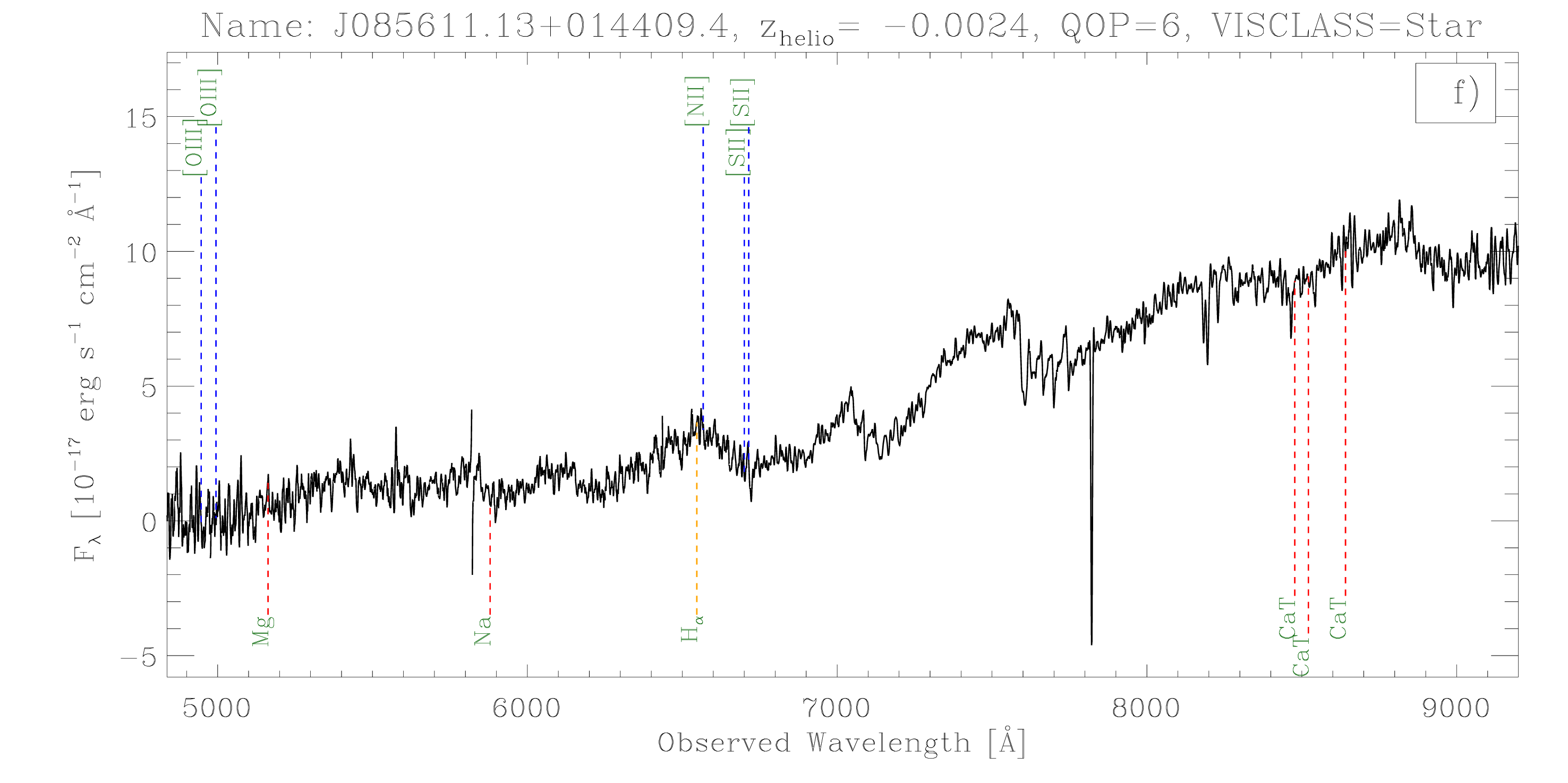}%Star
\end{minipage}
\caption[Example spectra for different visual classifications]{Example spectra for different visual classifications ({\tt VISCLASS}): a) {\bf Aa}, strong absorption lines and no emission lines visible; b) {\bf Aae}, strong absorption lines and weak emission lines; c) {\bf Ae}, strong emission lines with line ratios characteristic of an AGN;  d) {\bf AeB}, AGN with broad emission lines; e) {\bf SF}, strong emission lines with line ratios characteristic of star-forming galaxies; f) {\bf Star}, Galactic star spectrum. Note that there are bad pixels or cosmic ray contamination in some spectra.}
\label{fig:VISCLASS}
\end{figure*}

During the re-redshifting process, we assigned a spectral class based on a visual inspection ({\tt VISCLASS}) for each object with a reliable redshift (i.e. $\emph{Q}\geq3$). 
This visual classification was based on the scheme used by \cite{sadler02}. Example spectra of the main classes are shown in Figure \ref{fig:VISCLASS}, and the classification criteria are as follows:
\begin{description}
\item[\bf Aa] Stellar continuum with no apparent emission lines.
\item[\bf Aae] Stellar continuum with weak emission lines (e.g. [OIII]$\lambda5007$, H$\alpha$, [NII]$\lambda6583$ etc.).
\item[\bf Ae] Stellar continuum with strong narrow emission lines where the [OIII]$\lambda5007$ and [NII]$\lambda6583$ emission lines appear stronger than, or comparable to, the H$\beta$ and H$\alpha$ emission lines respectively.
\item[\bf AeB] Broad emission lines typical of a quasar spectrum.
\item[\bf SF] Similar to Ae, but instead the Balmer emission lines appear stronger than the forbidden lines.
\item[\bf Star] Galactic star.
\item[\bf BL Lac] Featureless optical spectrum with no obvious emission or absorption lines.
\item[\bf Unusual] Any object with a spectrum  that does not fit into the categories above. This category includes some radio-source hosts with weak emission lines superimposed on a strong, featureless continuum as well as broad-absorption line quasars (BAL QSOs and a few post-starburst galaxies with strong Balmer absorption lines.   
\end{description}

%Figure 10

The SDSS database also includes flags for different spectral classifications, based on the best-fitting template to each SDSS spectrum. SDSS {\tt specClass} flags of 3 and 4 correspond to QSO and high-redshift ($z>2.3$) QSO respectively \citep{stoughton02}, and for SDSS spectra with {\tt specClass} flag of 3 or 4, we set {\tt VISCLASS} to AeB. 
Similarly, we incorporate the visual classification of QSO provided by the 2SLAQ-QSO and 2QZ surveys into our {\tt VISCLASS} flag. Finally, we set the {\tt VISCLASS} to NA (null) for the  remaining sources where no visual classification was made. 

\subsection{Quantitative (automated) spectral  classification}\label{subsec:AUTOCLASS}
\citet*[][hereafter BPT]{baldwin81} devised a method to distinguish between the emission lines originating from an AGN and star formation by comparing the ratio of specific forbidden lines to neighbouring Balmer lines.

We used the BPT technique to carry  out an automated spectral classification for objects with $\emph{QOP}\geq3$ if 
the [OIII]$\lambda5007$ emission line fell within the observed spectral range. In practice, this imposes a redshift limit of $z<0.768$ for GAMA spectra and $z<0.838$ for WiggleZ and SDSS spectra. 75\% of the 10,856 LARGESS objects with reliable redshifts also have an automated spectral classification.

\subsubsection{Classifying WiggleZ spectra}

As explained in \S5.6.2, we did not measure absorption-corrected emission-line ratios for objects with WiggleZ spectra because of difficulties fitting the underlying stellar continuum in a reliable way. As a result, we could not use the BPT diagram to classify our WiggleZ spectra because we were not able to correct the Balmer emission lines for any underlying stellar absorption. 

For this reason, we only used the WiggleZ spectra to classify objects with $L_{\rm FIRST}>10^{24}$ W Hz$^{-1}$ or $z>0.3$ -- which can be assumed to be radio-loud AGN (see \S7.2.3 below). For these objects, we used the [OIII] equivalent width (which is not significantly affected by stellar absorption) to separate low-and high-excitation radio galaxies. For galaxies with spectra from GAMA and SDSS, we used the BPT diagram as described below. 

\subsubsection{Galactic stars}
Objects with a reliable redshift of $z<0.002$ are classified as Galactic stars. They may be either a genuine association of a FIRST radio source with a Galactic object or (more likely) a random superposition of a foreground star against a background radio source.
They make up $\sim 2\%$ of LARGESS objects with a reliable redshift.

\subsubsection{Star-forming galaxies}
We expect that galaxies in our sample whose radio emission is dominated by processes related to star formation rather than an AGN will lie at redshift $z\leq0.3$ and have 1.4 GHz radio luminosity $L_{\rm FIRST}\leq 10^{24}$ W Hz$^{-1}$ \citep[similar limits were chosen by][]{best12}), since the inferred star formation rate needed  to produce the observed radio emission would otherwise be unrealistically high. We also expect galaxies whose radio emission arises mainly from star formation processes to obey the \cite{hopkins03}\ relation between H$\alpha$ and radio luminosity, since both these quantities are proxies for the star-formation rate. 

\begin{figure}
\centering
\includegraphics[width=0.49\textwidth]{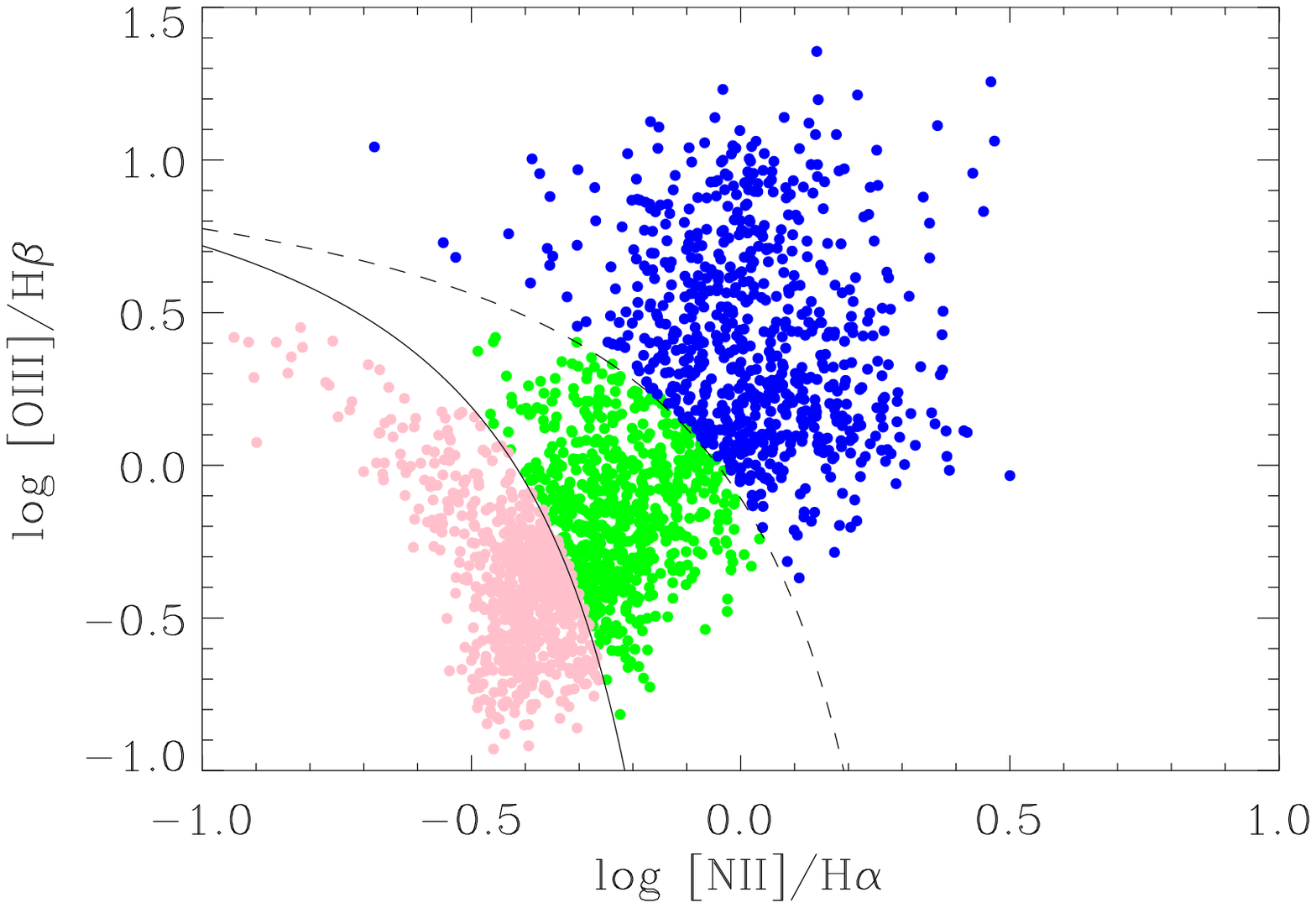}
\caption[The BPT emission line diagnostic diagram for the LARGESS sample]{The \citet{baldwin81} (BPT) emission-line diagnostic diagram for LARGESS sources with ${\rm SNR}>3$ in all four relevant emission lines. All objects below the \citet{kauffmann03a} limit (solid line; pink points) are assumed to have emission lines produced by ionizing radiation from star-formation regions. Galaxies above the \citet{kewley01} limit (dashed line) have emission lines arising from gas ionized by an AGN. Galaxies in between the \citet{kewley01} and \citet{kauffmann03a} limits (green points) are generally considered to be composite galaxies with some ongoing star formation.}
\label{fig:bpt}
\end{figure}

We first used the BPT diagnostic plot to identify objects with star-forming (SF) optical spectra (see Figure \ref{fig:bpt}). For galaxies with a signal-to-noise ratio (SNR) $> 3$ in each of the [OIII]$\lambda5007$, [NII]$\lambda6583$, H$\alpha$ and H$\beta$ lines, we define SF galaxies (pink points in Figure \ref{fig:bpt}) as those in the ``pure'' star-forming region of \cite{kauffmann03a}. 

For all galaxies not already classified as SF in the BPT diagnostic plot, we then compared the star-formation rate estimates inferred from the H$\alpha$ line with the star-formation rate estimates inferred from the 1.4\,GHz radio luminosity using the relations from \cite{hopkins03}. 

\begin{figure}
\centering
\includegraphics[width=0.49\textwidth]{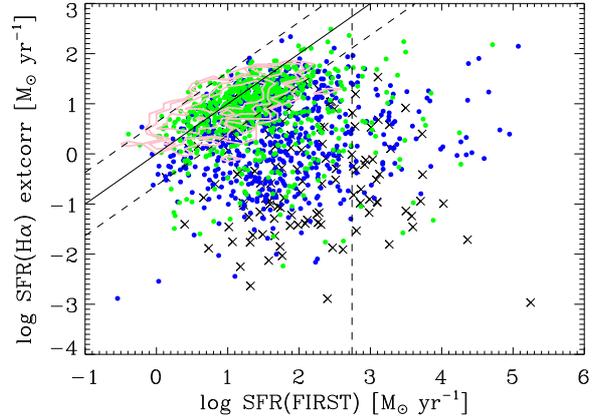}
\caption[Comparison between the SFR estimated from H$\alpha$ to the SFR derived from the total FIRST flux density]{Comparison between the SFR derived from H$\alpha$ to the SFR derived from the total FIRST flux density using the relation in \citet{hopkins03}. The green and blue points are respectively AGN and composite galaxies as defined by a BPT emission line diagnostic. We overlay contours for SF galaxies in pink. The crosses indicate galaxies that were not able to go on a BPT diagram, but have significant H$\alpha$ and H$\beta$ to estimate the SFR(H$\alpha$). The solid diagonal line is a one-to-one line, and the 3$\sigma$ limits to the relationship from \citet{hopkins03} are shown as the diagonal dashed lines. The vertical dashed line is the SFR inferred at a radio flux of 10$^{24}$ W Hz$^{-1}$, beyond which star-formation processes are unlikely to dominate the observed radio emission in our sample.}
\label{fig:sfr_radiovha}
\end{figure}

Galaxies where the optical spectrum was classified as an AGN in the BPT diagram were reclassified as star-forming galaxies if their radio luminosity placed them within 3$\sigma$ of the one-to-one relation in Figure \ref{fig:sfr_radiovha}, based on the methodology used in \cite{bardelli10}. This allows us to identify the dominant process for the radio emission from star-forming galaxies that also contain a radio-quiet AGN. 
The remaining objects with an AGN spectral classification in the BPT diagram constitute a robust sample of radio-loud AGN.

%Figure 11

%Figure 12

\subsubsection{Radio-loud AGN: separating HERGs and LERGs}

From this robust sample of radio-loud AGN, we separated low- and high- excitation radio galaxies using a cut in [OIII]$\lambda5007$ equivalent width (EW([OIII]$\lambda5007$)). 

\begin{figure}
\centering
\includegraphics[width=0.49\textwidth]{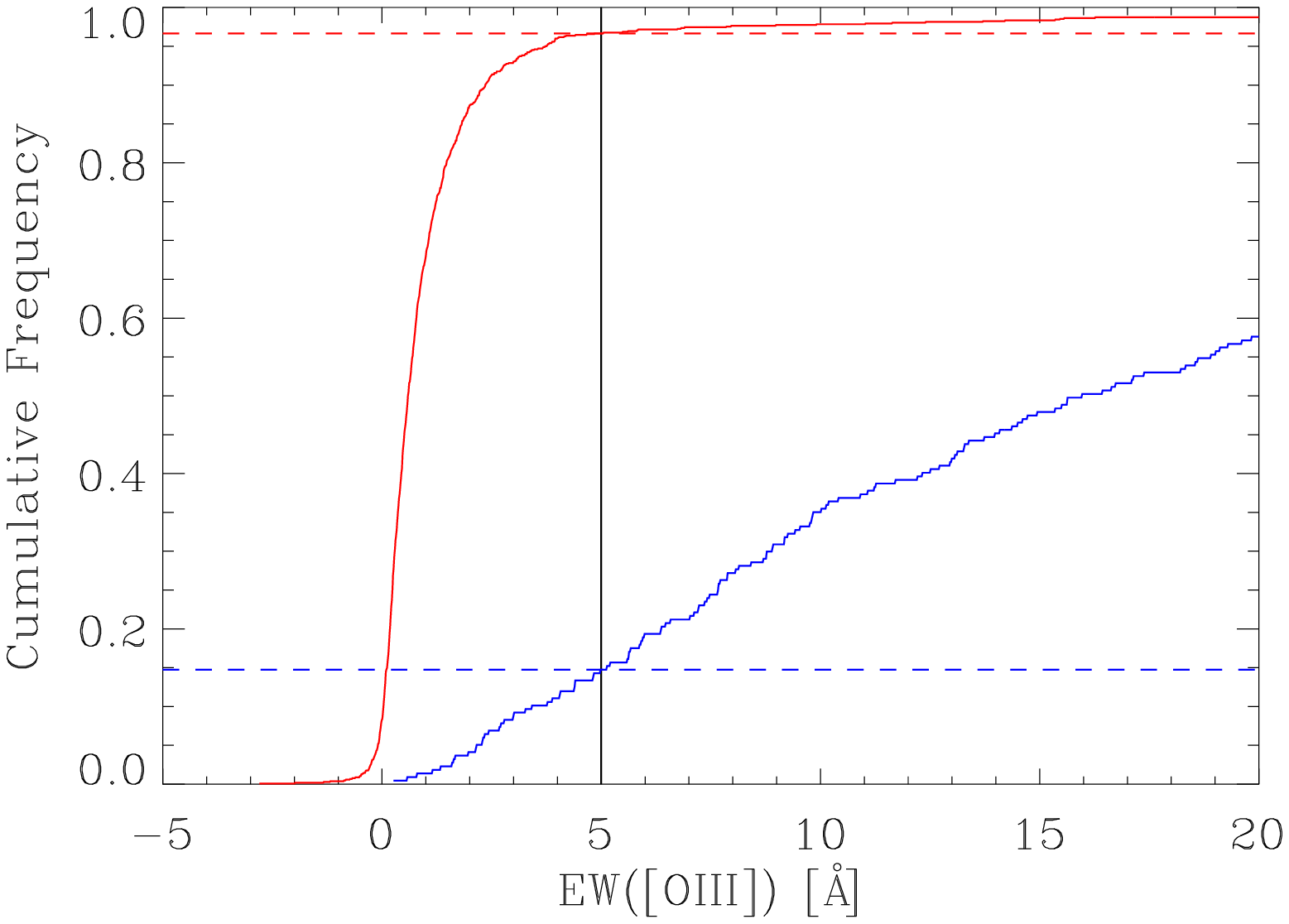}
\caption[Cumulative distribution of {[OIII]$\lambda5007$} equivalent width for Ae and Aa spectroscopic classes]{Cumulative distribution of [OIII]$\lambda5007$ equivalent width for galaxies with {\tt VISCLASS} Aa (absorption-line spectrum, red line) and Ae (spectrum with strong optical emission lines, blue line). The vertical line shows the 5\AA\ EW([OIII]$\lambda5007$) limit used to separate HERGs and LERGs. The horizontal red and blue dashed lines are the fraction of sources below this limit for Aa and Ae galaxies respectively. 
}
\label{fig:o3_cumdn}
\end{figure}

We defined high-excitation radio galaxies (HERGs) as those with SNR([OIII]$\lambda5007$) $> 3$ and EW([OIII]$\lambda5007$) $> 5$\AA. The choice of an EW([OIII]$\lambda5007$) $> 5$\AA~cutoff is based on a comparison of EW([OIII]$\lambda5007$) with the visual classification (see Figure \ref{fig:o3_cumdn}), and is the same cutoff value used by \cite{best12} to separate HERGs and LERGs in their SDSS sample.  
All other radio-loud AGN were classified as  low-excitation radio galaxies (LERGs). 
As can be seen from Figure \ref{fig:o3_cumdn}, this dividing line at EW([OIII]$\lambda5007$) $> 5$\AA\ also gives results that are generally consistent with our visual Aa and Ae classification.

%Figure 13

\subsection{Comparison of the automated and visual spectral classifications}

Table \ref{tab:full_specclass} shows a pairwise comparison between the qualitative (visual) and quantitative (automated) classifications for the 4,058 LARGESS objects for which both measures are available. 

For objects with weak or no emission lines, there is very good agreement between the Aa visual classification and the LERG automated classification.  
Some objects that were visually identified as having optical emission lines (i.e. class Aae, Ae and SF) are also classified as LERGs by the automated criteria. This is probably because the human eye is able to recognize weak emission lines that fall below the 5\AA\ EW([OIII]$\lambda5007$) limit used to separate HERGs and LERGs in the automated classification. Most visual Aae objects ($\sim81\%$) have an automated classification as LERGs, while the great majority of visual Ae objects ($\sim80\%$) are classified as HERGs. We therefore find a strong consistency between the visual and automated classifications for the optical spectra of radio AGN. 

For objects with stronger emission lines, the AGN/SF classification also appears robust in most cases. Of 591 objects classified visually as emission-line AGN (class Ae), only 23 (3.9\%) were reclassified as star-forming galaxies (SF) based on the BPT diagram and H$\alpha$/radio continuum comparison. For the visual AGN sample as a whole (classes Aa, Aae and Ae combined) the fraction reclassified as star-forming is $<1$\%.  We therefore conclude that the overall level of contamination of our AGN sample by star-forming objects is very low, and that our separation of AGN and SF radio sources is generally self-consistent and reliable. 

\subsection{Final spectroscopic classifications}

The best spectral classification ({\tt BESTCLASS}) is a combination of both the visual and automated classifications. All sources flagged as either Star, AeB, Unusual or BLLac in the {\tt VISCLASS} are also flagged as such in the {\tt BESTCLASS}. In all other cases, the  {\tt BESTCLASS} is set to the automated classification as described above, or set to ``NA'' if an automated classification is unavailable.

%Table 7
\ctable[
notespar,
cap = {Comparison between the visual and automated classification schemes},
caption = {Comparison between the visual spectral classification and the automated classification for the 4,058 objects with both types of classifications.},
label = {tab:full_specclass}
]{l rrrrr r}%
{ %\tnote[a]{footnote}
}{
\FL
Automated \ & \multicolumn{5}{c}{Visual Classification}  & Total\\
\cline{2-6}
classification &     Aa &  Aae &   Ae & SF & Star \\
\hline\hline
LERG & 2,474 & 360 & 121 & 132 & 0 & 3,087\\
HERG & 31 & 78 & 447 & 108 & 0 & 664\\
SF & 1 & 6 & 23 & 185 & 0 & 215\\
Star & 0 & 0 & 0 & 0 & 92 & 92\\
\hline
Total & 2,506 & 444 & 591 & 425 & 92 & 4,058
\LL
}

%Table 8
\ctable[
notespar,
cap = {Summary of final spectroscopic classifications},
caption = {Final spectroscopic classifications for the 10,856 LARGESS objects with a reliable ($Q\geq3$) redshift measurement. },
label = {tab:final_specclass}
]{l rrrrr r}%
{ %\tnote[a]{footnote}
}{
\FL
Class &  \multicolumn{3}{c}{Redshift}  \\
          &   All       & $z\leq0.8$  & $z>0.8$ \\
\hline\hline
LERG  & 5881 &   5864   &    17  \\  % 69\%
HERG  &   839 &    827   &     12  \\  % 10\%
AeB     &  1615 &    397   & 1218  \\  %  5% 
SF       &  1415 &  1415   &    ..    \\  % 17\%
Star     &   196  &    196   &   ..   \\
BL Lac &    19  &        5   &   14 \\
Unusual & 61  &       29   &  32  \\
Unclassified (NA) & 830 & 729 & 101 \\
\hline
Total & 10856 &  9462  & 1394 
\LL
}

Table \ref{tab:final_specclass} summarizes the final spectroscopic classifications for the full sample. The combination of radio and optical flux limits for our sample means that most of the objects detected above redshift $z=0.8$ are radio-loud QSOs (class AeB), so we also list the classifications of objects with $z\leq0.8$ and $z>0.8$ separately.  Low-excitation radio AGN (LERGs) are the dominant population, accounting for almost 70\% of the objects at $z\leq0.8$. The optical and mid-infrared properties of the spectroscopic sample are discussed in more detail in \S 9. 

%Table 9
\ctable[
star,
notespar,
cap = {Description of the columns in the LARGESS data table},
caption = {Description of the columns in the main LARGESS data table (Table \ref{tab:sample_data}). The format codes are \textsc{fortran} format descriptors. All SDSS photometric data are from the SDSS sixth data release.},
label = {tab:data_fmt}
]{l l rrp{8.5cm}}%
{ %\tnote[a]{footnote}
}{
\FL
Col. & Field & Format & Units & Description \\
\hline\hline
1 & NAME & a19 & - & IAU format object name \\
2 & SDSSID & i18 & - & SDSS photometric ID \\
3 & RA & f9.5 & deg & SDSS RA J2000 in decimal degrees \\
4 & DEC & f9.5 & deg & SDSS Dec J2000 in decimal degrees \\
5 & R\_PET & f6.3 & mag & SDSS Petrosian magnitude in $r$ band (extinction corrected) \\
6 & R\_PET\_ERR & f7.3 & mag & SDSS Petrosian magnitude error in $r$ band \\
7 & I\_MOD & f6.3 & mag & SDSS Model magnitude in $i$ band (extinction corrected) \\
8 & I\_MOD\_ERR & f6.3 & mag & SDSS Model magnitude error in $i$ band \\
9 & G\_MOD & f6.3 & mag & SDSS Model magnitude in $g$ band (extinction corrected) \\
10 & G\_MOD\_ERR & f6.3 & mag & SDSS Model magnitude error in $g$ band \\
11 & N\_FIRST & i2 & - & Number of FIRST components \\
12 & FIRST\_TOT & f7.2 & mJy & FIRST total integrated flux \\
13 & N\_NVSS & i1 & - & Number of NVSS components; $-1$ = Null value \\
14 & NVSS\_TOT & f9.3 & mJy & NVSS total integrated flux; $-99.000$ = Null value \\
15 & Z & f9.5 & - & Final best redshift; NaN = Null value \\
16 & QOP & i3 & - & Redshift reliability flag \\
17 & ZSOURCE & a12 & - & Survey source for the best redshift; NA = Null value \\
18 & OIII\_SN & e9.2 & - & [OIII]$\lambda5007$ SNR; NaN = Null value \\
19 & EW\_OIII & e9.2 & \AA & [OIII]$\lambda5007$ equivalent width; NaN = Null value \\
20 & VISCLASS & a13 & - & Best visual classification; NA = Null value \\
21 & BESTCLASS & a13 & - & Best spectroscopic classification; NA = Null value \\
22 & HI\_COMP & i1 & - & High-completeness region flag; 1 = in region; 0 = not in region \\
23 & ZSOURCE\_TARGET & i2 & - & Radio filler target flag for GAMA/WiggleZ;  1 = filler; 0 = not filler; $-1$ = {\tt ZSOURCE} is not GAMA/WiggleZ \\
24 & DISAGREE\_GAMA & i2 & - & Flag to indicate whether the listed redshift and quality code agree with the GAMA \textsc{autoz} (internal data: AATSpecAutozAllv22) estimate; 1 = disagree; 0 = agree; $-1$ = {\tt ZSOURCE} is not GAMA
\LL
}

%Table 10 
\begin{sidewaystable*}
\vspace*{-16cm}
\caption{A list of 20 sample objects from the main LARGESS data table -- the full table contains 19,179 sources. For a description of the data in each column, see Table \ref{tab:data_fmt}. }\label{tab:sample_data} 
%\resizebox{\textheight}{!}{
\begin{threeparttable}
\footnotesize
\begin{tabular}{c c  c  c  r  r  r  r  r  c  c  r  }
\hline
\mc{1}{c}{(1)} & \mc{1}{c}{(2)} & \mc{1}{c}{(3)} & \mc{1}{c}{(4)} & \mc{1}{c}{(5)} & \mc{1}{c}{(6)} & \mc{1}{c}{(7)} & \mc{1}{c}{(8)} & \mc{1}{c}{(9)} & \mc{1}{c}{(10)} & \mc{1}{c}{(11)} & \mc{1}{c}{(12)}  \\
\hline
\hline
J090001.05-000852.8  & 588848899892969855  & 135.00441  & -0.14802  & 19.507  &  0.063 &  18.814  &  0.020  &  21.270  &  0.084  & 1 &    4.77  \\  
J090001.28+053602.1 & 587732703391777311  & 135.00534  &  5.60060  &  19.676  &  0.059 &  19.120  &  0.027  &  21.212  &  0.079  & 1 &    4.12  \\ 
J090001.85+022231.7 & 587727944564277289  & 135.00772  &  2.37549  &  17.587  &  0.332 &  17.288  &  0.010  &  19.749  &  0.036  & 1 &    3.94  \\ 
J090002.65-003338.6  & 588848899356098923  & 135.01106  & -0.56073  &  20.494  &  0.061 &  20.443  &  0.045  &  20.871  &  0.037  & 1 &  14.49  \\ 
J090003.71+073056.7 & 587735343184937333  & 135.01547  &  7.51575  &  18.323  &  0.034 &  17.565  &  0.012  &  19.887  &  0.043  & 3 &  36.79  \\ 
J090004.24+033318.4 & 588010359603463018  & 135.01768  &  3.55514  &  20.802  &  0.185 &  19.544  &  0.034  &  22.483  &  0.212  & 1 &    6.21  \\  
J090004.52-002548.7  & 588848899356099571  & 135.01883  & -0.43020  &  20.953  &  0.185 &  19.836  &  0.043  &  23.021  &  0.356  & 1 &  14.36  \\  
J090004.66+000332.1 & 587725074990235700  & 135.01943  &  0.05893  &  18.129  &  0.023 &  17.541  &  0.009  &  19.487  &  0.020  & 1 &   12.89 \\  
J090005.05+000446.7 & 587725074990235725  & 135.02106  &  0.07966  &  15.146  &  0.009 &  14.621  &  0.003  &  15.721  &  0.003  & 1 &    5.33  \\ 
J090005.85+073634.0 & 587735343185003255  & 135.02439  &  7.60945  &  21.164  &  0.220 &  20.481  &  0.075  &  24.924  &  1.026  & 2 &  24.37  \\
J090005.87+072548.4 & 587734948047356113  & 135.02448  &  7.43012  &  17.793  &  0.021 &  17.123  &  0.009  &  18.817  &  0.019  & 1 &    6.92  \\  
J090006.26+021537.3 & 587727944564277524  & 135.02612  &  2.26038  &  20.031  &  0.045 &  19.751  &  0.029  &  19.973  &  0.020  & 1 &   24.36 \\  
J090006.43+022404.2 & 587727944564277661  & 135.02681  &  2.40119  &  19.238  &  0.045 &  18.528  &  0.019  &  20.823  &  0.057  & 2 &  15.75  \\  
J090008.02+033945.3 & 587728880868852173  & 135.03342  &  3.66259  &  20.226  &  0.069 &  19.692  &  0.033  &  21.559  &  0.077  & 1 &    9.86  \\  
J090010.12+023643.8 & 588010358529720530  & 135.04218  &  2.61218  &  19.518  &  0.218 &  19.308  &  0.027  &  21.353  &  0.080  & 2 &  12.06  \\  
J090010.57+080548.0 & 587735343721874462  & 135.04404  &  8.09667  &  20.998  &  0.231 &  19.721  &  0.053  &  22.750  &  0.407  & 1 &    1.06  \\  
J090011.14+050257.6 & 587732578298888642  & 135.04642  &  5.04936  &  20.094  &  0.066 &  18.955  &  0.025  &  21.739  &  0.131  & 1 &    5.02  \\  
J090012.53+023539.1 & 588010358529720551  & 135.05221  &  2.59422  &  17.335  &  0.015 &  16.922  &  0.007  &  18.148  &  0.010  & 1 &    1.08  \\ 
J090013.92+024717.0 & 588010358529720589  & 135.05803  &  2.78806  &  19.558  &  0.028 &  19.392  &  0.019  &  19.858  &  0.018  & 1 & 144.76 \\  
J090014.01+053549.7 & 587732703391777613  & 135.05840  &  5.59715  &  21.996  &  0.291 &  20.418  &  0.084  &  23.186  &  0.449  & 1 &    1.45  \\  
\hline
\hline
 \mc{1}{c}{(13)} & \mc{1}{c}{(14)} & \mc{1}{c}{(15)} & \mc{1}{c}{(16)} & \mc{1}{c}{(17)} & \mc{1}{c}{(18)} & \mc{1}{c}{(19)} & \mc{1}{c}{(20)} & \mc{1}{c}{(21)} & \mc{1}{c}{(22)} & \mc{1}{c}{(23)} & \mc{1}{c}{(24)}\\
\hline
 1 &     5.051  &  0.40906   & 4    & WiggleZ &      0.52  &    0.57  &  Aa        & LERG & 1 &  1 & -1 \\
 1 &     3.100  &  0.30267   & 4    & WiggleZ &    23.20  &   31.90 &  Ae        & HERG & 1 &  1 & -1 \\
 1 &     6.000  &  0.25091   & 4    & SDSS    &      2.27  &    0.73  & Aa        & LERG & 1 & -1 & -1 \\
 1 &   13.700  &  1.00783   & 4    & GAMA   &      2.21  &  161.00 &  AeB      &  AeB & 1 &  1 &  1 \\
 1 &   41.200  &  0.38434   & 5    & SDSS   &      1.46   &   0.23    &   NA        & LERG & 1 & -1 & -1 \\
 1 &    7.846   &  0.64148   & 3    & WiggleZ &     -0.73  &   -3.18   & Aa         & LERG & 1 &  1 & -1 \\
 1 &  14.600   &  0.61092   & 4    & WiggleZ &      1.43  &     ..       & Aa         & LERG & 1 &  1 & -1 \\
 1 &  56.000   &  0.26221   & 4     &  GAMA  &      6.65  &    3.05   & Aa         & LERG & 1 &  0 &  0 \\
 0 & -99.000   &  0.05386   & 4     & SDSS   &     22.00  &    2.74   &   SF      &   SF & 1 & -1 &  -1 \\
 1 &  29.400   &  0.00000   & -1    &    NA    &          ..  &        ..  &      NA        &  NA & 1 & -1  & -1 \\
 1 &    7.500   &  0.00000   & -1    &    NA    &           ..  &        ..  &      NA       &   NA & 1 & -1 & -1 \\
 1 &  23.647   & 0.61672    & 4     & SDSS  &           ..  &        ..  &     AeB      &   AeB & 1 & -1 & -1 \\
 1 &  21.600   & 0.34938    & 4     &  GAMA  &   0.54    &    0.27  &    Aa       & LERG & 1 &  0 &  0 \\
 1 & 10.400    & 0.35708    & 4     & WiggleZ &   1.39   &    1.17   &  Aae      & LERG & 1 &  1 & -1 \\
 1 &  13.500   & 0.20071    & 3      & GAMA  &    1.12   &         ..  &    Aa       & LERG & 1 &  0 &  0 \\
 0 & -99.000   &  0.00000   & -1   &    NA      &         ..  &        ..  &     NA       &   NA & 1 & -1 & -1 \\
 1 &    5.300   &  0.00000   & -1    &     NA     &         ..  &        ..  &    NA       &   NA & 1 & -1 & -1 \\
  0 &  -99.000  &  0.20000   & 4     & SDSS   &      6.08 &      1.86 &   SF      &    SF & 1 & -1 &  -1 \\
 1 &  140.700 &  1.18978   & 4     & SDSS   &          ..  &        ..  &    AeB     &    AeB & 1 & -1 &  -1 \\
 0 &  -99.000  &  0.75797   & 4    & WiggleZ &     8.04  &   21.90 &   Ae      &   HERG & 1 &  1 & -1 \\
\hline
\end{tabular}
\end{threeparttable}
%}
\end{sidewaystable*}

\section{The spectroscopic data table}

Tables \ref{tab:data_fmt} and \ref{tab:sample_data} present the final LARGESS spectroscopic data catalogue, ordered by Right Ascension. 
Table \ref{tab:data_fmt} describes each column of the data table, 20 lines of which are shown in Table \ref{tab:sample_data}. These are the first 20 objects after RA 09:00:00, an RA range that lies within one of our high-completeness GAMA fields (see Figure 1). 
The table includes optical positions and unique identifiers for each radio target. We also include extinction corrected SDSS $g,r$ and $i$ photometry and their errors, along with the total radio flux from FIRST (and NVSS if available) and the number of radio components associated with each optical object. The spectroscopic parameters included are the best redshift, quality code and the origin of the spectrum used for the final redshift. 

As explained in \S7, there are two spectroscopic classifications designed to separate star-forming galaxies from low- and high-excitation AGN. 
The first is a qualitative method based on visual inspection of each spectrum ({\tt VISCLASS}), and the second is a final best classification based on both the automated and visual methods ({\tt BESTCLASS}). Additionally we provide three flags: {\tt HI\_COMP} to indicate if a source is in a region with high spectroscopic completeness, {\tt ZSOURCE\_TARGET} to indicate if a source is a filler target explicitly observed for the LARGESS sample by the GAMA or WiggleZ team, and {\tt DISAGREE\_GAMA}, which indicates that our best redshift/quality is estimated from a GAMA spectrum, but differs from the previous GAMA \textsc{autoz} (internal data: AATSpecAutozAllv22) redshift or quality.

%Figure 14
\begin{figure*}
\centering
\includegraphics[width=\textwidth]{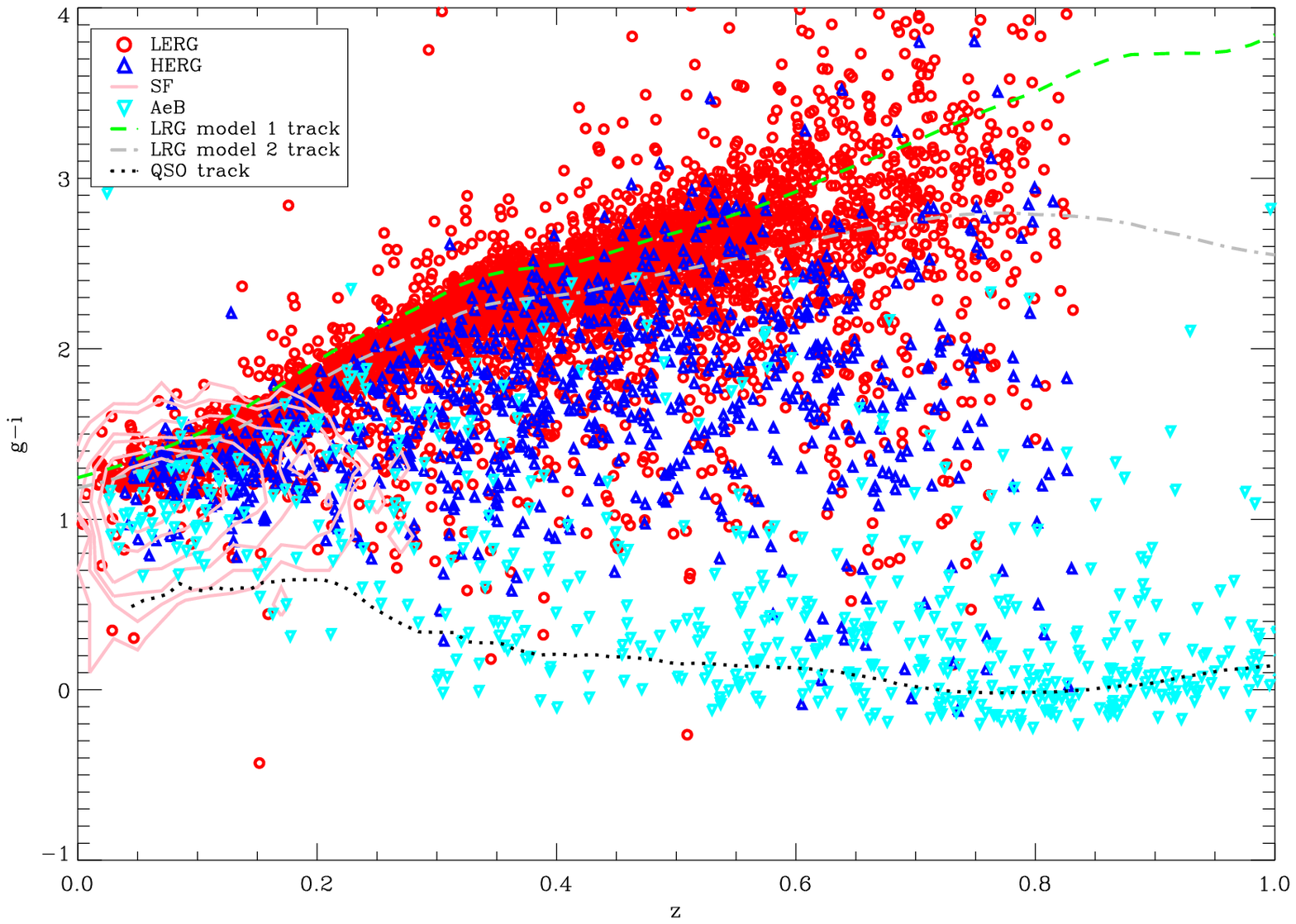}
\caption[$(g_{\rm mod}-i_{\rm mod})$ colour as a function of redshift for LARGESS galaxies separated by the spectroscopic classification]{The $(g_{\rm mod}-i_{\rm mod})$ colour as a function of redshift for the LERGs (red open circles), HERGs (blue open triangles), AeB objects with strong, broad Balmer emission lines (cyan open inverted triangles) and contours for SF galaxies. We show the tracks of two LRG models from \citep{wake06} (model 1, green dashed line; model 2, grey dash-dot line; see text for details) and median observed colours for QSOs (dotted line) with $i_{\rm PSF}<18.0$ from the SDSS DR3 QSO catalogue \citep{schneider05,croom09} on this plane. All three classes of radio galaxies have different spectral energy distributions (SEDs). The distribution of colour with redshift for AeB sources are quite flat with redshift suggesting a power-law SED, which is consistent with them following the QSO track. LERGs tend to follow the LRG track. The HERGs lie somewhere in between the LERG and AeB classes.}
\label{fig:giVz}
\end{figure*}

\section{LARGESS sample characteristics}\label{sec:sample_character}

We now discuss some general properties of the objects in the LARGESS data catalogue.  

\subsection{Optical colour versus redshift}\label{subsec:gi_vs_z}

Figure \ref{fig:giVz} shows the observed $(g_{\rm mod}-i_{\rm mod})$ optical colour as a function of redshift for all LARGESS objects with a reliable spectroscopic redshift and classification, split into the four main spectral classes (HERG, LERG, SF and AeB). This is a key plot for the LARGESS sample, and shows the relationship between optical colour and spectroscopic class for the full range of radio-selected AGN out to redshift $z>0.8$. 

The broad emission-line (AeB) objects have the bluest colours at all redshifts.  
At redshifts above $z\sim0.25$ their $(g_{\rm mod}-i_{\rm mod})$ colour is relatively flat as a function of redshift, and most AeB objects lie close to the track for optically-selected SDSS QSOs (the dotted line in Figure \ref{fig:giVz}), implying that their optical light is dominated by a non-stellar power-law spectrum \citep{peterson97}.
At lower redshift ($z<0.25$), most of our AeB objects have significantly redder colours than the track for optically-selected QSOs. This is probably due to a higher contribution from the host galaxy stellar light, since the AeB objects in our sample are selected via their radio emission (and are spectroscopically-defined), rather than being colour-selected like the SDSS QSOs. 
In summary, most of the AeB objects in the LARGESS catalogue have $(g_{\rm mod}-i_{\rm mod})$ colours similar to those of optically-selected QSOs. 

The LERGs in our sample are typically the reddest objects at all redshifts, with redder $(g_{\rm mod}-i_{\rm mod})$ colours at higher redshift. This 
is consistent with most LERG hosts being galaxies with an old, passively-evolving stellar population, i.e. Luminous Red Galaxies \citep[LRGs; ][]{eisenstein01}. Figure \ref{fig:giVz} also shows colour-redshift tracks for the two LRG models used by \cite{wake06} and derived using the \cite{bruzual93} stellar population synthesis code. Model 1 (green dashed line) is for a single 10 Gyr starburst, and evolves passively without any further star formation. Model 2 (grey dash-dot line) has 95\% of the final mass in a single burst and 5\% as a continuous level of star formation. Most LERGs lie near or in between these tracks, though a few LERGs scatter to much redder and bluer colours (especially at higher redshift) than models 1 and 2 respectively.

The HERGs generally lie in between the colours of the AeB and LERG classes at all redshifts, with bluer colours than the LRG model 2 and redder colours than typical QSOs. There are several plausible reasons why HERG host galaxies might have bluer optical colours than LERG hosts at the same redshift:  

\begin{enumerate} 
\item
Some optical light may come from a blue AGN continuum, at a lower level than seen in the the AeB objects, even though broad Balmer emission lines are not observed. In unified AGN models \citep[e.g.][]{antonucci93}, these would be objects where the central dusty torus has a smaller opening angle than in the AeB systems but some AGN light can still be seen. 
\item 
The HERG host galaxies may have some ongoing star formation, and contain a substantial young or intermediate-age stellar population, even though their observed radio emission is produced mainly by the central AGN. 
Some supporting evidence for this comes from stellar-population studies of local \citep{best12} and more distant \citep{johnston08}\ radio AGN, which find consistent evidence for a younger stellar population in radio galaxies with strong emission lines. \cite{johnston08} also found that the composite spectra of emission-radio AGN at $z\sim0.55$ were better fitted by a mixture of an old plus intermediate-age stellar population, rather than an old population with a non-stellar AGN continuum.  
\item
If the HERG hosts are typically lower-mass galaxies than the LERG hosts \citep[as found for low-redshift systems by e.g.][]{best12}, then they would also be expected to have lower metallicity and slightly bluer colours \citep{tremonti04}.  
\end{enumerate} 

Figure \ref{fig:giVz} does not allow us to distinguish between these possibilities, and in the next section (\S 9.2) we consider the additional information provided by mid-infrared photometry.  We also discuss the relationship between HERGs and AeB objects further in \S 10.1, and in \S10.2 and \$10.3 we compare the properties of HERG and LERG host galaxies using sub-samples matched in stellar mass, radio luminosity and redshift. 

Finally the SF class, which are only present in the LARGESS sample at low redshift ($z<0.3$), show a spread of $(g-i)$ values which is consistent with a range in both current star-formation rate and dominant stellar populations in these systems \cite[see e.g.][]{taylor11}.

\subsection{WISE mid-infrared colours}\label{sec:wise_data}

The Wide-field Infrared Survey Explorer \citep[WISE;][]{wright10} is an all-sky mid-infrared imaging survey. It provides photometry in four bands (W1--4) centred at 3.4, 4.6, 12, and 22 $\mu$m, with angular resolutions (FWHMs) of 6.1, 6.4, 6.5, and 12.0 arcsec in each band respectively. 

For extragalactic objects, the WISE data can provide useful information about recent star formation activity (vibrationally excited polycyclic aromatic hydrocarbon emission; rest-frame 6.2, 7.7, 8.6, and 11.3 $\mu$m), silicate absorption (rest-frame 10$\mu$m) and dust heated by strong radiation fields (e.g. AGN accretion disks; rest-frame 14--16$\mu$m). At shorter rest-wavelengths (1.6--4.5$\mu$m), the WISE W1 and W2 bands typically trace the Rayleigh-Jeans ($F_{\nu} \propto \nu^{2}$) tail of the old stellar population, which may be used to estimate the stellar mass of galaxies out to moderate redshifts \citep{xilouris04,wilman08,hwang12}. The W3 band in particular may also be strongly affected by the power-law emission ($F_{\nu} \propto \nu^{-\alpha}$) of an AGN \citep{jarrett11}.

To examine the mid-infrared properties of the LARGESS sample, we cross-matched the full data catalogue with the all-sky AllWISE data release\footnote{{http://wise2.ipac.caltech.edu/docs/release/allwise/}} \citep{cutri13}, using a 3\,arcsec matching radius. 
The great majority of LARGESS objects (18,124, or 94.5\%) have a WISE match, and 11,814 of these (95.8\% of the the LARGESS objects with good-quality spectra) have an reliable redshift. 
The reliability of the WISE cross-matching is estimated as $\sim99$\%.

In the following analysis we only consider objects with a reliable detection in each WISE band, i.e. we require that the SNR$_{\rm WISE} \geq 2$, reduced chi-square of the profile fit $\chi_{\rm WISE}^{2} \leq 3$, and there is no contamination and confusion flag (i.e. {\tt cc\_flag} = 0), for that band. Table \ref{tab:wise_good} shows the number of objects that satisfy this requirement for each WISE band. The highest number of reliable detections is in the W2 band, which is typically dominated by emission from the stellar galaxy. Although the W1 band is the most sensitive, it does not have the highest detection rate because the spectral energy distribution of AGN has a positive gradient (i.e. higher flux at longer wavelength) for wavelengths between 1 and 13 $\mu$m \citep[see][]{assef10}, so some AGNs may be detected in W2 but not W1. The lowest number of detections is in the W4 band, which is the least sensitive band and mainly detects emission from hot dust.

%Table 11
\ctable[
notespar,
cap = {Number of reliable WISE detections in each band},
caption = {Number of LARGESS sources with reliable WISE photometry in each band. This is split for: the full sample, those with spectra, and those with reliable redshifts. The final row gives the number with reliable WISE photometry in all four bands.},
label = {tab:wise_good}
]{l rrr}%
{ %\tnote[a]{footnote}
}{\FL
WISE band &  & \multicolumn{2}{c}{With Spectra} \\
\cline{3-4}
        &     Full sample & All & $\emph{QOP}\geq3$\\
\hline\hline
W1 & 15,926 & 10,075 & 8,784\\
W2 & 17,213 & 11,194 & 9,902\\
W3 & 7,782 & 5,609 & 5,317\\
W4 & 4,908 & 3,599 & 3,461\\
All & 3,585 & 2,597 & 2,503
\LL
}

\subsection{The WISE two-colour diagram}\label{subsec:mir_colours}
Since the release of the WISE data catalogue \citep{wright10}, many groups have used the unique wavelength coverage and sensitivity of WISE to identify and classify objects of interest for further study, including both Galactic \citep[e.g.][]{kirkpatrick11,cushing11} and extragalactic objects \citep[e.g.][]{lake12}. 
For QSOs, the WISE data can be used to identify both obscured and unobscured objects \citep[e.g.][]{jarrett11,bridge12,edelson12,mateos12,stern12}. 

\begin{figure*}
%\centering
\begin{minipage}\textwidth
\includegraphics[bb = 15 7 490 357,width=0.49\textwidth]{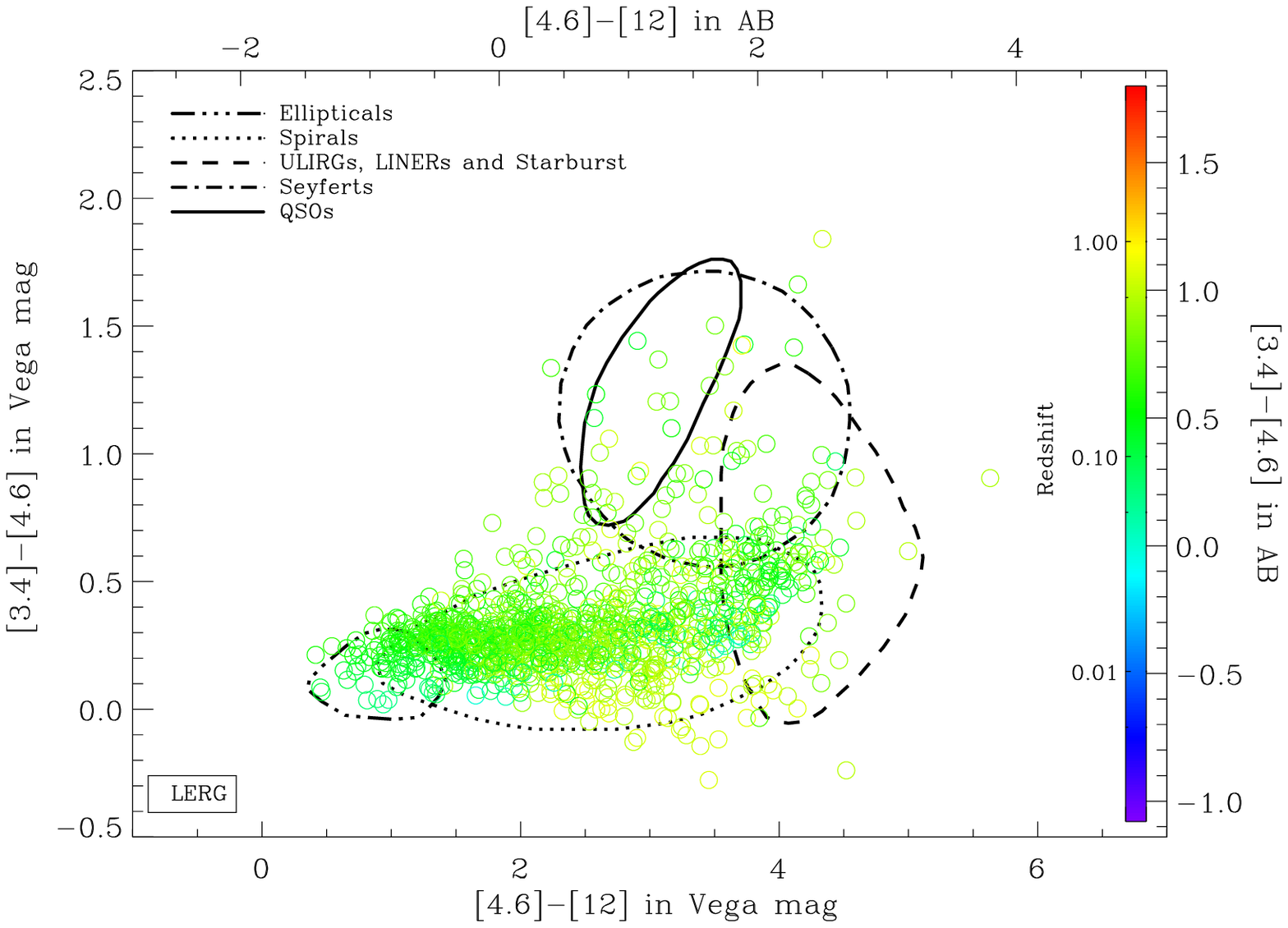}
\includegraphics[bb = 15 7 490 357,width=0.49\textwidth]{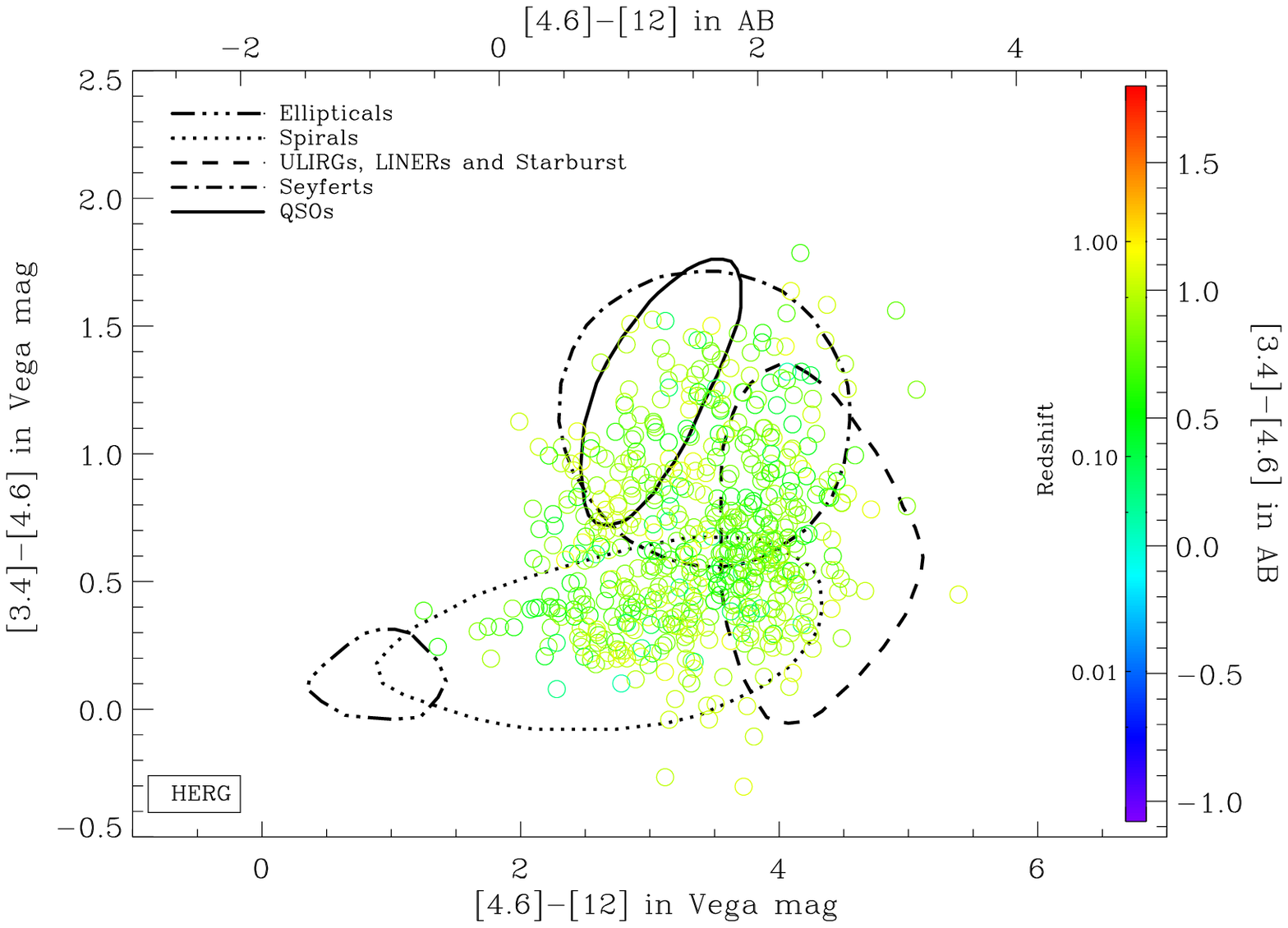}
\end{minipage}
\begin{minipage}\textwidth
\includegraphics[bb = 15 7 490 357,width=0.49\textwidth]{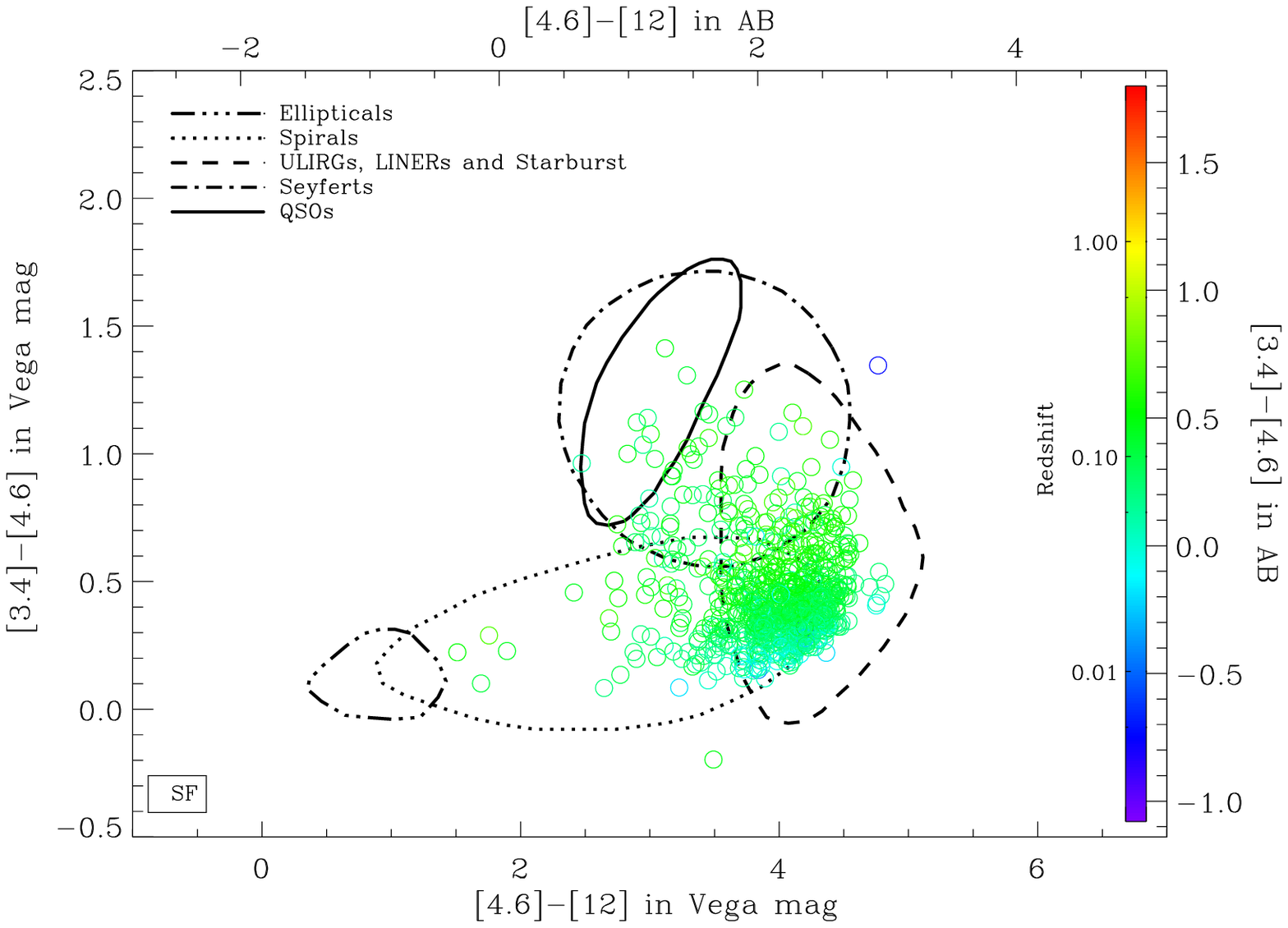}
\includegraphics[bb = 15 7 490 357,width=0.49\textwidth]{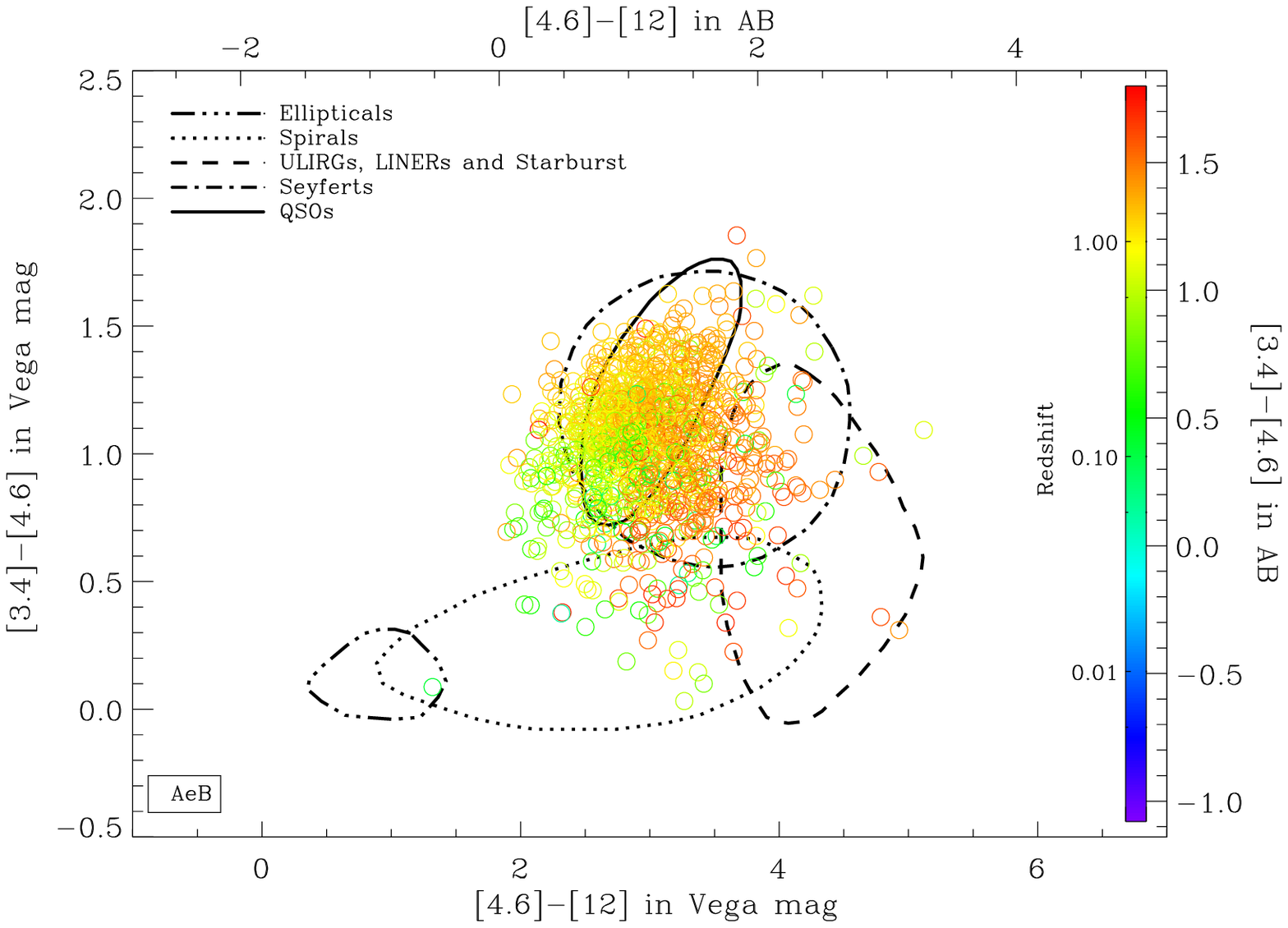}
\end{minipage}
\caption[WISE colour-colour digram of different spectroscopic classifications and colour coded by the redshift]{WISE mid-infrared two-colour diagrams for LARGESS objects split by spectral classification. Each plot shows the WISE (W1$-$W2) colour against (W2$-$W3) colour. The conversion from Vega magnitude to the AB magnitude system assumes a constant power-law for the WISE bands \citep{jarrett11}. The colour of individual data points is scaled with redshift as indicted on the right of each plot, and dashed and dotted lines highlight the expected locations of different classes of astronomical objects \citep{wright10}. } 
\label{fig:wise_W12W23_classz}
\end{figure*}

The WISE (W1$-$W2) versus (W2$-$W3) two-colour diagram is a powerful diagnostic tool for identifying the dominant source of MIR emission in individual objects. The diagram was first introduced by \cite{wright10}, who mapped out the regions occupied by different types of extragalactic sources by combining data from the Spitzer Wide-area InfraRed Extragalactic (SWIRE) survey with the models generated from the \textsc{grasil} \citep[GRAphite and SILicate;][]{silva98} code. These regions are highlighted by dashed and dotted lines in Figure \ref{fig:wise_W12W23_classz}. 

The power of this diagram lies in the fact that the WISE  (W2$-$W3) colour is a useful proxy for specific star-formation rate \citep{donoso12}, while the orthogonal (W1$-$W2) colour reflects the fractional contribution of non-stellar emission from an AGN. 
\cite{stern12} have shown that most AGN with a classical accretion disk can be reliably selected using a single colour cut of (W1$-$W2)$\geq$0.8, for WISE detections with ${\rm SNR(W2)}>10$ and ${\rm W2}<15.05$\,mag. 
In the following analysis, we only consider LARGESS objects with a reliable detection in each of the W1, W2 and W3 bands. 

Figure \ref{fig:wise_W12W23_classz shows the LARGESS objects in the WISE two-colour diagram, separated by their spectral classification and colour-coded by redshift. 
In these plots, } LERGs are the only objects found in the ``Ellipticals'' region, as expected for radio sources hosted by a massive galaxy with an old stellar population.  
The LERGs also extend well into the ``Spirals'' region, especially at higher redshift, implying that many of them contain a warm dust component in addition to an old stellar population. This is not too surprising, since modest amounts of dust are known to be common in normal elliptical galaxies \citep[e.g.][]{sadler85,rowlands12}. 
A warm dust component can arise from a range of processes (e.g. weak star formation or heating by post-AGB stars) that are not dominant at radio wavelengths nor related to an accretion disk, and so do not contradict the LERG classification. 

The SF and AeB objects fall mainly within the ``Starburst'' and ``QSO'' regions respectively.  For all three of the LERG, SF and AeB classes, there is a high degree of consistency between our optical  spectroscopic classification and the position of most objects in the WISE two-colour plot. 

The situation is rather different for the HERGs.  As can be seen from Figure \ref{fig:wise_W12W23_classz}, these objects are very scattered in the colour-colour plane, and span most of the the ``Seyfert'', ``Spiral'' and ``Starburst'' regions. The diverse mid-infrared (MIR) colours of the HERG population strongly suggest that this is a heterogeneous class of objects in which the physical process dominating the optical/MIR light is not the same for all members. 

The broad dispersion of HERG host-galaxy colours seen in Figure  \ref{fig:wise_W12W23_classz} is perhaps not surprising, since 
the W1$-$W2 colours of Seyfert galaxies are known to depend on the relative contribution of starlight by the host galaxy compared to the AGN contribution as well as the amount of dust extinction \cite[e.g.][and references therein]{stern12}. It therefore seems likely that the position of individual HERGS in the WISE two-colour plot depends on both the star-formation rate and obscuration within the host galaxy and the current accretion rate of gas onto the central black hole. 

%Figure 15

As in Figure \ref{fig:giVz}, LERGs are shown by red open circles, HERGs are blue open triangles and AeB objects by cyan inverted triangles. SF galaxies are plotted as pink stars.as indicated by the colour-coded labels on the left of the plot. The solid black box is the region defined by \citet{jarrett11} to select QSOs. 

\begin{figure}
\centering
\includegraphics[width=0.5\textwidth]{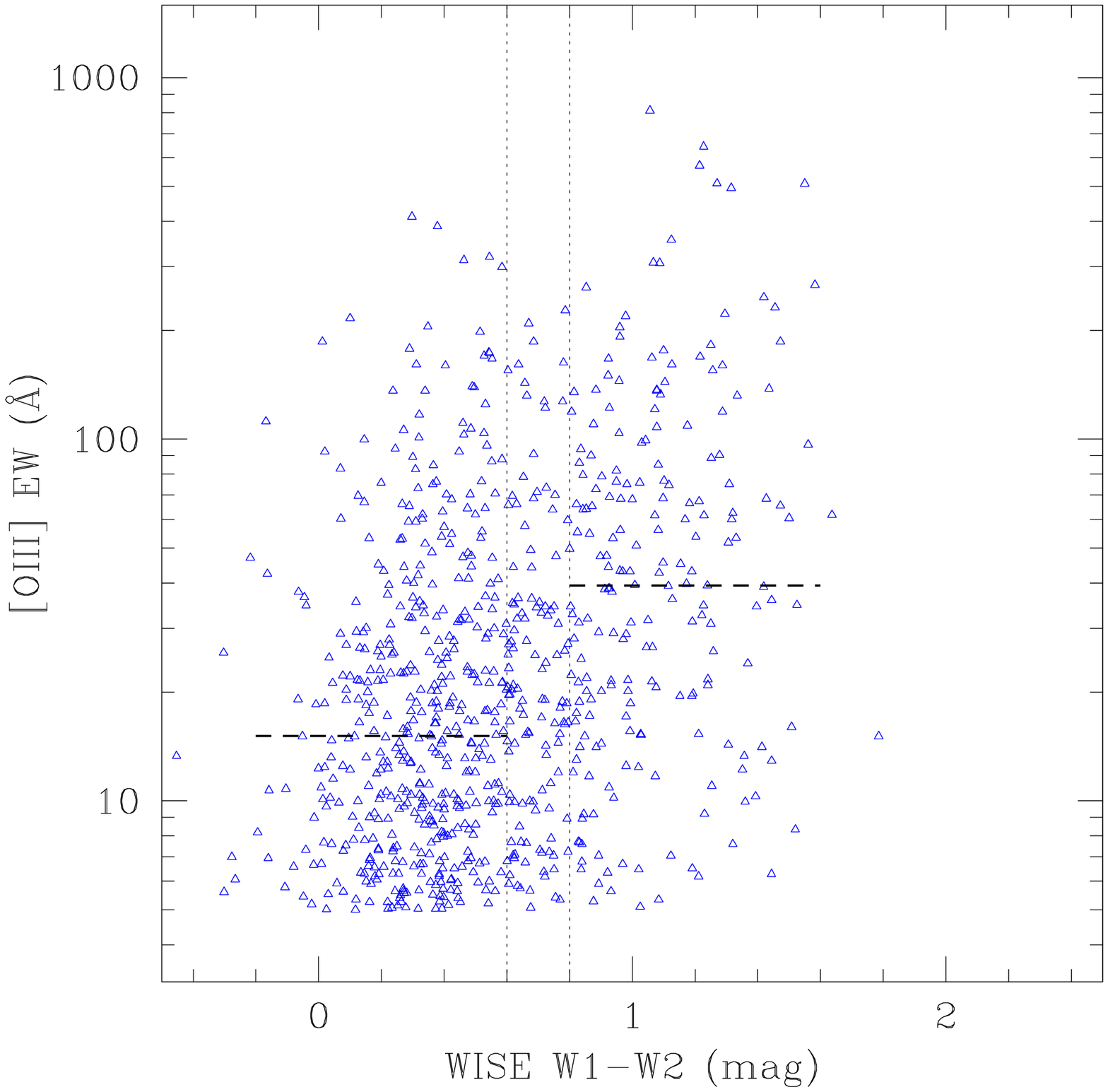}
\caption[]{Comparison of the optical [OIII] emission-line width and WISE W1-W2 colour for HERG host galaxies in the LARGESS sample, The vertical dotted lines separate objects in which the mid-IR light is dominated by the stellar galaxy (leftwards of the (W1-W2)=0.6 line) from those where the AGN dominates (to the right of the (W1-W2)=0.8 line). Horizontal dashed lines show the median value of EW [OIII] for these two sub-groups. 
}
\label{fig:em2_herg}
\end{figure}

Some additional supporting evidence for this picture comes from Figure \ref{fig:em2_herg}. This plots the WISE (W1-W2) colour for HERGs in our sample against the equivalent width of the [OIII] 5007\AA\ emission line - which can be used as a rough proxy for AGN accretion rate \citep[][]{kauffmann03a,heckman05,trouille10}. In general, the light of the underlying stellar population dominates the mid-IR SED in objects with W1-W2$<0.6$\, mag, while the AGN light dominates in objects with (W1-W2)$>0.8$\,mag \citep{stern12}.  

%Figure 16

We find a median [OIII] equivalent width of 15.1\AA\ for the stellar-dominated objects and 39.4\AA\ for the AGN-dominated objects, implying that the objects with a higher AGN accretion rate also have a higher average contribution from AGN light  in the mid-infrared. As can be seen from Figure \ref{fig:o3_cumdn}, 
the distribution of [OIII] equivalent widths in the LARGESS sample spans a broad continuum rather than having a bimodal distribution. It is therefore not surprising to see a broad diversity in the optical and mid-IR colours of HERGs, reflecting the underlying broad distribution of AGN accretion rates and the differing relative contributions of AGN and galaxy light. Since almost all the HERGs in which the galaxy light dominates have WISE colours characteristic of star-forming galaxies rather than quiescent red galaxies (see e.g. Figure 16), it seems likely that there is some ongoing star formation in most HERG host galaxies.

\section{The host galaxies of High- and Low- Excitation radio AGN}\label{sec:rg_prop}

We now look in  more detail at the host galaxies of distant radio AGN. Earlier studies (mainly of local galaxies at redshift $z<0.2$) have found that the host galaxies of HERGs and LERGs differ in their typical stellar mass \citep[e.g.][]{smolcic09a,best12}, star formation rate/history \citep[e.g.][]{herbert10,hardcastle13}, environment \citep[e.g.][]{best04,sabater13}, and many other properties \citep[][]{smolcic09a,best12,janssen12}. 

These differences cannot be accommodated within a single AGN unification model where the observed properties depend only on the orientation of a dusty torus. Instead, it is now generally accepted that a dichotomy exists between HERGs and LERGs, in which the key parameter is the accretion rate of gas onto the supermassive black hole \citep{hardcastle07,heckman14}. Since a dusty torus is not present in most LERGs, their observed optical properties are not expected to be orientation-dependent. 
In contrast, HERG host galaxies are expected to show orientation-dependent properties and we discuss this briefly in the next section. 

\subsection{AGN unification and the host galaxies of HERG and AeB objects} 

In AGN unification models \citep{antonucci93,urry95}, we would expect the strong narrow-line radio AGN in our sample (HERGs) to have the same underlying host-galaxy population as the broad-line radio AGN (AeB) and differ only in orientation. 
In this picture, the HERG optical spectra lack broad Balmer emission wings because the line-of-sight to the central broad-line region is obscured by a dusty torus. The optical--UV radiation from the accretion disk is absorbed by this torus and re-emitted at longer wavelengths \citep{congdon89}. In this scenario, the accretion disk continuum will not be seen at optical wavelengths in HERGs. Instead, light from the host galaxy of HERGs is expected to dominate the emission at optical wavelengths --  with the spectra of AeB objects having an additional optical contribution from the AGN accretion disk. 

\begin{figure}
\centering
\includegraphics[width=0.5\textwidth]{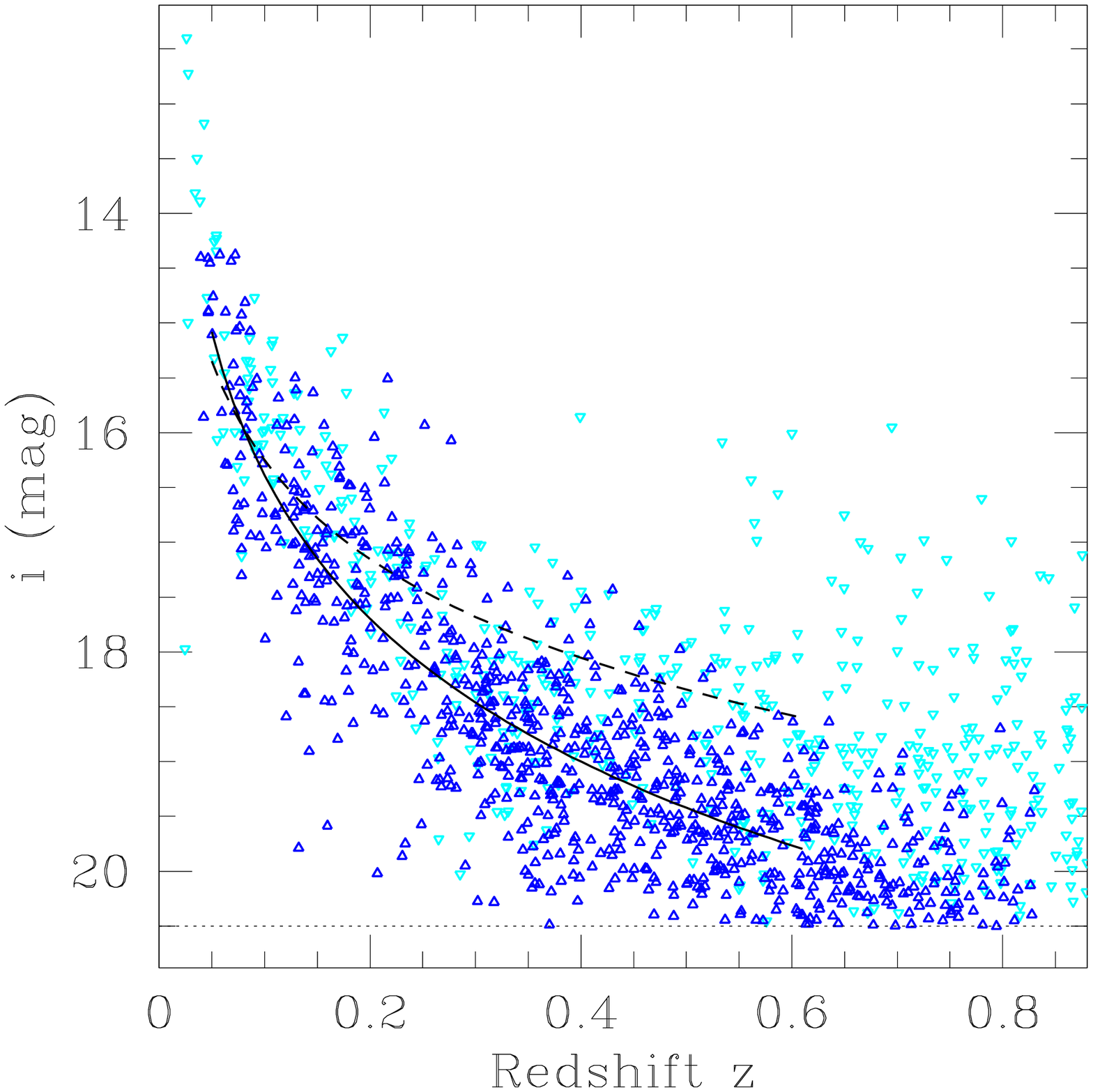}
\caption[]{A comparison of the $i_{\rm mod}$\ apparent magnitudes of the host galaxies of HERGs (dark blue points) and broad-line radio AGN (spectral class AeB; cyan points). The solid and dashed lines are least-squares fits to the HERG and AeB data points respectively, and the horizontal dotted line shows the optical limit of the LARGESS catalogue at  $i_{\rm mod}=20.5$\,mag. 
}
\label{fig:zmag}
\end{figure}

The LARGESS catalogue includes both HERG and AeB objects across a common range in redshift, and Figure \ref{fig:zmag} compares the observed $i_{\rm mod}$\ magnitudes of these two classes over the redshift range $0<z<0.8$. For redshifts out to $z\sim0.4$, most HERG and AeB objects lie well above the $i_{\rm mod}=20.5$\,mag cutoff of the catalogue and so we can compare the two classes directly.  As can be seen from  Figure \ref{fig:zmag}, the magnitude difference is small at $z<0.1$, but at higher redshifts the AeB objects start to become systematically brighter. 

%Figure 17

%Figure 18
\begin{figure}
\centering
\includegraphics[width=0.5\textwidth]{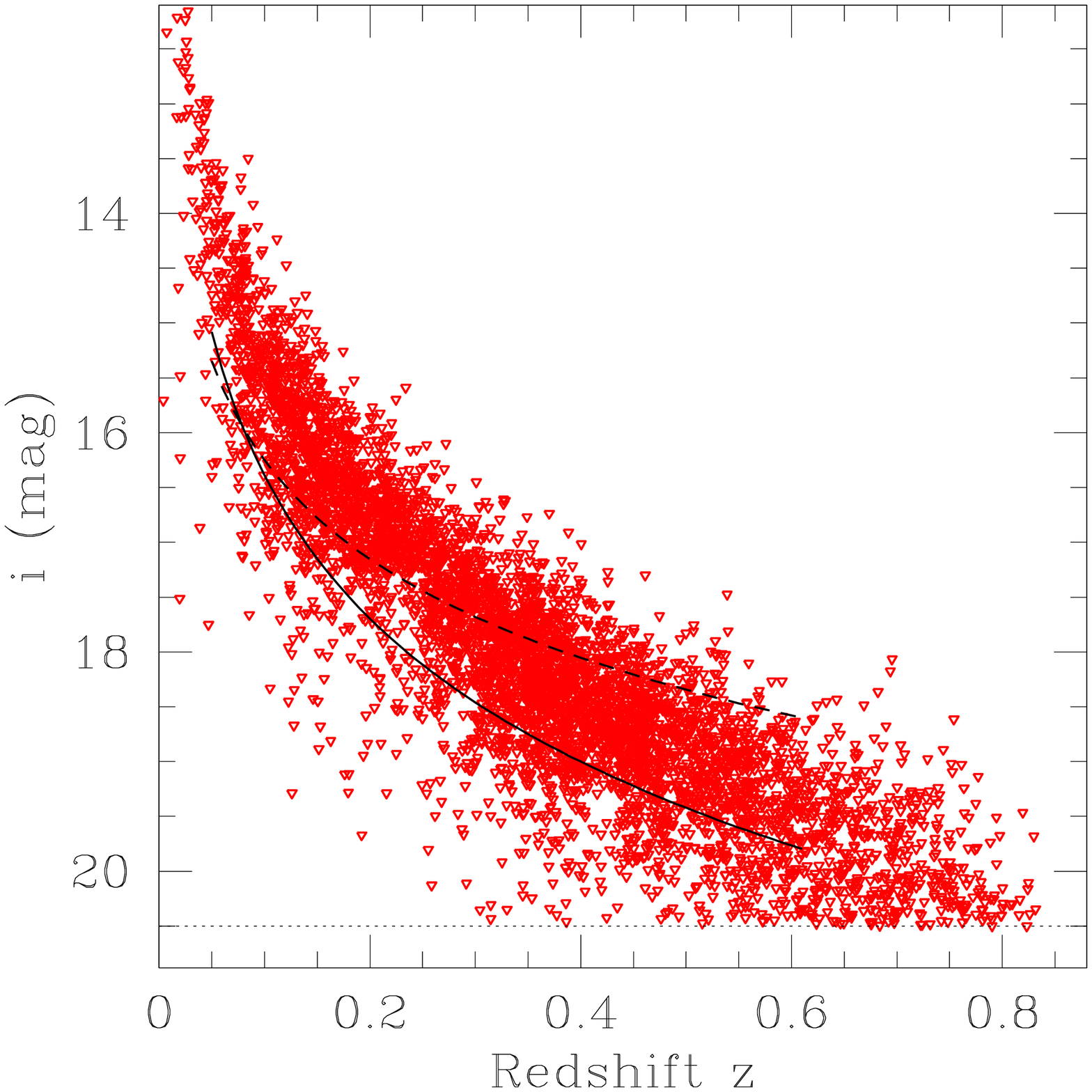}
\caption[]{The $i_{\rm mod}$\ apparent magnitudes of the host galaxies of LERGs (red points). For comparison with the HERG population, the solid and dashed lines are the same least-squares fits to the HERG and AeB data points shown in Figure \ref{fig:zmag}. The horizontal dotted line again shows the optical limit of the LARGESS catalogue at  $i_{\rm mod}=20.5$\,mag. } 
\label{fig:zmag4}
\end{figure}

At $z=0.4$, the AeB objects are typically 0.95 mag brighter than the HERGs in the SDSS $i$\ band. If both classes are drawn from the same parent galaxy population, this would imply that the AGN contributes around 60\% of the $i$-band light in the AeB objects at $z\sim0.4$. At higher redshifts than this, a direct comparison with the HERGs becomes more difficult because lower-luminosity HERGs will be excluded by the optical magnitude cut of our catalogue, but it is clear that a class of very optically-bright ($i_{\rm mod}<18$\,mag) AeB objects starts to appear at $z>0.5$. 

The relative fractions of HERGs and AeB objects are similar across the whole redshift range sampled by our survey ($0.1<z<0.8$), with 20--30\% of the combined (HERG$+$AeB) population being AeB objects and the remaining 70--80\% HERGs. This is broadly consistent with unified models in which HERGs and AeB radio sources are members of the same underlying galaxy population seen at different orientations. 

From a comparison of Figures \ref{fig:zmag} and \ref{fig:zmag4}, we can see  that LERG host galaxies are significantly brighter (by $\sim$0.5\,mag on average in $i$) than HERG hosts across the full redshift range, in agreement with the findings of earlier studies {\citep[e.g.][]{smolcic09a,best12}}. 

\subsection{What is missed by colour-selected Luminous Red Galaxy samples at redshift $0.4<z<0.8$?}  

We can now answer one of the questions raised in the introduction to this paper -- did earlier, colour-selected studies of radio AGN in the distant Universe \citep{sadler07,donoso09} miss a significant population of `blue' radio galaxies? 

The 2SLAQ Luminous Red Galaxy (LRG) survey \citep{cannon06}\ was designed for cosmological studies of  large-scale structure at intermediate redshifts ($0.4<z<0.8$), and photometrically pre-selected to contain luminous ($>2$L$_*$) galaxies with red colours characteristic of an old stellar population. Massive red galaxies of this kind are the hosts of almost all radio AGN in the local Universe \citep[e.g.][]{best05,mauch07}, and \cite{sadler07} used spectroscopy from the 2SLAQ LRG survey to compile one of the largest and most uniform spectroscopic samples of radio AGN beyond the local Universe --- providing strong evidence for the cosmic evolution of low-power radio galaxies. \cite{sadler07} did however note that the rate of cosmic evolution measured for low-power radio galaxies in their study was a lower limit, since any additional population of blue radio galaxies at redshift $z>0.4$ might be removed by the 2SLAQ LRG colour selection and so missed from their analysis. 

%Figure 19
\begin{figure}
\centering
\includegraphics[width=0.5\textwidth]{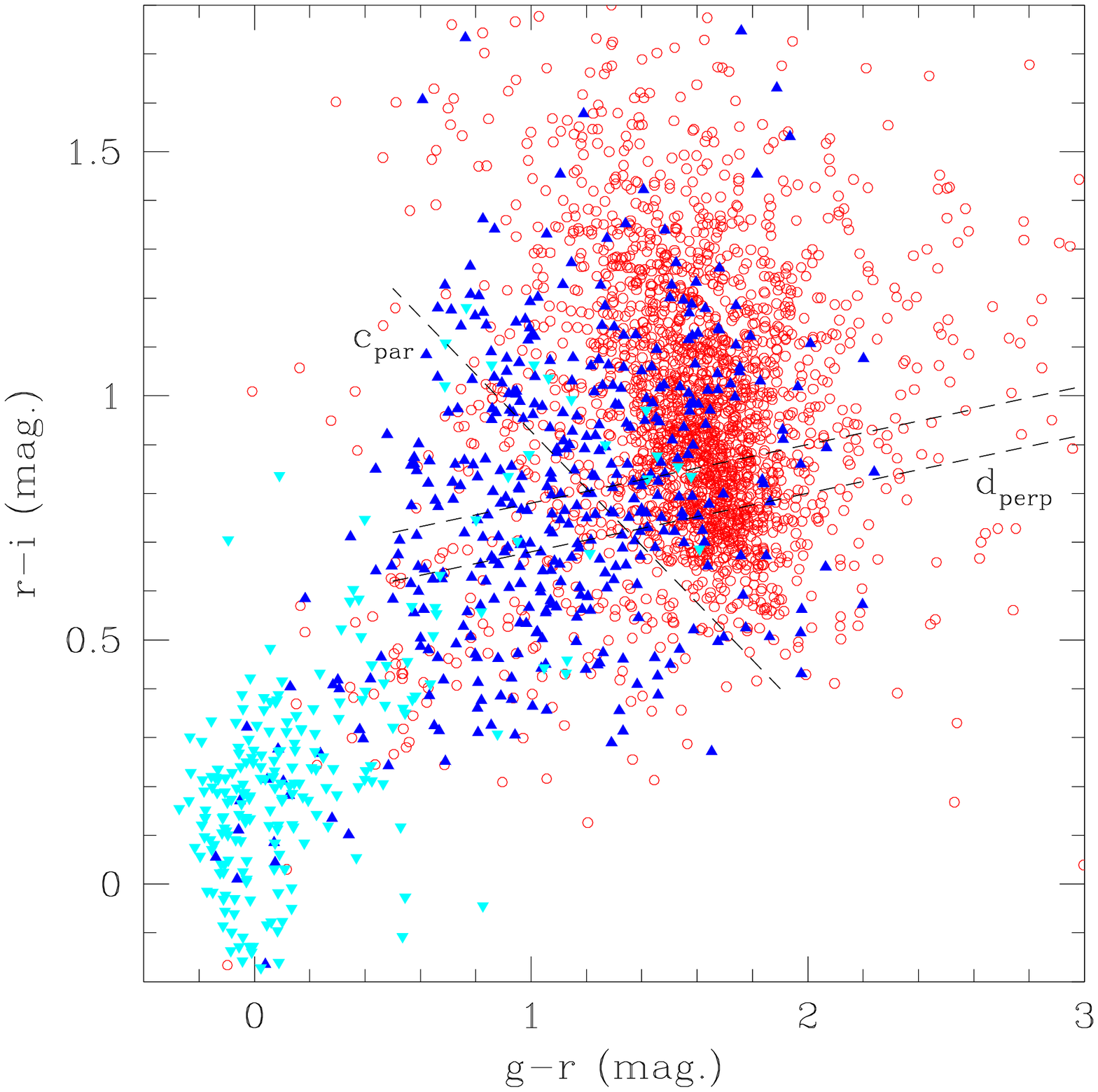}
\caption[]{Optical two-colour plot for radio AGN in the LARGESS sample with spectroscopic redshifts in the range $0.4<z<0.8$.  As in earlier plots, LERGs are shown by red dots, HERGs by dark blue triangles and AeB objects by cyan triangles. The two diagonal lines marked C$_{\rm par}$ and d$_{\rm perp}$ are the limits used by \cite{cannon06} (based on earlier work  by \cite{eisenstein01}) to select luminous red galaxies (LRGs) in this redshift range. }
\label{fig:phot3}
\end{figure}

Figure \ref{fig:phot3} shows an optical two-colour diagram for LARGESS radio AGN in the redshift range $0.4<z<0.8$. The colour cuts used by \cite{cannon06} to select the Luminous Red Galaxies (LRGs) used as the basis of the \cite{sadler07} radio sample are shown by the  dashed lines marked C$_{\rm par}$ and d$_{\rm perp}$. The first of these (C$_{\rm par}$) divides star-forming galaxies from galaxies with old, passively-evolving stellar populations \citep{eisenstein01}, while the two lines marked d$_{\rm perp}$ select out galaxies above a fixed limit in stellar mass. 

There are several points to note in Figure \ref{fig:phot3}. 
As expected (see e.g. Figure \ref{fig:giVz}), the LARGESS AeB (QSO) objects (cyan points) have very blue colours and fall well to the left of the C$_{\rm par}$ line. 
Most of the LERGs in the LARGESS sample (red points in Figure \ref{fig:phot3}) fall above the C$_{\rm par}$ and D$_{\rm perp}$ lines, and so would have been included in the 2SLAQ LRG sample. However, the LARGESS sample has a fainter optical cut-off ($i<20.5$\,mag) than the 2SLAQ LRG sample  ($i<19.8$\,mag), and so the LARGESS LERG distribution extends to lower-mass galaxies that fall below the 2SLAQ D$_{\rm perp}$ lines. 

In contrast to the LERGs, only about half the HERGs in the LARGESS sample (dark blue points) lie to the right of the C$_{\rm par}$ line. It is clear that while some HERGS at $0.4<z<0.8$ are hosted by luminous  red galaxies (3--4\% of the \cite{sadler07} LRGs showed high-excitation [NeIII] and [NeV] lines in their optical spectra), a significant number are blue enough to be excluded by the 2SLAQ LRG colour cut.  

As can be seen from Table \ref{tab:2slaq}, over 90\% of LERGs at $0.4<z<0.8$ satisfy the 2SLAQ C$_{\rm par}$ colour cut, but more than half of the 
HERG objects would be excluded. The overall completeness of the sample is only mildly affected (only 16\% of all radio AGN in this redshift range would be excluded by the C$_{\rm par}$ colour cut), so the \cite{sadler07} and \cite{donoso09} measurements remain a reasonable guide to the AGN radio luminosity function in the distant Universe. The effect is, however, much more dramatic for the HERG population, and it is clear that a substantial fraction of the HERG population ($\sim50$\%) was missed by the 2SLAQ LRG colour cut. 
This underlines the importance of using a colour-unbiased sample to measure the HERG radio luminosity function (as distinct from the overall luminosity function for radio AGN) beyond the local Universe. 

%Table 12
\ctable[
notespar,
caption = {Photometric properties of LARGESS HERG and LERG host galaxies at redshift $0.4<z<0.8$},
label = {tab:2slaq}
]{lrrrr}%
{ %\tnote[a]{footnote}
}{\hline
Class & All & \multicolumn{1}{c}{C$_{\rm par}\geq1.6$}  & \multicolumn{1}{c}{C$_{\rm par}<1.6$} \\
          &     & \multicolumn{1}{c}{(`passive')}  & \multicolumn{1}{c}{(`star-forming')} \\
\hline
LERG & 2236  & 2021 & 215 \\ 
HERG &  413  &   203 & 210 \\ 
Total   &  2649 & 2224 & 425 \\ 
\hline
HERG fraction & 16\% & 9\% & 49\%  
\LL
}

\subsection{A matched sample of HERG and LERG host galaxies} 
We now turn to a comparison of the host-galaxy properties of HERGs and LERGs across the redshift range spanned by the LARGESS sample. To do this, we need to take into account the likelihood that the host galaxies of these two classes span different ranges in stellar mass. 

In the following analysis, we will therefore compare the properties of samples of LERGs and HERGs that have been carefully matched in stellar mass, redshift and radio luminosity. 
We will also split these samples into two coarse redshift bins at $0.01<z\leq0.3$ and $z>0.3$. The lower-redshift bin is well matched to the median redshift of the sample used in \cite{best12}, while the higher-redshift bin allows us to see whether there is any evidence for redshift evolution in the properties of LERG and HERG host galaxies. The galaxy stellar masses used in this analysis were derived from rest-frame $(g-i)$ colours and $i$-band luminosities using the relationship defined by \cite{taylor11}. 

The LERGs in our sample greatly outnumber the HERGs, so for each HERG we identified three LERGs matched in stellar mass ($|\Delta \log M_{\star}| < 0.1$ dex), redshift ($|\Delta z| < 0.01$) and radio luminosity ($|\Delta \log L_{\rm NVSS}| < 0.25$ dex) to act as control galaxies for the corresponding HERG. Once a LERG has been assigned as a control galaxy, it is removed from subsequent matches so that the final list of controls contain a unique list of LERGs. HERGs are not used if we are not able to find three control LERGs. These limits ensure that we have a fair comparison between the properties of LERGs and HERGs, without being biased by other known differences which may influence the properties that we are interested in. We repeated the matching process to create 100 realisations so that our results are more robust. This approach is similar to the one used by \cite{best12}. 

%Table 13
\ctable[
%star,
notespar,
cap = {KS-test probability of rejecting the null hypothesis that the properties of the HERG are the same as a matched sample of LERGs},
caption = {Table of KS-test probability of rejecting the null hypothesis that the LERG and HERG matched sample are the same for each parameter. The values are the median from the 100 realisations.},
label = {tab:HERGLERG_dn}
]{l rr}%
{ %\tnote[a]{footnote}
}{\FL
Parameter & $0.01<z\leq0.3$ & $z>0.3$\\
\hline\hline
\multicolumn{3}{c}{ {\bf (i) Parameters used to match the samples} } \\
Redshift ($z$) & $<1\%$ & $<1\%$ \\
Stellar mass ($M_{\star}$) & 69\% & 12\% \\
Radio luminosity ($L_{\rm NVSS}$) & $<1\%$ & 6\% \\
\multicolumn{3}{c}{\bf (ii) Parameters not used for matching}  \\
Rest-frame $(g_{\rm mod}-i_{\rm mod})_{0}$ colour & $>99\%$ & $>99\%$ \\
Size ($R_{50}$) & 41\% & 86\% \\
Concentration index ($C$) & $>99\%$ & $>99\%$
\LL
}

Table \ref{tab:HERGLERG_dn} shows the median KS-test probability from the 100 realisations. These are the probability of rejecting the null hypothesis that the distribution of a property is the same between the HERG and LERG matched samples. Table \ref{tab:HERGLERG_dn} confirms that there is no significant difference between the HERG and LERG samples for the three parameters (stellar mass, redshift and radio luminosity) used to match them. 

\subsubsection{Comparing LERG and HERG host galaxies of similar stellar mass: optical properties}

\begin{figure*}
\begin{minipage}\textwidth
\includegraphics[width=0.49\textwidth]{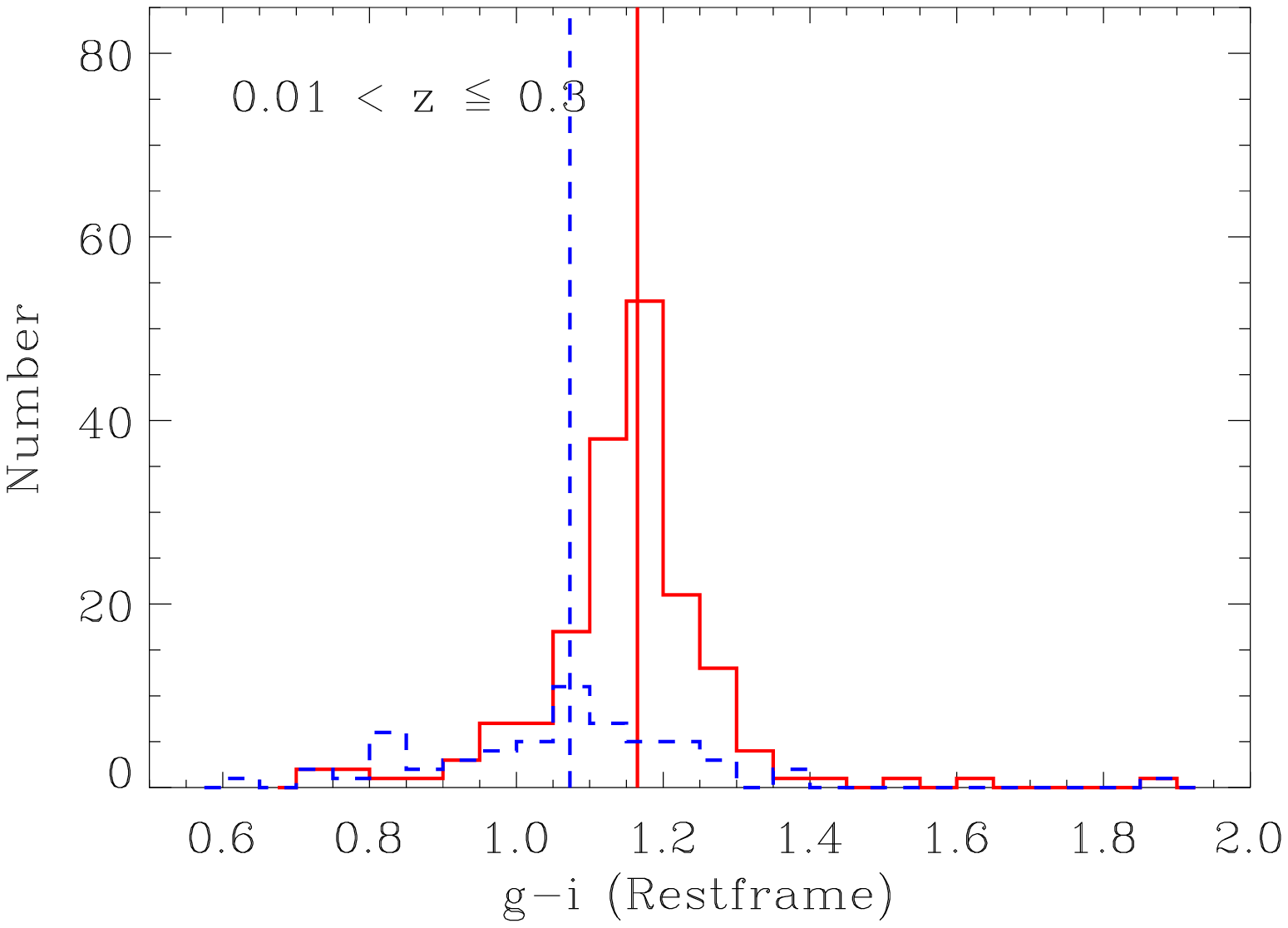}
\includegraphics[width=0.49\textwidth]{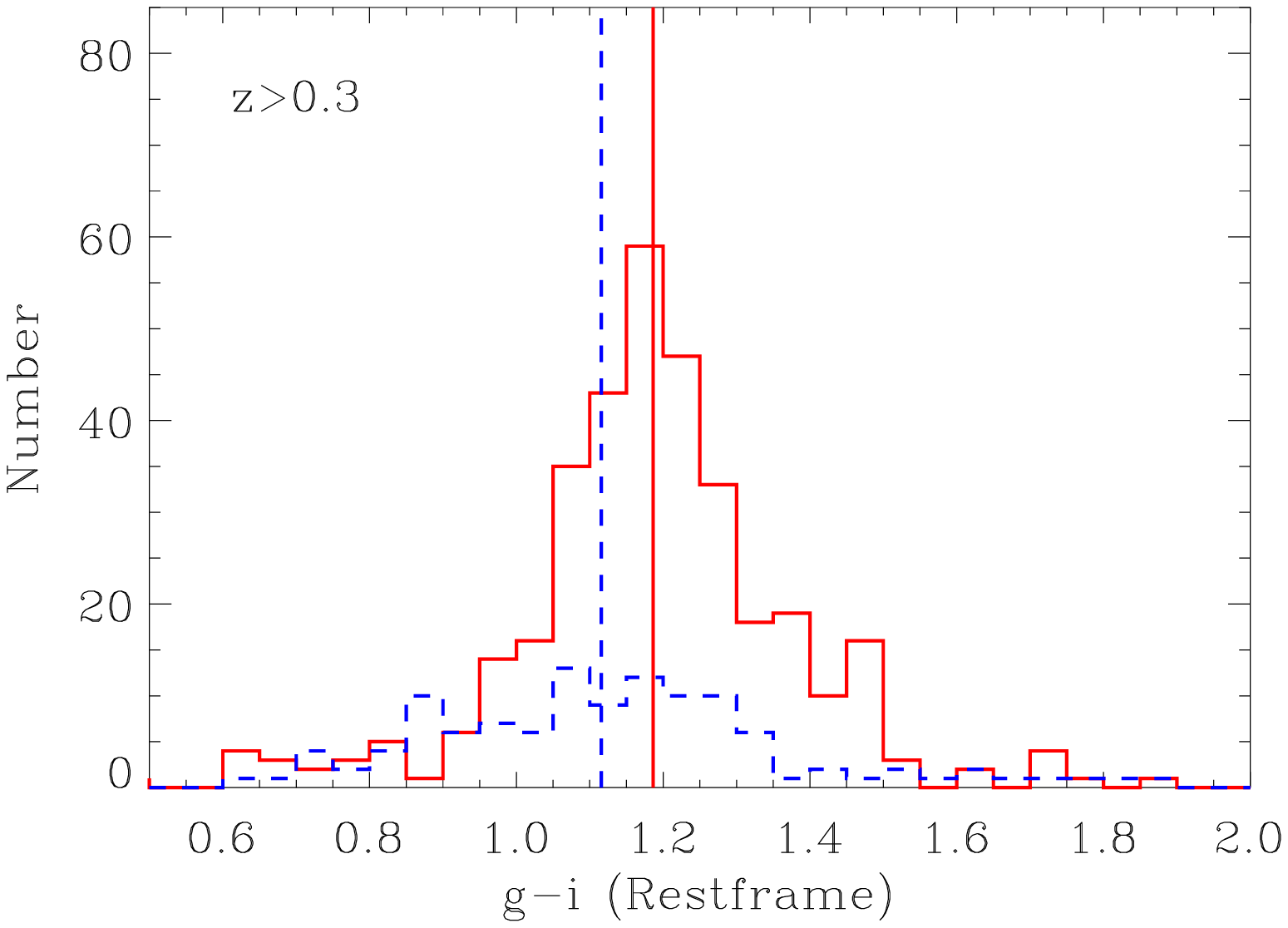}
\end{minipage}
\begin{minipage}\textwidth
\includegraphics[width=0.49\textwidth]{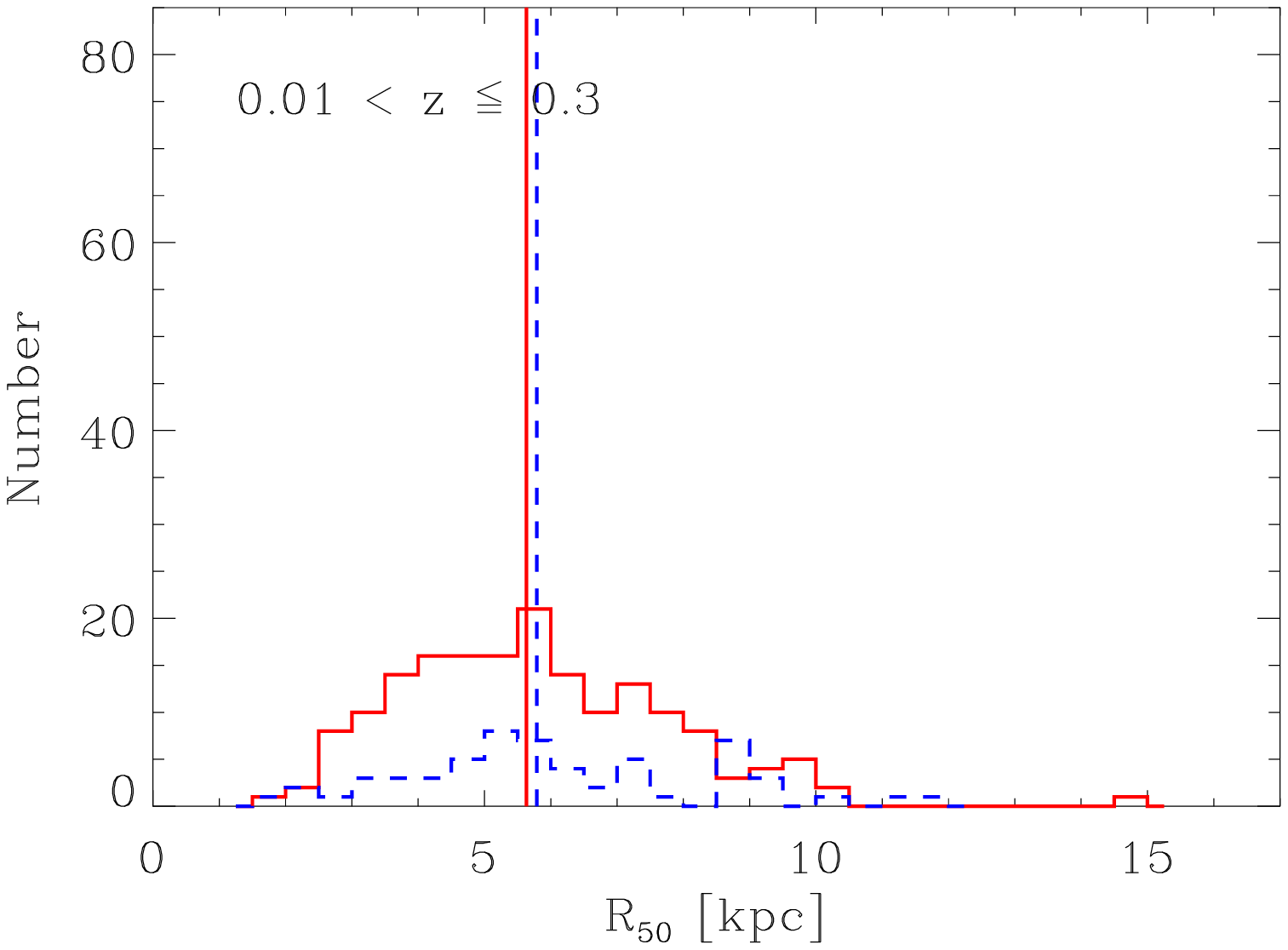}
\includegraphics[width=0.49\textwidth]{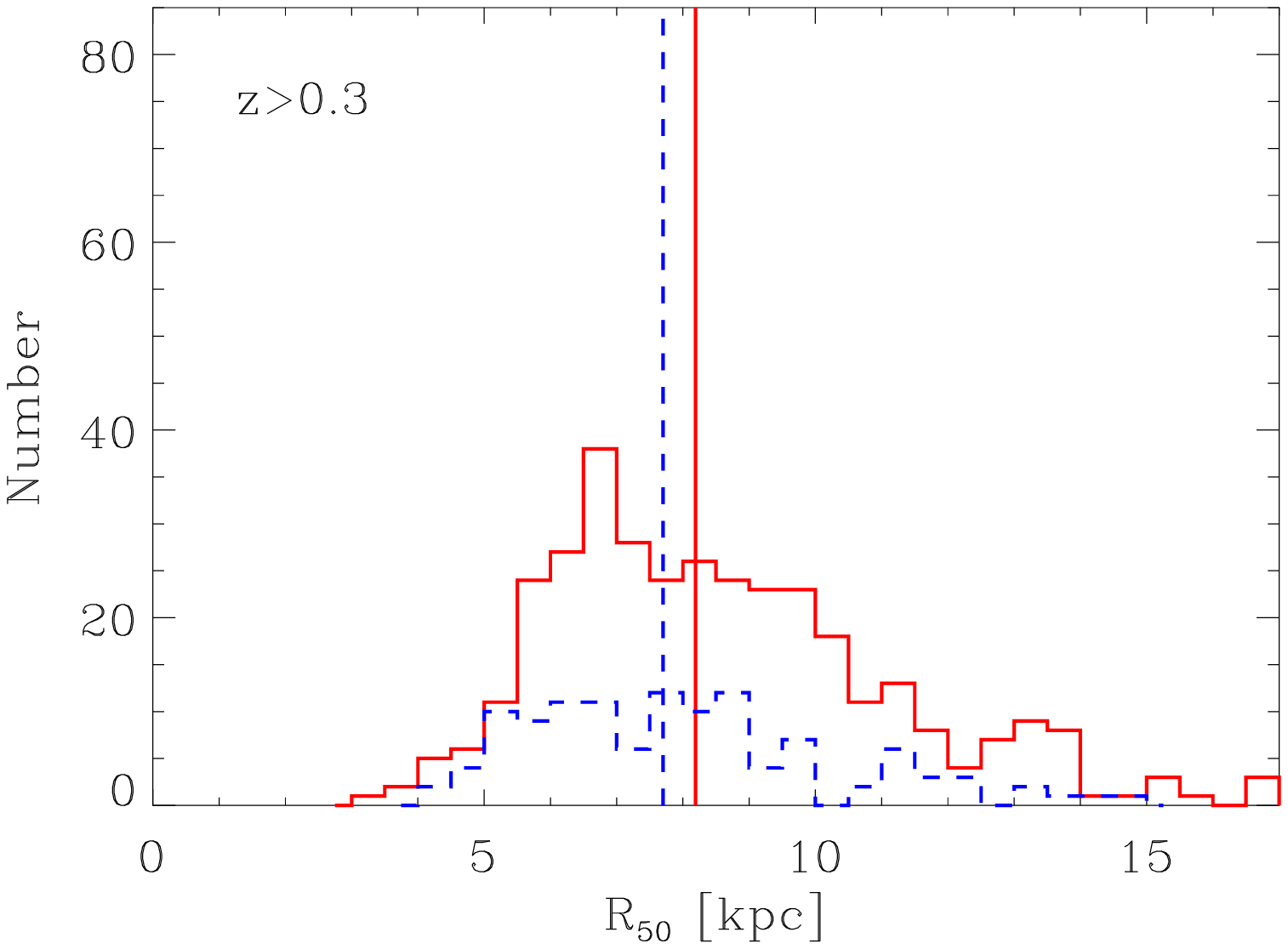}
\end{minipage}
\begin{minipage}\textwidth
\includegraphics[width=0.49\textwidth]{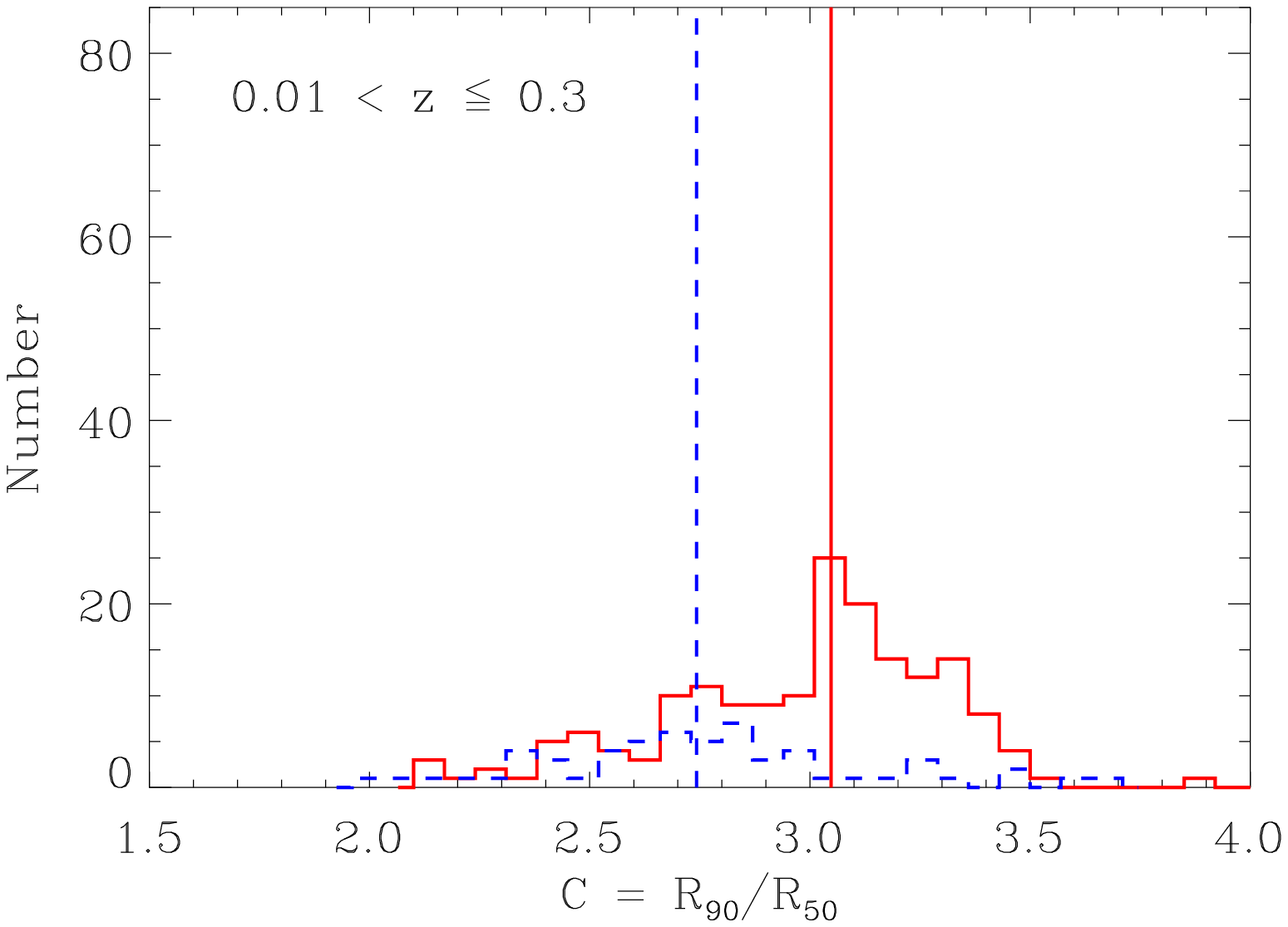}
\includegraphics[width=0.49\textwidth]{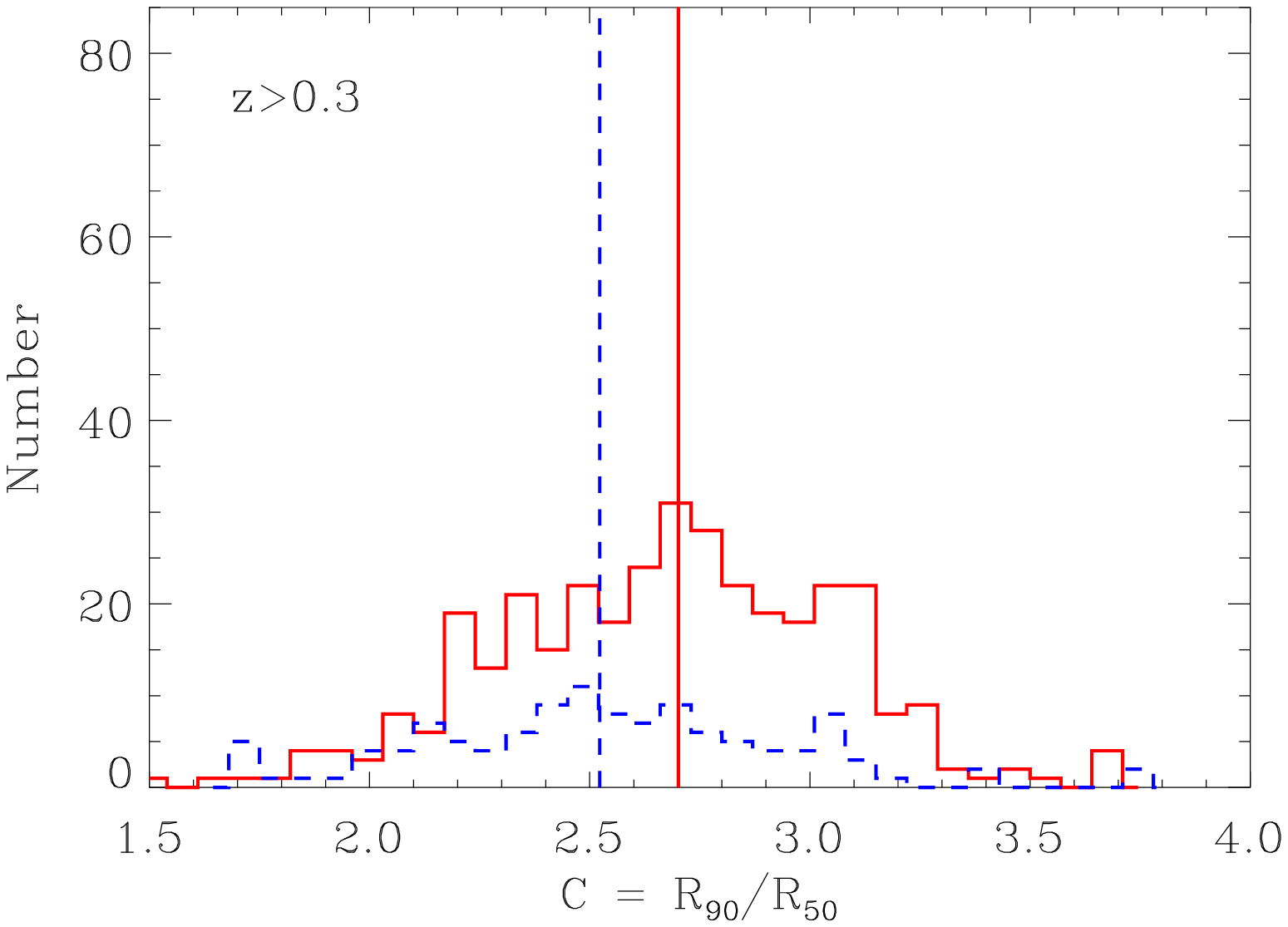}
\end{minipage}

\caption[Distribution of properties between matched sample of LERGs and HERGs]{Comparison of properties for matched samples of LERGs (red solid line) and HERGs (blue dashed line). The samples have been selected to have the same distribution in redshift, radio luminosity and stellar mass. From top to bottom, the plots show: (i) the rest-frame $(g_{\rm mod}-i_{\rm mod})_{0}$ colour, (ii) the radius containing 50\% of the galaxy light ($R_{50}$; in physical units), and (iii) the concentration index ($C =R_{90}/R_{50}$, where $R_{90}$ is the radius containing 90\% of the galaxy light). The left-hand plots show the results for objects with redshift $0.01<z\leq0.3$, and the right-hand column row is for $z>0.3$. Vertical lines in each plot indicate the median values of the distributions.}  
\label{fig:HERGLERG_dn}
\end{figure*}

Figure \ref{fig:HERGLERG_dn} shows the distributions of redshift, radio luminosity, stellar mass, $(g_{\rm mod}-i_{\rm mod})_{0}$ rest-frame colour, linear projected half light radius ($R_{50}$, calculated using the half-light Petrosian angular size in the $r$-band $r_{\rm pet,50}$ and the angular diameter distance equation) and the concentration index ($C = r_{\rm pet,90}/r_{\rm pet,50}$, where $r_{\rm pet,90}$ is the radius containing 90\% of the Petrosian flux), for LERGs (red) and HERGs (blue) in our matched sample. The top six plots are for the low-redshift sample and the lower six plots for the higher-redshift sample. 

%Figure 20 

For the three parameters that were not matched (colour, concentration index and linear size), we find that LERGs have slightly redder $(g-i)_{0}$ colour and a slightly higher concentration index than the HERGs. Although the differences are small, they are significant at the $>99\%$ confidence level, and are similar ($\sim0.1$\,mag in rest-frame $g-i$ colour) in both redshift bins. We interpret these differences as evidence that LERGs are dominated by an older stellar population, and hosted by a higher fraction of elliptical galaxies (in the morphological sense), than HERGs of similar stellar mass. The linear sizes of HERG and LERG host galaxies were not significantly different in either redshift bin. The results shown in Figure \ref{fig:HERGLERG_dn} imply that since the colour difference between the host galaxies of HERGs and LERGs persists when we compare galaxies that are matched in stellar mass, it is unlikely to be caused by metallicity differences and so must arise from differences in star-formation rate (or possibly the presence of non-stellar continuum light in some HERG hosts).

A colour difference between HERG and LERG host galaxies has also been seen in earlier studies. \cite{smolcic09} found  a dichotomy in many of the optical properties of weak and powerful radio AGN both in the local Universe and out to redshifts as high as $z\sim1.3$, while  \cite{best12} \citep[but see also][]{janssen12} found similar results for sample of LERGs and HERGs with $z<0.3$. 

In the picture developed from earlier X-ray studies \cite[]{allen06,evans06,hardcastle06,hardcastle07} the central AGN in HERGs and LERGs accrete gas at different rates, with HERGs being fuelled at relatively high rates by cold gas (some of which may be available for associated star formation) while the LERGs are fuelled at a much lower rate through radiatively-inefficient accretion flows - most likely from a surrounding hot gas halo. In the latter case, `radio-mode feedback'  \citep{croton06}, in which mechanical energy from the radio jet is energetically coupled  to the surrounding hot gas, operates in the LERG galaxies to keep the star formation rate low.  Also, because elliptical galaxies are typically found in environments (sufficiently massive haloes) where cold-mode accretion is less likely to  occur, the higher fraction of elliptical for LERGs again supports the idea that they are accreting in the hot-mode. 
 
Our results are in broad agreement with this picture, though we note that in some HERGs the bluer colours may be due at least in part to the presence of a non-stellar AGN continuum rather than ongoing star formation, as discussed earlier in \S9.3.

\subsubsection{Comparing LERG and HERG host galaxies of similar stellar mass: WISE mid-IR data} 

\begin{figure*}
\begin{minipage}\textwidth
\includegraphics[width=0.49\textwidth]{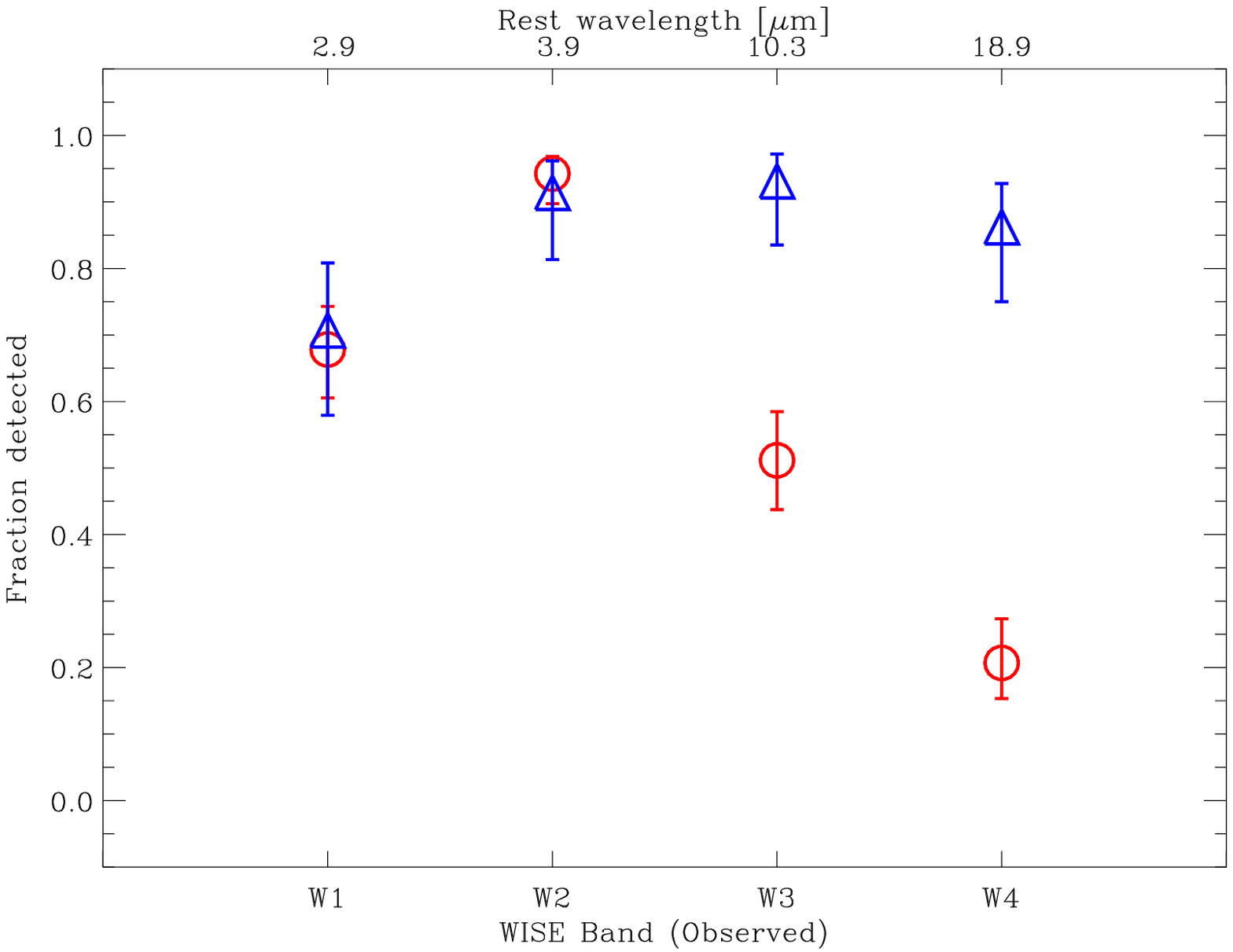}
\includegraphics[width=0.49\textwidth]{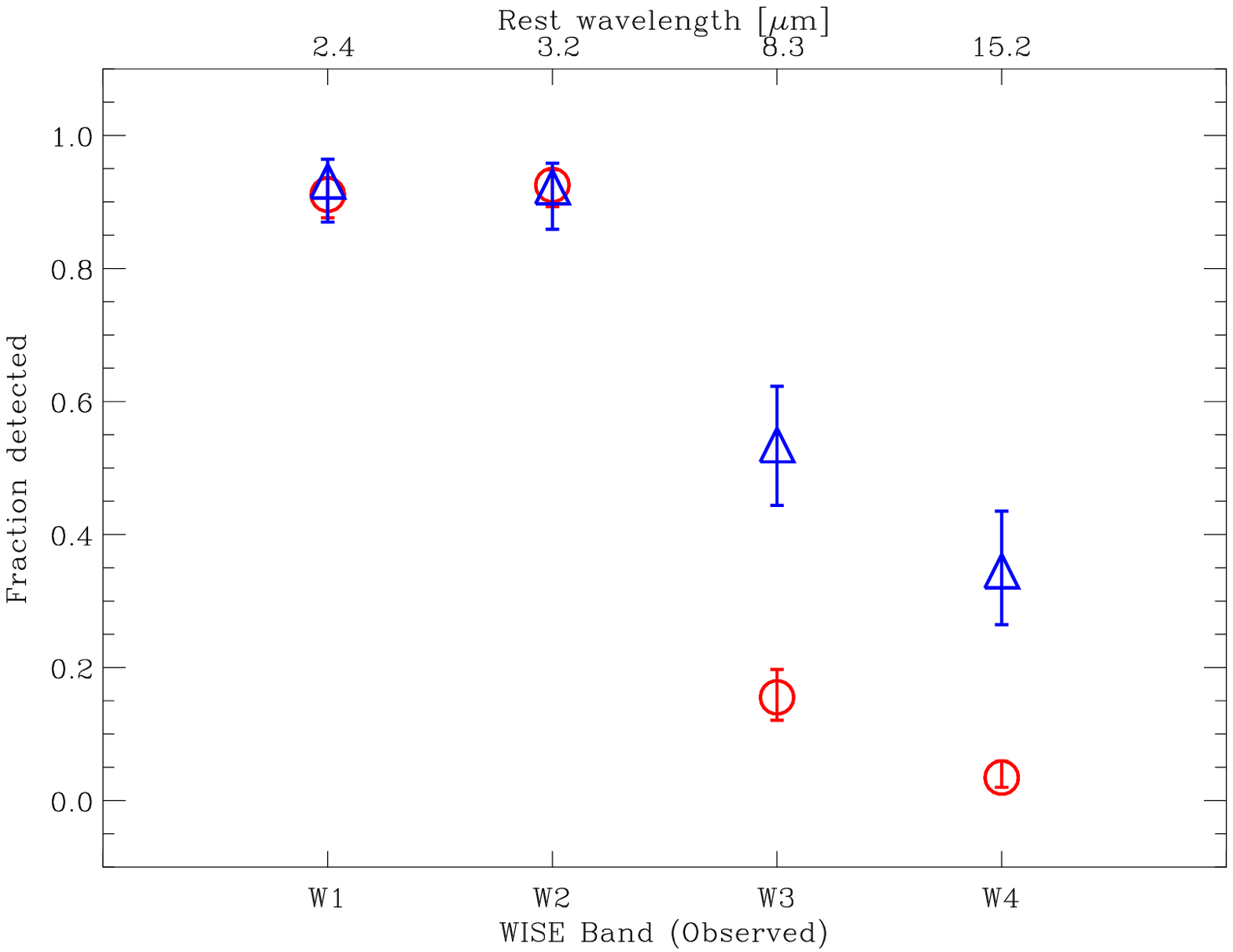}
\end{minipage}
\caption[Fraction with a reliable WISE photometry in each band for matched sample of LERGs and HERGs]{Fraction detected in each WISE band for HERGs (blue triangles) and LERGs (red circles) that are matched in redshift, radio luminosity and stellar mass. Left panel is for those with $0.01<z\leq0.3$ and right panel are for those with $z>0.3$.}
\label{fig:HERGLERG_wisedetection}
\end{figure*}

Figure \ref{fig:HERGLERG_wisedetection} shows the fraction of HERGs (blue) and LERGs (red) from the matched sample with a reliable detection in each WISE band. As in \S10.3.1, the two samples were matched in stellar mass, redshift and radio luminosity.  The equivalent rest-frame wavelengths for each band, calculated from the median redshift, are also shown on the top axis. Once again these plots are for a single realisation of our Monte Carlo simulations. The errors are estimated assuming a Beta distribution adopting a 95\% confidence interval \citep{cameron11}. 

There are large differences in the fraction detected in the W3 and W4 bands in both redshift bins, in contrast to the detection rates in the W1 and W2 bands. We use a $Z$-test to quantify the significance between two fractions. Since the errors are asymmetric, we assume the pooled error is the quadrature sum of the lower error in the larger fraction and the upper error in the smaller fraction. The null hypothesis is that the two fractions are the same (i.e. a two-tailed test). The median values of the $Z$-test from the 100 iterations are listed in Table \ref{tab:HERGLERG_wisedetection}.

From the values in Table \ref{tab:HERGLERG_wisedetection}, we conclude that the detection fractions detected in the W1 and W2 bands are not significantly different 
for HERGs and LERGs. As mentioned earlier, the emission in the WISE W1 and W2 bands arises mainly from the old stellar population, so is expected to correlate with galaxy stellar mass \citep[e.g.][]{hwang12,wen13}. Since this is one of the parameters used to match the the two samples, we would not expect to see a difference in these two bands. 

%Figure 21

When we turn to the W3 and W4 bands, we find that HERGs are significantly ($\sim3\sigma$ in W3 and $\sim4\sigma$ in W4) more likely to have a reliable detection than LERGs in both bands. This difference is seen in both redshift bins. 

The mid-infrared emission in the WISE W3 and W4 bands is most likely reprocessed emission from a starburst (W3 band in particular) and/or an optical-AGN (W4 band in particular) \citep[e.g.][]{donoso12,hwang12}. The fact that we see differences in both the W3 and W4 detection rates is consistent with the presence of a radiatively-efficient accretion disk and dusty torus in HERGs, which is absent in LERGs. \cite{hardcastle06} found similar results in their study of X-ray properties in LERGs and HERGs. They found that HERGs have a nuclear X-ray absorption component, which was rarely observed in the LERGs, and noted that the absence of such a component in LERGs is consistent with these objects lacking a radiatively-efficient accretion disk.

%Table 14
\ctable[
%star,
notespar,
cap = {The median $\sigma$ significance when comparing the fraction of LERGs and HERGs that have a reliable WISE detection in different bands},
caption = {The median $\sigma$ significance when comparing the fraction of LERGs and HERGs that have a reliable WISE detection in different bands.},
label = {tab:HERGLERG_wisedetection}
]{lcc}%
{ %\tnote[a]{footnote}
}{\FL
 & \multicolumn{2}{c}{$\sigma$ significance}\\
 \cline{2-3}
WISE band & $0.01<z\leq0.3$ & $z>0.3$\\
\hline\hline
W1 & 0.2 & 0.3\\
W2 & 0.5 & 0.1\\
W3 & 3.3 & 3.9\\
W4 & 5.0 & 3.8
\LL
}

\section{Summary}\label{sec:summary1}

We have compiled the Large Area Radio Galaxy Evolution Spectroscopic Survey (LARGESS) catalogue, a new, large dataset designed to study the evolution and environments of high- and low- excitation radio galaxies (HERGs and LERGs) out to $z\sim0.8$. The catalogue contains 19,179 radio sources detected by the FIRST survey that have an SDSS optical counterpart with $i_{\rm mod} < 20.5$.  We have obtained new good-quality optical spectra for over 5,000 LARGESS objects, and adding these to lower-redshift archival data means that 10,856 LARGESS objects currently have a reliable optical redshift. The spectroscopic completeness varies across the survey area, and is highest ($\sim90$\%) in the three equatorial GAMA fields.  The median redshift of radio AGN in the LARGESS catalogue is $z\sim0.44$, and the bright end of the radio luminosity function (P$_{1.4}> 10^{24}$\,W\,Hz$^{-1}$) is well-sampled for all the main radio AGN populations over the redshift range $0<z<0.8$. The catalogue also contains a minority population of nearby ($z<0.3$) star-forming galaxies. 

The great majority of LARGESS radio sources are compact (only 10\% have complex or multi-component radio structures on scales larger than a few arcsec), which makes the optical matching process straightforward in most cases. We estimate that our final matched sample is 95\% complete and has 94\% reliability. 

The LARGESS sources typically have between 5\% and 25\% of their 1.4\,GHz radio emission in a diffuse component that is seen by the NVSS survey but missed by the higher-resolution FIRST images. This needs to be taken into account in any subsequent analysis. In particular, as discussed in \S4.3, any analysis requiring accurate flux densities for extended radio sources should impose a $S^{\rm FIRST}_{\rm tot} \geq 3.5$ mJy limit for the LARGESS catalogue. 

For objects with a reliable redshift from the SDSS, GAMA and WiggleZ surveys, we made both a visual and an automated classification of the spectrum. These two classification schemes were combined to provide a single classification, and allow us to distinguish between (i) Galactic stars, (ii) radio galaxies with weak or no emission lines (LERGs), (iii) galaxies with emission lines or radio emission generated from star formation (SF), and (iv) radio galaxies where both emission line and radio emission arise from the AGN (broad emission lines, AeB and narrow emission lines, HERGs).

While the optical counterparts of most radio AGN are massive red galaxies at all redshifts out to $z=0.8$, we find that at least 15\% of radio AGN at $0.4<z<0.8$ are hosted by blue galaxies which appear likely to have ongoing star formation. It is notable that at least half of the HERG population in this redshift range are `blue radio galaxies' rather than passively-evolving red galaxies. Roughly 8\% of the radio AGN population in the same redshift range are radio-loud QSOs with broad Balmer emission lines in their optical spectra. 

We find that LERGs, HERGs and AeB (QSO) objects have very different broad-band spectral-energy distributions (SEDs). The optical and mid-IR properties of LERGs are generally similar to those of luminous red galaxies dominated by an old stellar population, while the AeB (QSO) objects are consistent with a power-law SED. In contrast, the HERGs are a much more heterogeneous population in terms of both their optical and mid-IR colours. 

We used a matched sample of HERGs and LERGs to compare their physical properties, and found that that HERGs have bluer colours (probably indicative of a younger stellar population) compared to LERGs of similar mass, redshift, and radio luminosity. The concentration index is also higher for LERGs than HERGs, suggesting  that LERGs are  hosted by a higher fraction of elliptical galaxies than the HERGs. Both of these results are independent of redshift. The matched sample of LERGs and HERGs show no significant difference in the linear sizes in any redshift bin. These results confirm results of previous studies, but also extend them into a higher redshift range.

By comparing the fraction with a reliable detection in various WISE mid-infrared bands, we found that HERGs are significantly more likely to be detected in the W3 and W4 bands than a matched sample of LERGs, both locally and at higher redshifts.. From this, we inferred the presence of a radiatively efficient accretion disk and dusty torus in the HERGs, which is absent in the LERGs. This is consistent with theories that suggest a dichotomy in the accretion mode of HERGs and LERGs. 

\vspace{1cm}
\noindent{\bf ACKNOWLEDGEMENTS}

\noindent
JHYC would like to acknowledge the funding provided by the University of Sydney via an Australian Postgraduate Award, the Postgraduate Research Scholarship Scheme and the William \& Catherine McIlrath Scholarship. JHYC would also like to acknowledge the support provided by the Astronomical Society of Australia. 

EMS and SMC acknowledge the financial support of the Australian Research Council through Discovery Project grants DP1093086 and DP 130103198. SMC acknowledges the support of an Australian Research Council Future Fellowship (FT100100457). We thank Philip Best and Dick Hunstead for helpful comments.

GAMA is a joint European-Australasian project based around a spectroscopic campaign using the Anglo-Australian Telescope. The GAMA input catalogue is based on data taken from the Sloan Digital Sky Survey and the UKIRT Infrared Deep Sky Survey. Complementary imaging of the GAMA regions is being obtained by a number of independent survey programs including GALEX MIS, VST KIDS, VISTA VIKING, WISE, Herschel-ATLAS, GMRT and ASKAP providing UV to radio coverage. GAMA is funded by the STFC (UK), the ARC (Australia), the AAO, and the participating institutions. The GAMA website is http://www.gama-survey.org/.

Funding for the SDSS and SDSS-II has been provided by the Alfred P. Sloan Foundation, the Participating Institutions, the National Science Foundation, the U.S. Department of Energy, the National Aeronautics and Space Administration, the Japanese Monbukagakusho, the Max Planck Society, and the Higher Education Funding Council for England. The SDSS Web Site is http://www.sdss.org/.

The SDSS is managed by the Astrophysical Research Consortium for the Participating Institutions. The Participating Institutions are the American Museum of Natural History, Astrophysical Institute Potsdam, University of Basel, University of Cambridge, Case Western Reserve University, University of Chicago, Drexel University, Fermilab, the Institute for Advanced Study, the Japan Participation Group, Johns Hopkins University, the Joint Institute for Nuclear Astrophysics, the Kavli Institute for Particle Astrophysics and Cosmology, the Korean Scientist Group, the Chinese Academy of Sciences (LAMOST), Los Alamos National Laboratory, the Max-Planck-Institute for Astronomy (MPIA), the Max- Planck-Institute for Astrophysics (MPA), New Mexico State University, Ohio State University, University of Pittsburgh, University of Portsmouth, Princeton University, the United States Naval Observatory, and the University of Washington.

This publication makes use of data products from the Wide-field Infrared Survey Explorer, which is a joint project of the University of California, Los Angeles, and the Jet Propulsion Laboratory/California Institute of Technology, funded by the National Aeronautics and Space Administration.

This research has made use of data obtained from or software provided by the US National Virtual Observatory, which is sponsored by the National Science Foundation.

\bsp

\bibliographystyle{mn2e}
\bibliography{references}

%%%%%%%%

\label{lastpage}

\end{document}